\pdfoutput=1
\documentclass[11pt,twoside,a4paper,cmspaper,final,collab]{cms-tdr}
\def\svnVersion{cfcb7bd}\def\svnDate{2023/06/13}\def\cmsCernNoTag{CERN-EP-2022-120}\def\cmsCernDate{\today}\def\cmsMessage{Published in the European Physical Journal C as \href{http://dx.doi.org/10.1140/epjc/s10052-023-11632-6}{\doi{10.1140/epjc/s10052-023-11632-6}.}}
\begin{document}\cmsNoteHeader{HIG-20-013}

\newcommand{\tautau}{\ensuremath{\PGt\PGt}\xspace}
\newcommand{\bbH}{\ensuremath{\PQb\PAQb\PH}\xspace}
\newcommand{\hww}{\ensuremath{\PH\to\PW\PW}\xspace}
\newcommand{\ggH}{\ensuremath{\Pg\Pg\PH}\xspace}
\newcommand{\ttH}{\ensuremath{\ttbar\PH}\xspace}
\newcommand{\VH}{\ensuremath{\PV\PH}\xspace}
\newcommand{\WW}{\ensuremath{\PW\PW}\xspace}
\newcommand{\WZ}{\ensuremath{\PW\PZ}\xspace}
\newcommand{\ZZ}{\ensuremath{\PZ\PZ}\xspace}
\newcommand{\ZH}{\ensuremath{\PZ\PH}\xspace}
\newcommand{\WH}{\ensuremath{\PW\PH}\xspace}
\newcommand{\wg}{\ensuremath{\PW\PGg}\xspace}
\newcommand{\wgs}{\ensuremath{\PW\PGg^{\ast}}\xspace}
\newcommand{\ptH}{\ensuremath{\pt^{\PH}}\xspace}
\newcommand{\njet}{\ensuremath{N_{\text{jet}}}\xspace}
\newcommand{\mTH}{\ensuremath{\mT^{\PH}}\xspace}
\newcommand{\mll}{\ensuremath{m_{\Pell\Pell}}\xspace}
\newcommand{\mjj}{\ensuremath{m_{\text{jj}}}\xspace}
\newcommand{\ptll}{\ensuremath{\pt^{\Pell\Pell}}\xspace}
\newcommand{\dR}{\ensuremath{\Delta R}\xspace}
\newcommand{\dphill}{\ensuremath{\Delta\phi_{\Pell\Pell}}\xspace}
\newcommand{\detajj}{\ensuremath{\Delta\eta_{\text{jj}}}\xspace}
\newcommand{\mHtil}{\ensuremath{\widetilde{m}_{\PH}}\xspace}
\newcommand{\ptone}{\ensuremath{\pt{}_1}\xspace}
\newcommand{\pttwo}{\ensuremath{\pt{}_2}\xspace}
\newcommand{\ptthr}{\ensuremath{\pt{}_3}\xspace}
\newcommand{\expt}{\ensuremath{\,\text{(exp)}}\xspace}
\providecommand{\cmsTable}[1]{\resizebox{\textwidth}{!}{#1}}
\newlength\cmsTabSkip\setlength{\cmsTabSkip}{1ex}
\newlength\cmsFigureWidth\ifthenelse{\boolean{cms@external}}{\setlength{\cmsFigureWidth}{\columnwidth}}{\setlength{\cmsFigureWidth}{0.7\textwidth}}
\hyphenation{re-com-men-ded}
\hyphenation{re-weight-ed}

\cmsNoteHeader{HIG-20-013}

\title{Measurements of the Higgs boson production cross section and couplings in the \texorpdfstring{\PW}{W} boson pair decay channel in proton-proton collisions at \texorpdfstring{$\sqrt{s}=13\TeV$}{sqrt(s) = 13 TeV}}
\titlerunning{Measurement of properties of the Higgs boson production in the \texorpdfstring{\PW}{W} boson pair decay channel}

\date{\today}

\abstract{
   Production cross sections of the standard model Higgs boson decaying to a pair of \PW bosons are measured in proton-proton collisions at a center-of-mass energy of 13\TeV. The analysis targets Higgs bosons produced via gluon fusion, vector boson fusion, and in association with a \PW or \PZ boson. Candidate events are required to have at least two charged leptons and moderate missing transverse momentum, targeting events with at least one leptonically decaying \PW boson originating from the Higgs boson. Results are presented in the form of inclusive and differential cross sections in the simplified template cross section framework, as well as couplings of the Higgs boson to vector bosons and fermions. The data set collected by the CMS detector during 2016--2018 is used, corresponding to an integrated luminosity of 138\fbinv. The signal strength modifier $\mu$, defined as the ratio of the observed production rate in a given decay channel to the standard model expectation, is measured to be $\mu = 0.95^{+0.10}_{-0.09}$. All results are found to be compatible with the standard model within the uncertainties.
}

\hypersetup{pdfauthor={CMS Collaboration},%
pdftitle={Measurements of the Higgs boson production cross section and couplings in the W boson pair decay channel in proton-proton collisions at sqrt(s)=13 TeV},%
pdfsubject={CMS},%
pdfkeywords={CMS, Higgs boson, WW}}

\maketitle 

\section{Introduction}
\label{sec:introduction}	

After the observation of a scalar particle compatible with the standard model (SM) Higgs boson by the ATLAS and CMS Collaborations in 2012~\cite{Aad:2012tfa,Chatrchyan:2012xdj,Chatrchyan:2013lba}, the two experiments have focused on 
performing precision measurements of the properties of the new particle. The large data sample collected at the CERN LHC during the data taking periods through 2018 allowed the 
measurement of the Higgs boson quantum numbers and couplings to other SM particles with an unprecedented level of accuracy~\cite{aad:2016}.
All results reported so far are compatible with the SM within the current uncertainties.

Among all the Higgs boson decay channels predicted by the SM, the one into a pair of \PW bosons has the second largest branching fraction ($\approx$22\%), while benefitting from a lower background with respect to the more probable decay in a pair of \PQb quarks. This combination makes this channel one of the most sensitive for measuring the production cross section of the Higgs boson and its couplings to SM particles. This paper presents the measurement of the Higgs boson properties in the \hww decay channel targeting the gluon fusion (\ggH) and vector boson fusion (VBF) production mechanisms, as well as associated production with a vector boson (\VH, where \PV stands for either a \PW or a \PZ boson). The measurement utilizes final states with at least two charged leptons arising either from the associated vector boson or from the products of the \hww decays. In all cases at least one of the \PW bosons originating from the Higgs boson is required to decay leptonically.

The properties of the Higgs boson are probed by measuring the inclusive cross sections for each production mechanism, as well as the production cross sections in finer phase spaces defined according to the simplified template cross section (STXS) framework~\cite{Berger:2019wnu}. In addition, measurements of the Higgs boson couplings to fermions and vector bosons are presented.

The analysis is based on proton-proton ($\Pp\Pp$) collision data produced at the LHC at $\sqrt{s}=13\TeV$ and collected by the CMS detector during 2016--2018, for a total integrated luminosity of about 138\fbinv. This paper builds on previous analyses published by the CMS Collaboration in the \hww channel focused on the inclusive production cross section and coupling measurements at $\sqrt{s}=7$, 8, and 13\TeV~\cite{Chatrchyan:2013iaa,Sirunyan:2018egh}, and on differential fiducial production cross section measurements at 8\TeV~\cite{Khachatryan:2016vnn} and 13\TeV~\cite{CMS:2020dvg}. Similar measurements have also been reported in several Higgs boson decay channels by the ATLAS and CMS Collaborations~\cite{CMS:2021gxc, CMS:2021kom, CMS:2021ugl, ATLAS:2020wny, ATLAS:2020rej}.

Results reported in this paper show an overall improvement of the measurement accuracy thanks to new analysis techniques specifically devised to increase the sensitivity to particular production mechanisms (\eg, VBF with a different-flavor lepton pair in the final state), to the inclusion of new channels that have not been investigated in Run 2 before, such as VBF and \VH production with a same-flavor pair of charged leptons and a hadronically decaying V, and \ZH production with a three-lepton final state, and to the larger integrated luminosity analyzed. Moreover, \WH production with two same sign leptons is measured for the first time in CMS. Tabulated results are provided in the HEPData record for this analysis~\cite{hepdata}.

This paper is organized as follows. A brief overview of the CMS apparatus is given in Section~\ref{sec:detector}. The data set and simulated samples used are described in Section~\ref{sec:datasets}. Sections~\ref{sec:event_selection}--\ref{sec:STXS} describe in detail the event selection and categorization strategy, as well as the discriminating variables used to target each final state. The estimation of the backgrounds is described in Section~\ref{sec:background}, and the sources of systematic uncertainty and their treatment are given in Section~\ref{sec:uncertainties}. Results are presented in Section~\ref{sec:results}. Finally, closing remarks are given in Section~\ref{sec:conclusions}.

\section{The CMS detector and event reconstruction}
\label{sec:detector}

The CMS apparatus is a general purpose detector designed to tackle a wide range of measurements. The central feature of CMS is a superconducting solenoid of 6\unit{m} internal diameter, providing a magnetic field of 3.8\unit{T}. Within the solenoid volume are a silicon pixel and strip tracker, a lead tungstate crystal electromagnetic calorimeter (ECAL), and a brass and scintillator hadron calorimeter (HCAL), each composed of a barrel and two endcap sections. Forward calorimeters extend the pseudorapidity ($\eta$) coverage provided by the barrel and endcap detectors. Muons are detected in gas-ionization chambers embedded in the steel flux-return yoke outside the solenoid.

The events of interest are selected using a two-tiered trigger system. The first level, composed of custom hardware processors, uses information from the calorimeters and muon detectors to select events at a rate of around 100\unit{kHz} within a fixed latency of about 4\mus~\cite{Sirunyan:2020zal}. The second level, known as the high-level trigger (HLT), consists of a farm of processors running a version of the full event reconstruction software optimized for fast processing, and reduces the event rate to around 1\unit{kHz} before data storage~\cite{Khachatryan:2016bia}. Events passing the trigger selection are stored for offline reconstruction. A more detailed description of the CMS detector, together with a definition of the coordinate system and the kinematic variables, can be found in Ref.~\cite{Chatrchyan:2008zzk}. 

Muons are identified and their momenta are measured in the range $\abs{\eta} < 2.4$ by matching tracks in the muon system and the silicon tracker. The single muon trigger efficiency exceeds  90\%  over  the  full $\eta$ range,  and  the  efficiency  to reconstruct and identify muons is greater than 96\%. The relative transverse momentum (\pt) resolution for muons with \pt up to 100\GeV is 1\% in the barrel and 3\% in the endcaps~\cite{Sirunyan:2018fpa,Sirunyan:2019yvv}.

Electrons are identified and their momenta are measured in the interval $\abs{\eta} < 2.5$ by combining tracks in the silicon tracker with spatially compatible energy deposits in the ECAL, also accounting for the energy of bremsstrahlung photons likely originating from the electron track. The single electron trigger efficiency exceeds  90\%  over  the  full $\eta$ range. The efficiency to reconstruct and identify electrons ranges between 60 and 80\% depending on the lepton \pt. The momentum resolution for electrons with $\pt\approx 45\GeV$ from $\PZ \to \Pe\Pe$ decays ranges from 1.7 to 4.5\% depending on the $\eta$ region. The resolution is generally better in the barrel than in the endcaps and also depends on the bremsstrahlung energy emitted by the electron as it traverses the material in front of the ECAL~\cite{Khachatryan:2015hwa}.

In order to achieve better rejection of nonprompt leptons, increasing the sensitivity of the analysis, leptons are required to be isolated and well reconstructed by imposing a set of requirements on the quality of the track reconstruction, shape of calorimetric deposits, and energy flux in the vicinity of the particle trajectory. On top of these criteria, a selection on a dedicated multivariate analysis (MVA) tagger developed for the CMS \ttH analysis~\cite{Sirunyan:2020icl}, referred to as ttHMVA, is added in all analysis categories for muon candidates. In categories targeting the VH production modes with leptonically decaying V boson, it is found that adding a selection on the ttHMVA tagger for electrons improves the sensitivity of the analysis.

Multiple $\Pp\Pp$ interaction vertices are identified from tracking information by use of the adaptive vertex fitting algorithm~\cite{Fruhwirth:2007hz}. The primary vertex is taken to be the vertex corresponding to the hardest scattering in the event, evaluated using tracking information alone, as described in Section 9.4.1 of Ref.~\cite{CMS-TDR-15-02}.

The particle-flow (PF) algorithm~\cite{CMS-PRF-14-001} aims to reconstruct and identify each individual particle in an event, with an optimized combination of information from the various elements of the CMS detector. The energy of muons is obtained from the curvature of the corresponding track. The energy of charged hadrons is determined from a combination of their momentum measured in the tracker and the matching ECAL and HCAL energy deposits, corrected for the response function of the calorimeters to hadronic showers. The energy of photons is obtained from the ECAL measurement. The energy of electrons is determined from a combination of the electron momentum at the primary interaction vertex as determined by the tracker, the energy of the corresponding ECAL cluster, and the energy sum of all bremsstrahlung photons spatially compatible with originating from the electron track. Finally, the energy of neutral hadrons is obtained from the corresponding corrected ECAL and HCAL energies.

Hadronic jets are reconstructed from PF objects using the infrared and collinear safe anti-\kt algorithm~\cite{Cacciari:2008gp, Cacciari:2011ma} with a distance parameter of 0.4. The jet momentum is determined from the vector sum of all PF candidate momenta in the jet. From simulation, reconstructed jet momentum is found to be, on average, within 5 to 10\% of the momentum of generator jets, which are jets clustered from all generator-level final-state particles excluding neutrinos, over the entire \pt spectrum and detector acceptance. Additional \Pp{}\Pp interactions within the same or nearby bunch crossings (pileup) can contribute additional tracks and calorimetric energy deposits to the jet momentum. To mitigate this effect, charged particles identified as originating from pileup vertices are discarded, and an offset correction is applied for remaining contributions from neutral pileup particles~\cite{CMS-PRF-14-001}. Jet energy corrections are derived from simulation studies so that the average measured response of jets becomes identical to that of generator jets. In situ measurements of the momentum imbalance in dijet, photon+jet, {\PZ}+jet, and multijet events are used to account for any residual differences in jet energy scale in data and simulation~\cite{Khachatryan:2016kdb,CMS-DP-2020-019}. The jet energy resolution amounts typically to 15\% at 10\GeV, 8\% at 100\GeV, and 4\% at 1\TeV. Additional selection criteria are applied to each jet to remove jets potentially dominated by anomalous contributions from various subdetector components or reconstruction failures. Jets are measured in the range $\abs{\eta}<4.7$. In the analysis of data recorded in 2017, to eliminate spurious jets caused by detector noise, all jets in the range $2.5<\abs{\eta}<3.0$ were excluded~\cite{CMS-PAS-JME-16-003}.

We refer to the identification of jets likely originating from \PQb quarks as \PQb tagging~\cite{Sirunyan:2017ezt, CMS-DP-2017-013}. For each jet in the event a score is calculated through a multivariate combination of different jet properties, making use of boosted decision trees (BDTs) and deep neural networks (DNNs). Jets are considered \PQb tagged if their associated score exceeds a threshold, tuned to achieve a certain tagging efficiency as measured in \ttbar events. Typically three thresholds, called working points (WPs) in the following, are provided, labeled loose, medium, and tight, corresponding to probabilities of mistagging a jet originating from a lighter quark as coming from a bottom quark of 10, 1, and 0.1\%, respectively. Unless otherwise specified, the loose WP of the DeepCSV tagger is used throughout this paper.

The missing transverse momentum vector \ptvecmiss is computed as the negative vector sum of the transverse momenta of all the PF candidates in an event, and its magnitude is denoted as \ptmiss~\cite{Sirunyan:2019kia}. The \ptvecmiss is modified to account for corrections to the energy scale of the reconstructed jets in the event. The pileup per particle identification algorithm~\cite{Bertolini:2014bba} is applied to reduce the pileup dependence of the \ptvecmiss observable. The \ptvecmiss is computed from the PF candidates weighted by their probability to originate from the primary interaction vertex~\cite{Sirunyan:2019kia}.

\section{Data sets and simulations}
\label{sec:datasets}

The data sets used in the analysis were recorded by the CMS detector in 2016, 2017, and 2018, corresponding to an integrated luminosity of 36.3, 41.5, and 59.7\fbinv, respectively~\cite{CMS-LUM-17-003,CMS-PAS-LUM-17-004,CMS-PAS-LUM-18-002}.

The events selected in the analysis are required to pass criteria based on HLT algorithms that require the presence of either one or two electrons or muons, satisfying isolation and identification requirements. For the 2016 data set, the single-electron trigger requires a \pt threshold of 25\GeV for electrons with $\abs{\eta}<2.1$ and 27\GeV for $2.1<\abs{\eta}<2.5$. For the single-muon trigger the \pt threshold is 24\GeV for $\abs{\eta}<2.4$. In the dielectron (dimuon) trigger the \pt thresholds of the leading (highest \pt) and trailing (second-highest \pt) electron (muon) are respectively 23 (17) and 12 (8)\GeV. In the dilepton $\Pe\PGm$ trigger, the \pt thresholds are 23 and 12\GeV for the leading and trailing lepton, respectively. For the first part of data taking in 2016, a lower \pt threshold of 8\GeV for the trailing muon was used.
In the 2017 data set, the \pt thresholds of the single electron and single muon triggers are raised respectively to 35 and 27\GeV, while they are set to 32 and 24\GeV in the 2018 data set. For both 2017 and 2018 data sets, the \pt thresholds of the dilepton triggers are kept the same as the last part of the 2016 data set. The trigger selection is summarized in Table~\ref{tab:trigger}.

\begin{table*}
    \centering
    \topcaption{Trigger requirements on the data set used in the analysis.}
    \begin{tabular}{lcc}
        \hline
        Trigger                           & Year & Requirements                \\
        \hline
        \multirow{3}{*}{Single electron}  & 2016 & $\pt>25\GeV$, $\abs{\eta}<2.1$ or $\pt>27\GeV$, $2.1<\abs{\eta}<2.5$ \\
                                          & 2017 & $\pt>35\GeV$, $\abs{\eta}<2.5$ \\
                                          & 2018 & $\pt>32\GeV$, $\abs{\eta}<2.5$ \\ [\cmsTabSkip]

        \multirow{3}{*}{Single muon}      & 2016 & $\pt>24\GeV$, $\abs{\eta}<2.4$ \\
                                          & 2017 & $\pt>27\GeV$, $\abs{\eta}<2.4$ \\
                                          & 2018 & $\pt>24\GeV$, $\abs{\eta}<2.4$ \\ [\cmsTabSkip]

        Double electron                   & All years & $\ptone>23\GeV$, $\pttwo>12\GeV$, $\abs{\eta_{1,2}}<2.5$ \\ [\cmsTabSkip]

        Double muon                       & All years & $\ptone>17\GeV$, $\pttwo>8\GeV$, $\abs{\eta_{1,2}}<2.4$ \\ [\cmsTabSkip]

        \multirow{3}{*}{Electron-muon}    & \multirow{3}{*}{All years} & $\ptone>23\GeV$, $\pttwo>12\GeV$ \\
                                          &                            & $\pttwo>8\GeV$ in first part of 2016 data taking \\
                                          &                            & $\abs{\eta_{\Pe}} < 2.5$, $\abs{\eta_{\PGm}} < 2.4$ \\                             
        \hline
    \end{tabular}
    \label{tab:trigger}
\end{table*}

Monte Carlo (MC) event generators are used in the analysis to model the signal and background processes. Three independent sets of simulated events, corresponding to the 2016, 2017, and 2018 data sets, are used for each process of interest, in order to take into account year-dependent effects in the CMS detector, data taking, and event reconstruction. Despite different matrix element generators being used for different processes, all simulated events corresponding to a given data set share the same set of parton distribution functions (PDFs), underlying event (UE) tune, and parton shower (PS) configuration. The PDF set used is NNPDF 3.0~\cite{Ball:2013hta,Ball:2011uy} at NLO for 2016 and NNPDF 3.1~\cite{Ball:2017nwa} at NNLO for 2017 and 2018. The CUETP8M1~\cite{Khachatryan:2015pea} tune is used to describe the UE in 2016 simulations, while the CP5~\cite{Sirunyan:2019dfx} tune is adopted in 2017 and 2018 simulated events. For all the simulations, the matrix-element event generators are interfaced with \PYTHIA~\cite{Sjostrand:2014zea} 8.226 in 2016, and 8.230 in 2017 and 2018, for the UE description, PS, and hadronization.

Simulated events are used in the analysis to model Higgs boson production through \ggH, VBF, \VH, and associated production with top quarks (\ttH) or bottom quarks (\bbH), although \ttH and \bbH have a negligible contribution in the analysis phase space. All Higgs boson production processes except \bbH are generated using the \POWHEG v2~\cite{Nason:2004rx,Frixione:2007vw,Alioli:2010xd,Bagnaschi:2011tu,Nason:2009ai,Luisoni:2013kna,Hartanto:2015uka} event generator, which describes Higgs boson production at next-to-leading order (NLO) accuracy in quantum chromodynamics (QCD), including finite quark mass effects. Instead, \bbH production is simulated using the \MGvATNLO v2.2.2 generator~\cite{Alwall:2014hca}. The \ZH production process is simulated including both gluon- and quark-induced contributions. The \textsc{minlo hvj}~\cite{Luisoni:2013kna} extension of \POWHEG v2 is used for the simulation of \WH and quark-induced \ZH production, providing a description of {\VH}+0- and 1-jet processes with NLO accuracy.
For \ggH production, the simulated events are reweighted to match the NNLOPS~\cite{Hamilton:2013fea,Hamilton:2015nsa} prediction in the hadronic jet multiplicity (\njet) and Higgs boson transverse momentum (\ptH) distributions, according to a two-dimensional map constructed using these observables. Moreover, for a better description of the phase space with more than one jet, the \textsc{minlo hjj}~\cite{Frederix:2015fyz} generator is used, giving NLO accuracy for $\njet \geq 2$ and leading order (LO) accuracy for $\njet \geq 3$. The simulated samples are normalized to the cross sections recommended in Ref.~\cite{deFlorian:2227475}; in particular, the next-to-next-to-next-to-leading order cross section is used to normalize the \ggH sample.
The Higgs boson mass ($m_\PH$) in the event generation is assumed to be 125\GeV, while the value of 125.38\GeV~\cite{HggMass} is used for the calculation of cross sections and branching fractions, yielding values of $48.31\unit{pb}$, $3.77\unit{pb}$, $1.36\unit{pb}$, $0.88\unit{pb}$, and $0.12\unit{pb}$ for the \ggH, VBF, \WH, quark-induced \ZH, and gluon-induced \ZH processes, respectively, and 22.0\% for the \hww branching ratio~\cite{deFlorian:2227475}. The decay to a pair of \PW bosons and subsequently to leptons or hadrons is performed using the \textsc{JHUgen}~\cite{Bolognesi:2012mm} v5.2.5 generator in 2016, and v7.1.4 in 2017 and 2018, for \ggH, VBF, and quark-induced \ZH samples. The Higgs boson and \PW boson decays are performed using \PYTHIA 8.212 for the other signal simulations.
For the \ggH, VBF, and \VH production mechanisms, additional Higgs boson simulations are produced using the \POWHEG v2 generator, where the Higgs boson decays to a pair of \PGt leptons. These events are treated as signal in the analysis, with the exception of the measurement in the STXS framework, in which they are treated as background.

The background processes are simulated using several event generators. The quark-initiated nonresonant \WW process is simulated using \POWHEG v2~\cite{Nason:2013ydw} with NLO accuracy for the inclusive production. The \MCFM v7.0~\cite{Campbell:1999ah,Campbell:2011bn,Campbell:2015qma} generator is used for the simulation of gluon-induced \WW production at LO accuracy, and the normalization is chosen to match the NLO cross section~\cite{Caola:2016trd}. The nonresonant electroweak (EW) production of \WW pairs with two additional jets (in the vector boson scattering topology) is simulated at LO accuracy with \MGvATNLO v2.4.2 using the MLM matching and merging scheme~\cite{MLM}. Top quark pair production (\ttbar), as well as single top quark processes, including $\PQt\PW$, $s$-, and $t$-channel contributions, are simulated with \POWHEG v2~\cite{Frixione:2007nw,Alioli:2009je,Re:2010bp}. The Drell--Yan (DY) production of charged-lepton pairs is simulated at NLO accuracy with \MGvATNLO v2.4.2 with up to two additional partons, using the FxFx matching and merging scheme~\cite{Frederix:2012ps}. Production of a \PW boson associated with an initial-state radiation photon (\wg) is simulated with \MGvATNLO v2.4.2 at NLO accuracy with up to one additional parton in the matrix element calculations and the FxFx merging scheme. Diboson processes containing at least one \PZ boson or a virtual photon ($\PGg^{\ast}$) with mass down to 100\MeV are generated with \POWHEG v2~\cite{Nason:2013ydw} at NLO accuracy. Production of a \PW boson in association with a $\PGg^{\ast}$ (\wgs) for masses below 100\MeV is simulated by \PYTHIA 8.212 in the parton showering of \wg events. Triboson processes with inclusive decays are also simulated at NLO accuracy with \MGvATNLO v2.4.2.

For all processes, the detector response is simulated using a detailed description of the CMS detector, based on the \GEANTfour package~\cite{Agostinelli:2002hh}.  The distribution of the number of pileup interactions in the simulation is reweighted to match the one observed in data. The average number of pileup interactions was 23 (32) in 2016 (2017 and 2018).

The efficiency of the trigger system is evaluated in data on a per-lepton basis by selecting dilepton events compatible with originating from a \PZ boson. The per-lepton efficiencies are then combined probabilistically (\ie, the overall efficiency for an event passing any of the triggers listed above is calculated) to obtain the overall efficiencies of the trigger selections used in the analysis. The procedure has been validated by comparing the resulting efficiencies with MC simulation of the trigger. A correction has been derived as a function of $\dR=\sqrt{\smash[b]{(\Delta\eta)^2+(\Delta\phi)^2}}$ between the two leptons to account for any residual discrepancy, which is found to be on average below 1\%. The resulting efficiencies are then applied directly on simulated events.

\section{Event selection and categorization}
\label{sec:event_selection}

The analysis targets events in which a Higgs boson is produced via \ggH, VBF, or \VH processes, and subsequently decays to a pair of \PW bosons. Events are selected by requiring at least two charged leptons (electrons or muons) with high \pt, high \ptmiss, and a varying number of hadronic jets. Throughout this paper, unless otherwise specified, only hadronic jets with $\pt > 30\GeV$ are considered. Categories targeting Higgs bosons produced via \ggH, VBF, and \VH with a hadronically decaying vector boson ({\VH}2j) are subdivided in different-flavor (DF) and same-flavor (SF) by selecting {\Pe}{\PGm}, and {\Pe}{\Pe}/{\PGm}{\PGm} pairs, respectively. Categories targeting \VH production with a leptonically decaying vector boson are subdivided in four subcategories based on the number of leptons and hadronic jets required: \WH{}SS (same sign), \WH{}3\Pell, \ZH{}3\Pell, and \ZH{}4\Pell targeting the $\WH\to \Pell^{\pm}\Pell^{\pm} 2\PGn \PQq\PQq$, $\WH\to 3\Pell 3\PGn$, $\ZH\to 3\Pell \PGn \PQq\PQq$, and $\ZH\to 4\Pell 2\PGn$ processes, respectively. In all cases events containing additional leptons with $\pt > 10\GeV$ are rejected. A summary of the different categories is given in Table~\ref{tab:categories_overview}, with a more detailed breakdown given in Table~\ref{tab:fit_overview}.

\begin{table*}[h]
    \centering
    \topcaption{Overview of the selection defining the analysis categories (a more detailed breakdown is given in Table~\ref{tab:fit_overview}).}
    \begin{tabular}{lccc}
        \hline
        Category & Number of leptons & Number of jets & Subcategorization \\
        \hline
        \ggH        & 2 & \NA     & (DF, SF) $\times$ (0 jets, 1 jet, $\geq$2 jets) \\
        VBF         & 2 & $\geq$2 & (DF, SF) \\
        \VH{}2j     & 2 & $\geq$2 & (DF, SF) \\
        \WH{}SS     & 2 & $\geq$1 & (DF, SF) $\times$ (1 jet, 2 jets) \\
        \WH{}3\Pell & 3 & 0       & SF lepton pair with opposite or same sign \\
        \ZH{}3\Pell & 3 & $\geq$1 & (1 jet, 2 jets) \\
        \ZH{}4\Pell & 4 & \NA     & (DF, SF) \\
        \hline
    \end{tabular}
    \label{tab:categories_overview}
\end{table*}

Across all categories, in the 2016 data set, events are required to pass single- or double-lepton triggers. An additional requirement is placed on the lepton \pt to be above 10\GeV, and the highest \pt (leading) lepton in the event is furthermore required to have $\pt>25\GeV$. In the 2017 and 2018 data sets the threshold for leptons is increased to 13\GeV because of a change in the trigger setup. Where yields suffice, events are further split according to the charge and \pt ordering of the dilepton system, \pt of the subleading lepton, and number of hadronic jets in the event, as detailed in following sections. The number of expected and observed events in each category are given in Section~\ref{sec:results}.

\section{Gluon fusion categories}
\label{sec:ggH}

This section describes the categories targeting the \ggH production mechanism, both in DF and SF final states. In DF final states, the main background processes are nonresonant \WW, top quark production (both single and pair), DY production of a pair of \PGt leptons that subsequently decay to an $\Pe\PGm$ pair and associated neutrinos, and {\PW}+jets events when a jet is misidentified as a lepton. Subdominant backgrounds include \WZ, \ZZ, $\PV\PGg$, $\PV\PGg^{\ast}$, and $\PV\PV\PV$ production. In SF final states, the dominant background contribution is given by DY events, with subdominant components arising from top quark and \WW production, as well as events with misidentified leptons.

\subsection{Different-flavor ggH categories}
\label{subsec:ggH_DF}

On top of the common selection, the leading leptons are required to form an $\Pe\PGm$ pair with opposite charge. Contributions arising from top quark production are reduced by rejecting events containing any jet with $\pt>20\GeV$ that is identified as originating from a bottom quark by the tagging algorithm. The dilepton invariant mass \mll is required to be above 12\GeV to suppress QCD events with multiple misidentified jets. Events with no genuine missing transverse momentum (arising from the presence of neutrinos in signal events), as well as \tautau events, are suppressed by requiring $\ptmiss>20\GeV$. The latter are further reduced by requiring the \pt of the dilepton system \ptll to exceed 30\GeV, as leptons arising from a \tautau pair are found to have on average lower \pt than those coming from a \WW pair. Finally, to further suppress contributions from \tautau and {\PW}+jets events, where the subleading lepton does not arise from a \PW boson decay, the transverse mass built with \ptvecmiss and the momentum of the subleading lepton $\mT(\Pell_2, \ptmiss)$ is required to be greater than 30\GeV, where \mT for a collection of particles $\{P_i\}$ with transverse momenta $\ptvec{}_{,i}$ is defined as:
\begin{linenomath}
\begin{equation}
    \mT(\{P_i\}) = \sqrt{ \left( \sum \abs{\ptvec{}_{,i}} \right)^2 - \left| \sum \ptvec{}_{,i} \right|^2 }.
    \label{eq:mth}
\end{equation}
\end{linenomath}
Selected events are further split into subcategories in order to exploit the peculiar kinematics of the target final state. Events with zero, one, and more than one hadronic jets are separated into distinct categories. In order to better constrain the {\PW}+jets background, the 0- and 1-jet categories are subdivided into two categories each according to the charge and \pt ordering of the dilepton pair. This subdivision exploits the fact that the signal is charge symmetric, while in {\PW}+jets events \PWp bosons are more abundant than \PWm bosons. Finally, these subcategories are further divided according to whether the \pt of the subleading lepton (\pttwo) is above or below 20\GeV. This results in a four-fold partitioning of the 0- and 1-jet DF \ggH categories. In categories with more than one hadronic jet, a selection on the invariant mass of the leading dijet pair \mjj is added to ensure that there is no overlap with the VBF and VH categories.

Given the presence of neutrinos in the final state, the mass of the Higgs boson candidate can not be reconstructed in the \WW channel. Nevertheless, specific features of the channel make it possible to achieve good sensitivity. In particular, the scalar nature of the Higgs boson results in the two final-state leptons being preferentially emitted in the same hemisphere. This fact compresses the distribution of \mll for signal events to lower values with respect to the nonresonant \WW process. This shape difference alone however is not sufficient to disentangle the signal from other background processes, such as DY production of \tautau pairs and $\PV\PGg$, that populate the low-\mll phase space. The Higgs boson transverse mass $\mTH = \mT(\Pell\Pell, \ptmiss)$ is thus introduced as a second discriminating variable. A selection on \mTH is applied by requiring its value to be above 60\GeV for signal events. It is found that signal and background events populate different regions of the $(\mll,\mTH)$ plane. The signal extraction fit is therefore performed on a two-dimensional $(\mll,\mTH)$ binned template, allowing for good signal-to-background discrimination.

\begin{figure*}[htbp]
    \centering
    \includegraphics[width=0.49\textwidth]{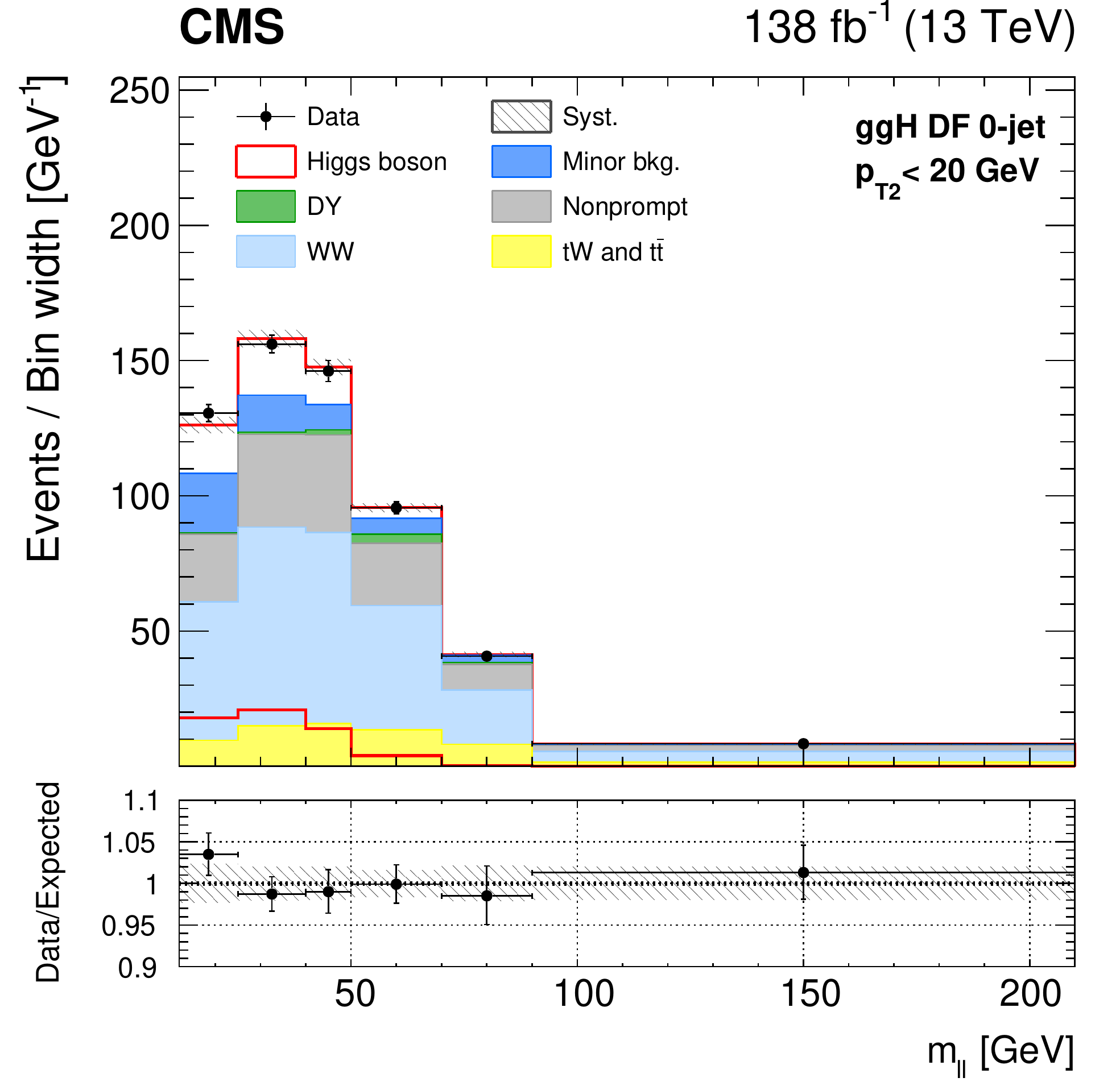}
    \includegraphics[width=0.49\textwidth]{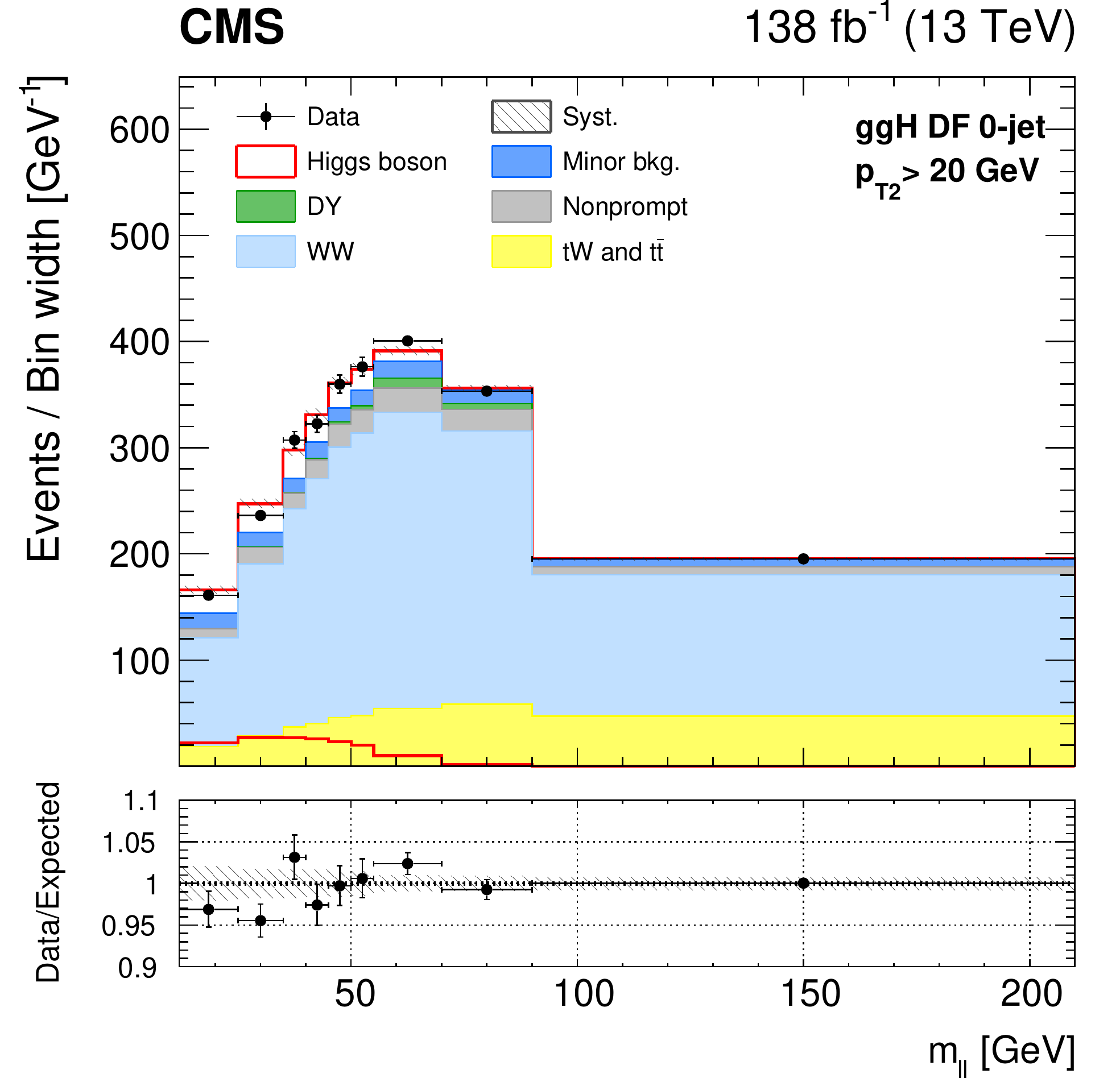}
    \\
    \includegraphics[width=0.49\textwidth]{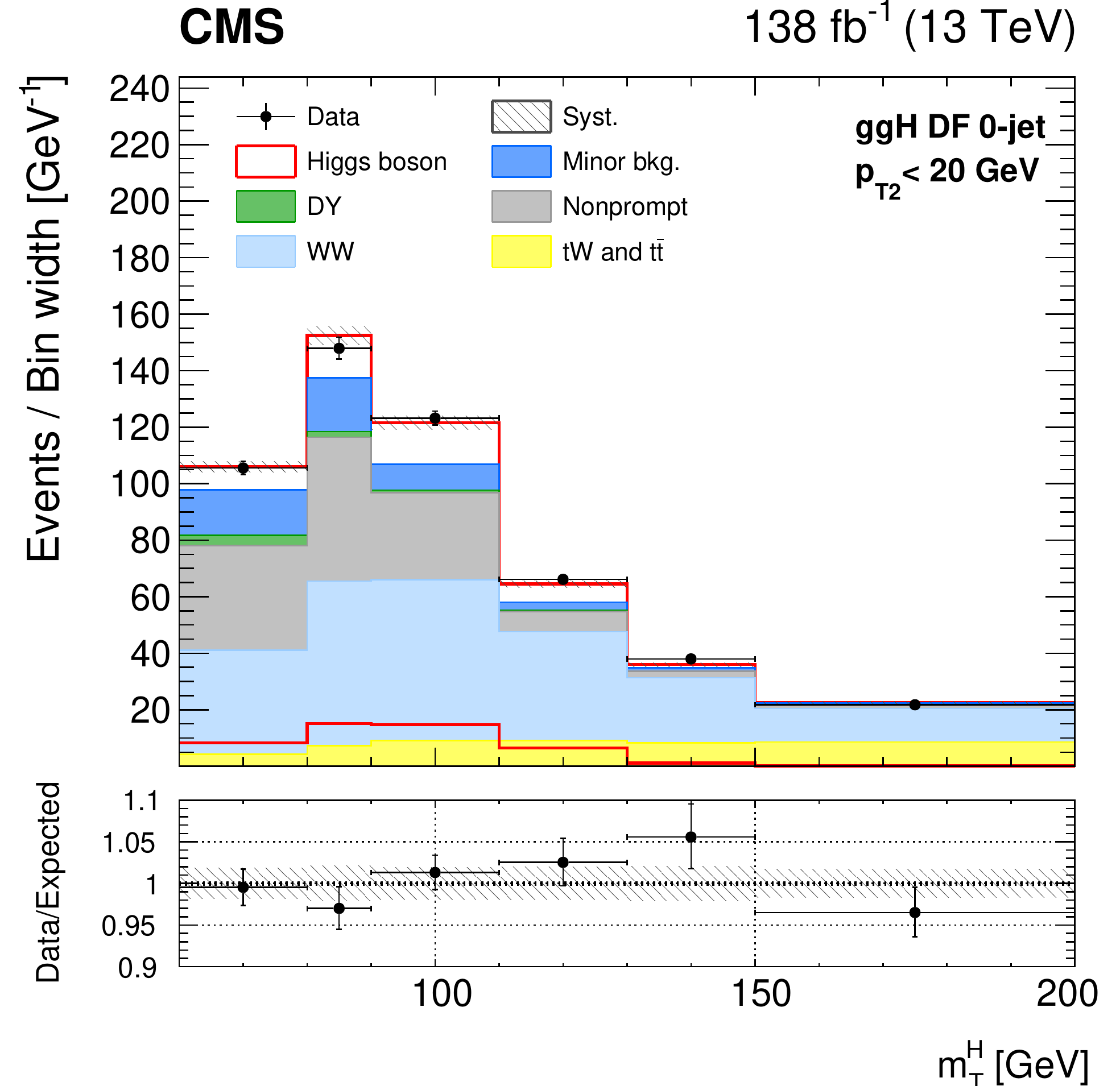}
    \includegraphics[width=0.49\textwidth]{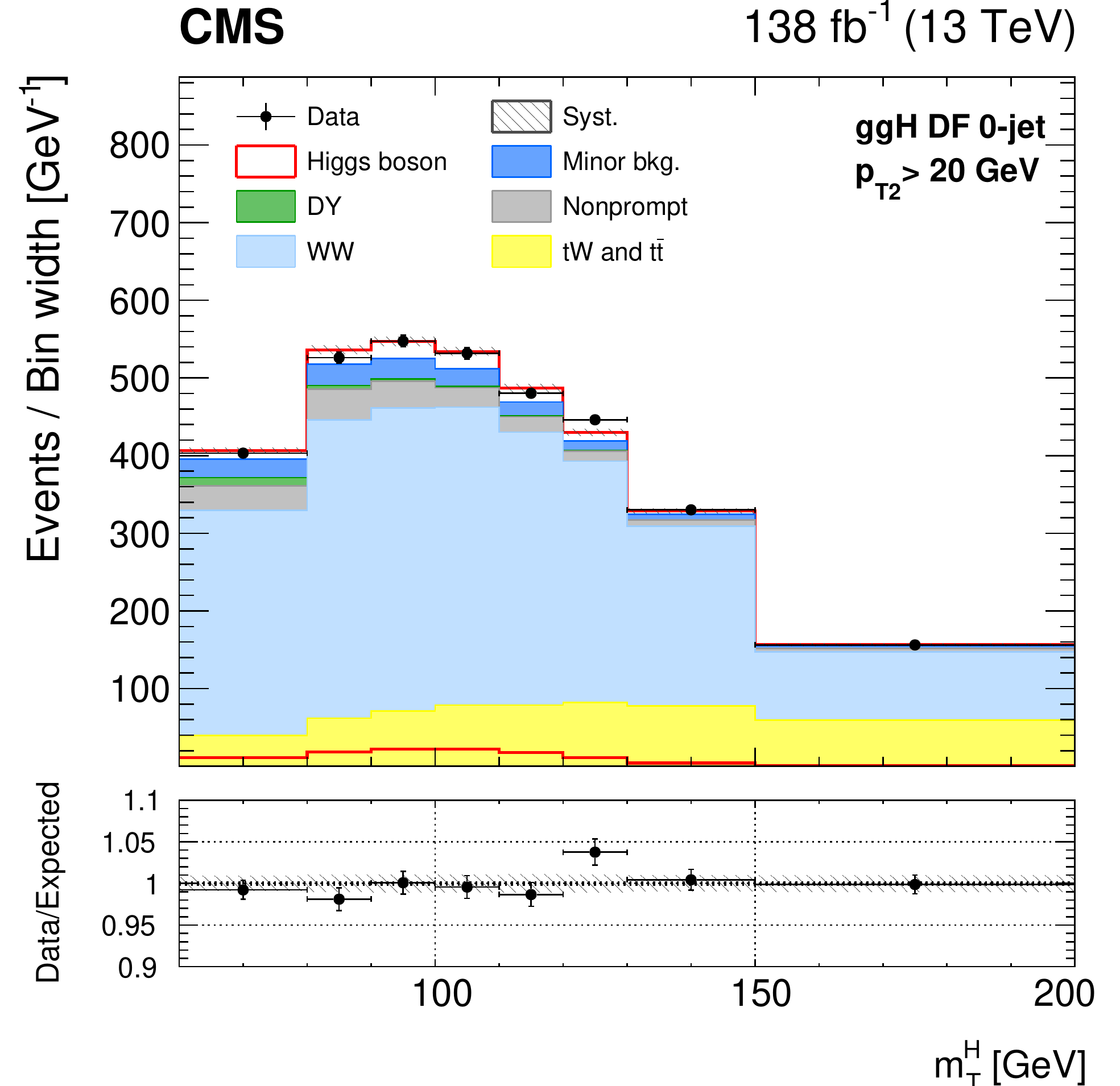}
    \caption{Observed distributions of the \mll (upper) and \mTH (lower) fit variables in the 0-jet \ggH $\pttwo<20\GeV$ (left) and $\pttwo>20\GeV$ (right) DF categories. The uncertainty band corresponds to the total systematic uncertainty in the templates after the fit to the data. The signal template is shown both stacked on top of the backgrounds, as well as superimposed. The yields are shown with their best fit normalizations from the simultaneous fit. Vertical bars on data points represent the statistical uncertainty in the data. The overflow is included in the last bin. The lower panel in each figure shows the ratio of the number of events observed in data to that of the total SM MC as extracted from the fit.}
    \label{fig:mll_mth_ggH_DF_0j}
\end{figure*}

\begin{figure*}[htbp]
    \centering
    \includegraphics[width=0.49\textwidth]{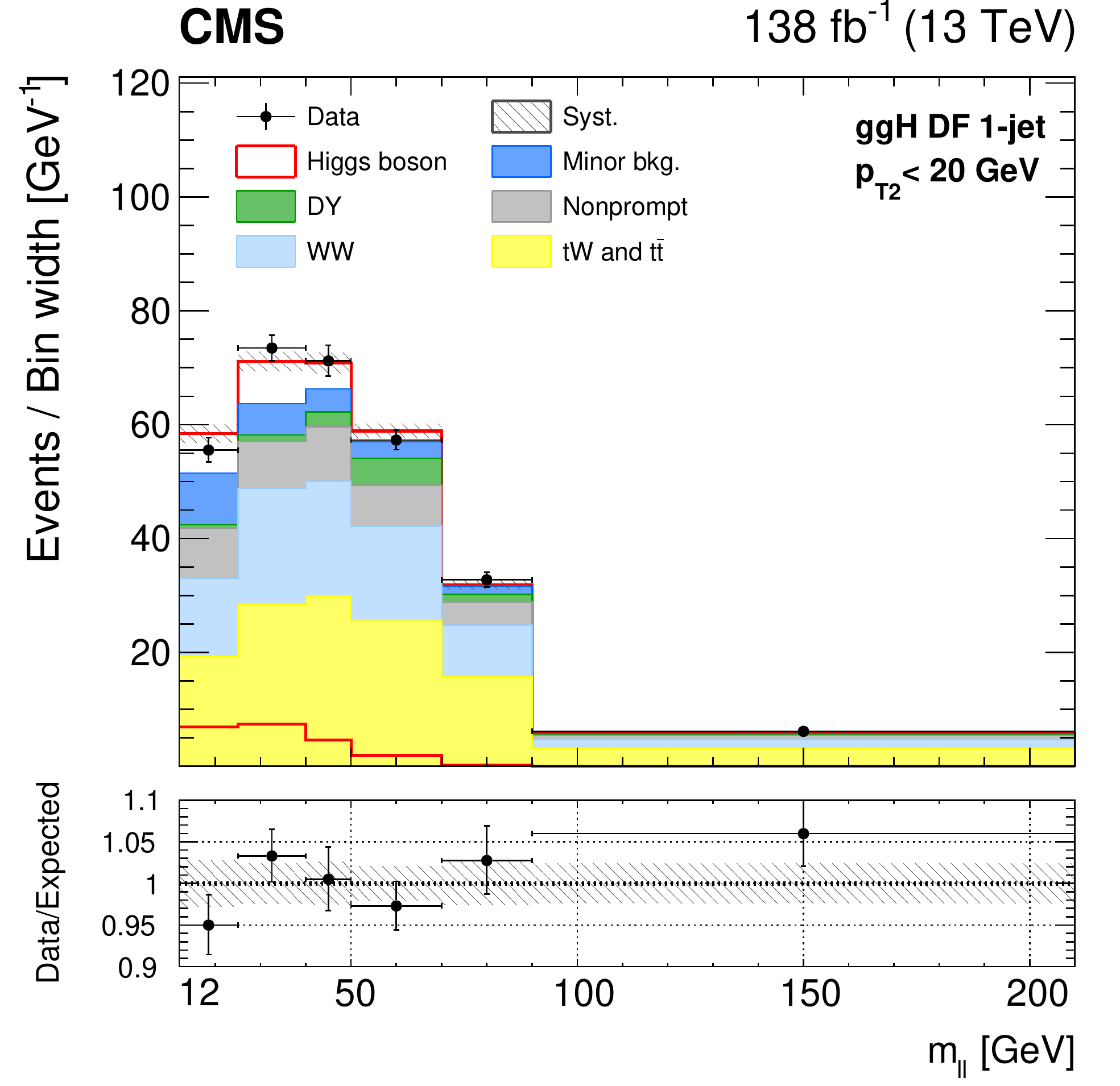}
    \includegraphics[width=0.49\textwidth]{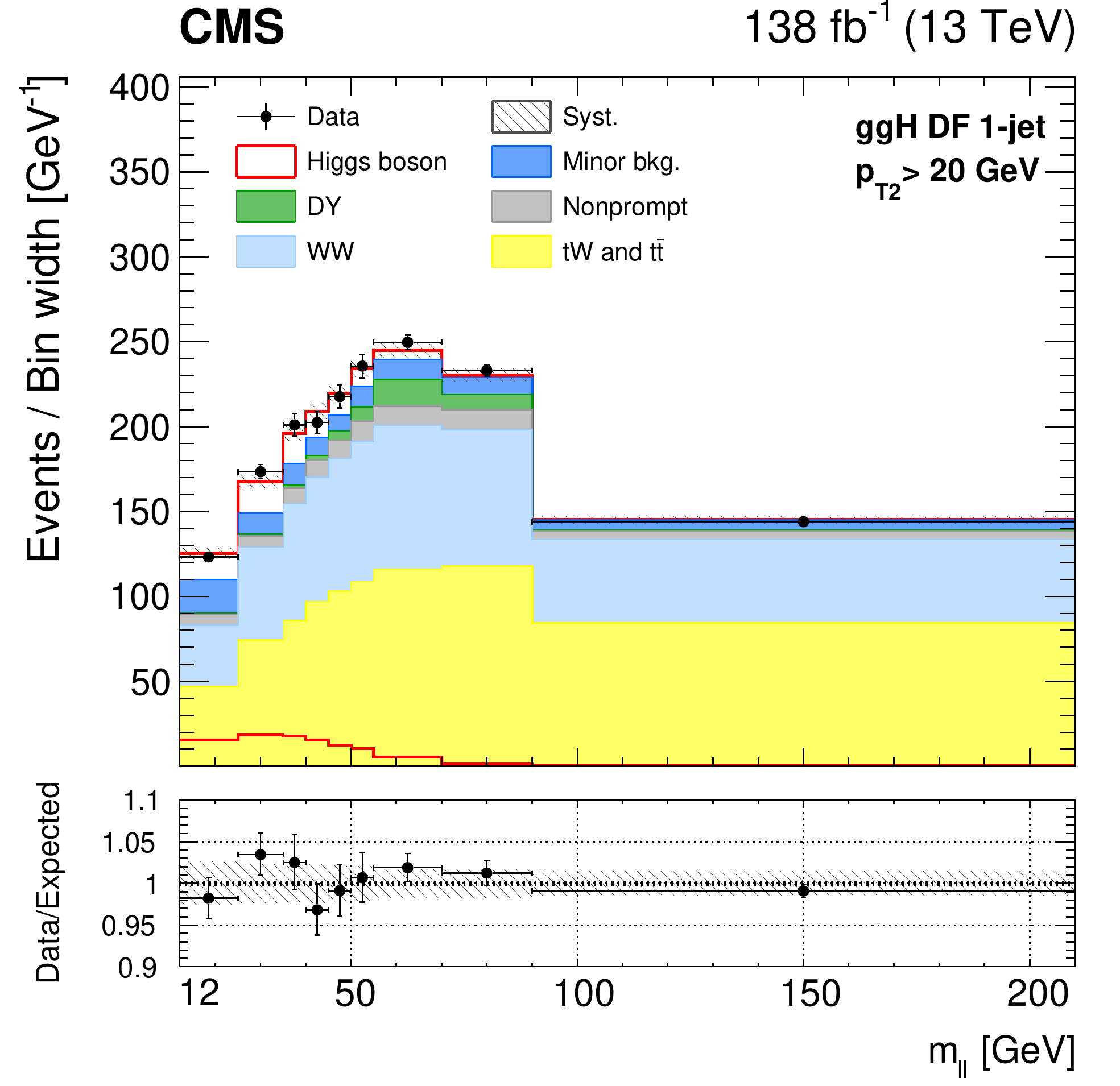}
    \\
    \includegraphics[width=0.49\textwidth]{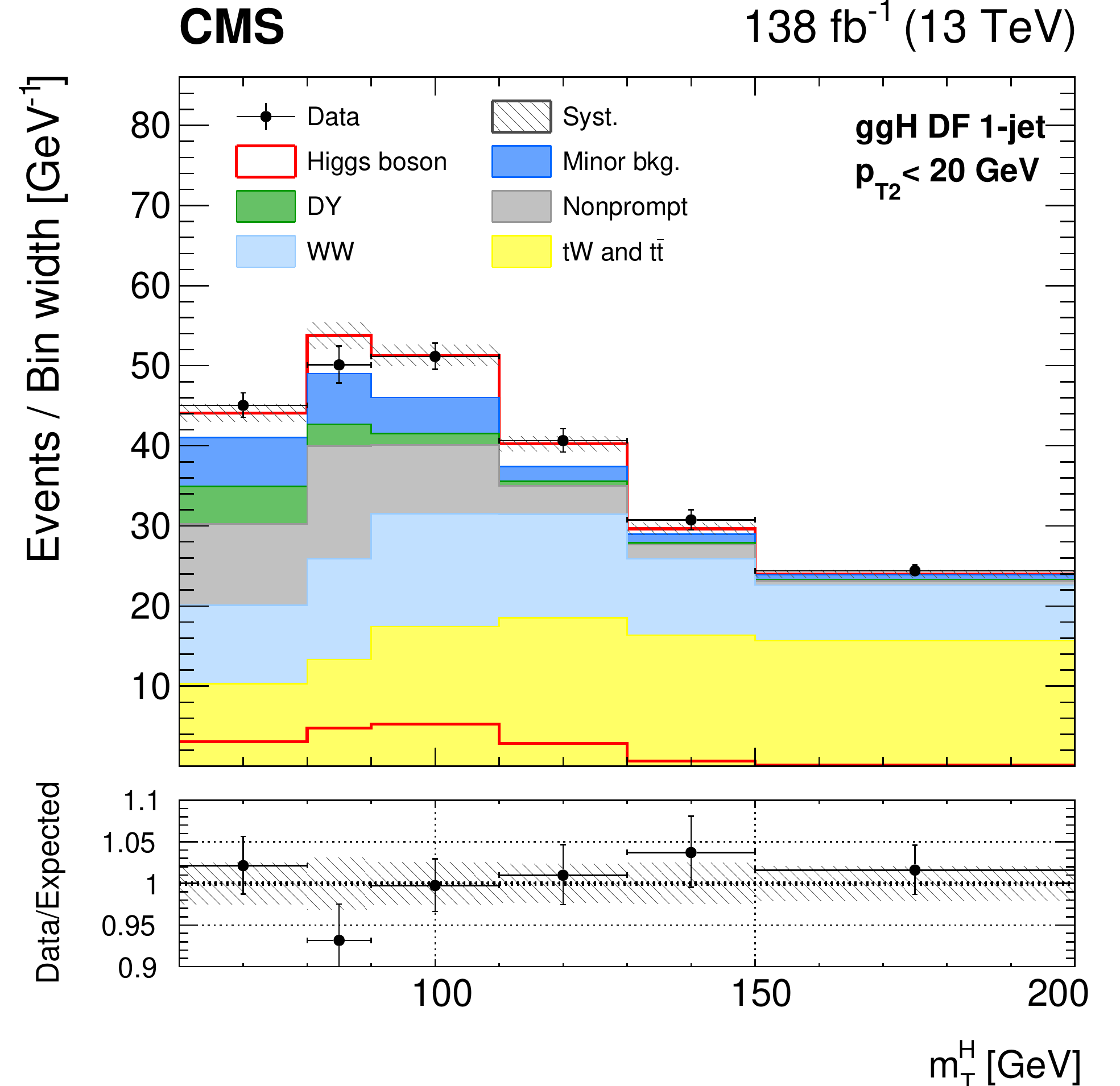}
    \includegraphics[width=0.49\textwidth]{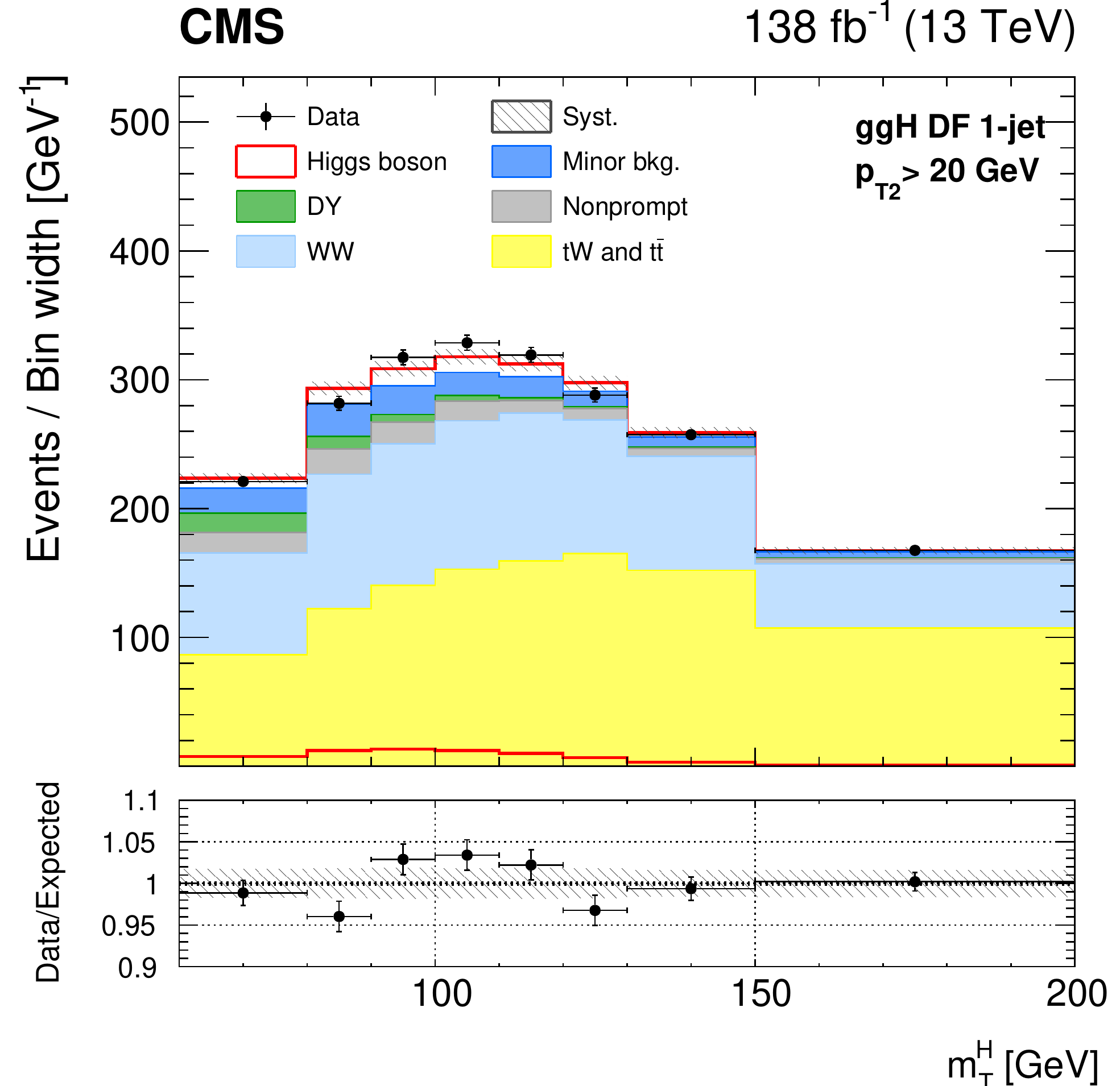}
    \caption{Observed distributions of the \mll (upper) and \mTH (lower) fit variables in the 1-jet \ggH $\pttwo<20\GeV$ (left) and $\pttwo>20\GeV$ (right) DF categories. A detailed description is given in the Fig.~\ref{fig:mll_mth_ggH_DF_0j} caption.}
    \label{fig:mll_mth_ggH_DF_1j}
\end{figure*}

\begin{figure*}[htbp]
    \centering
    \includegraphics[width=0.49\textwidth]{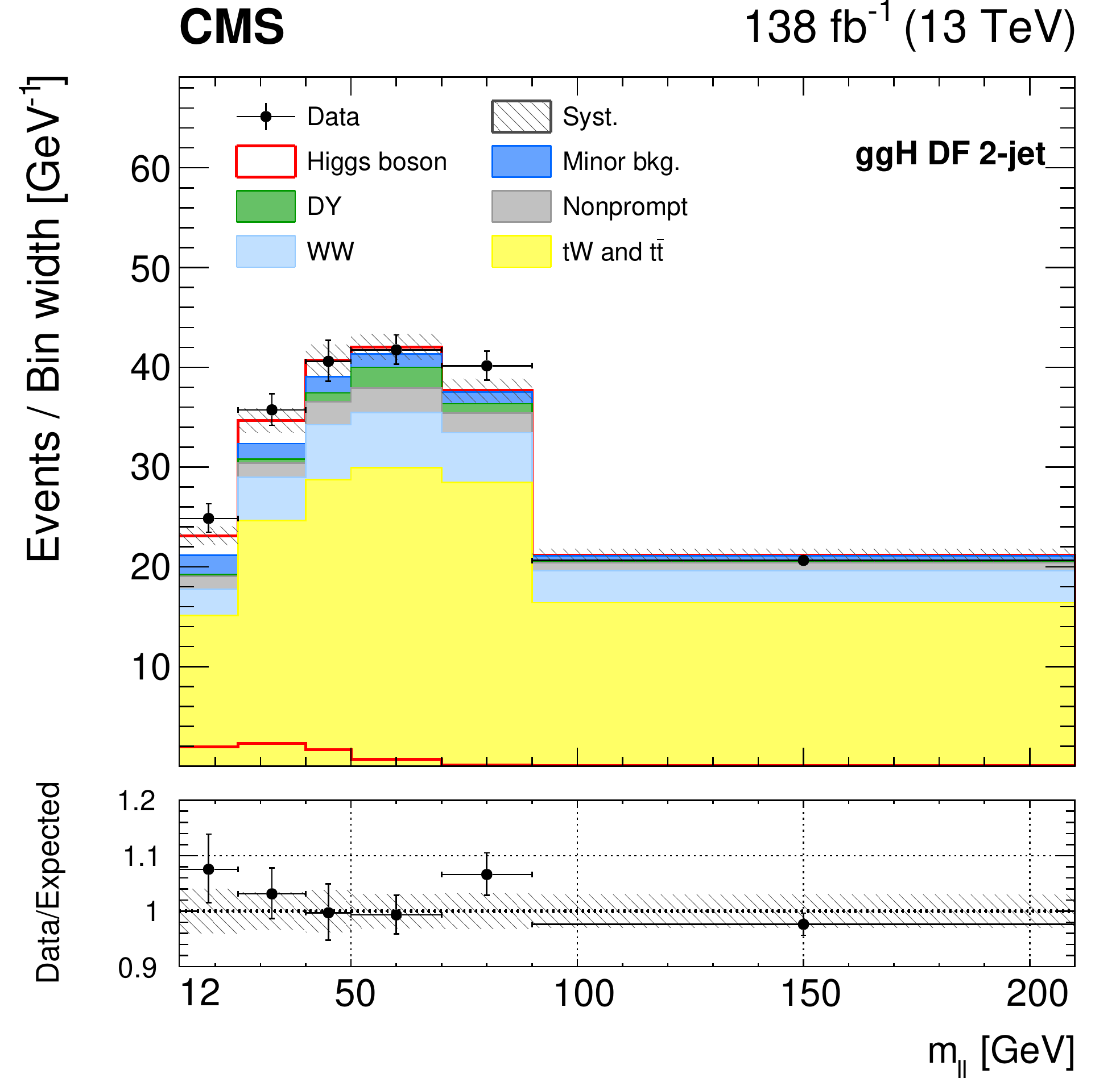}
    \includegraphics[width=0.49\textwidth]{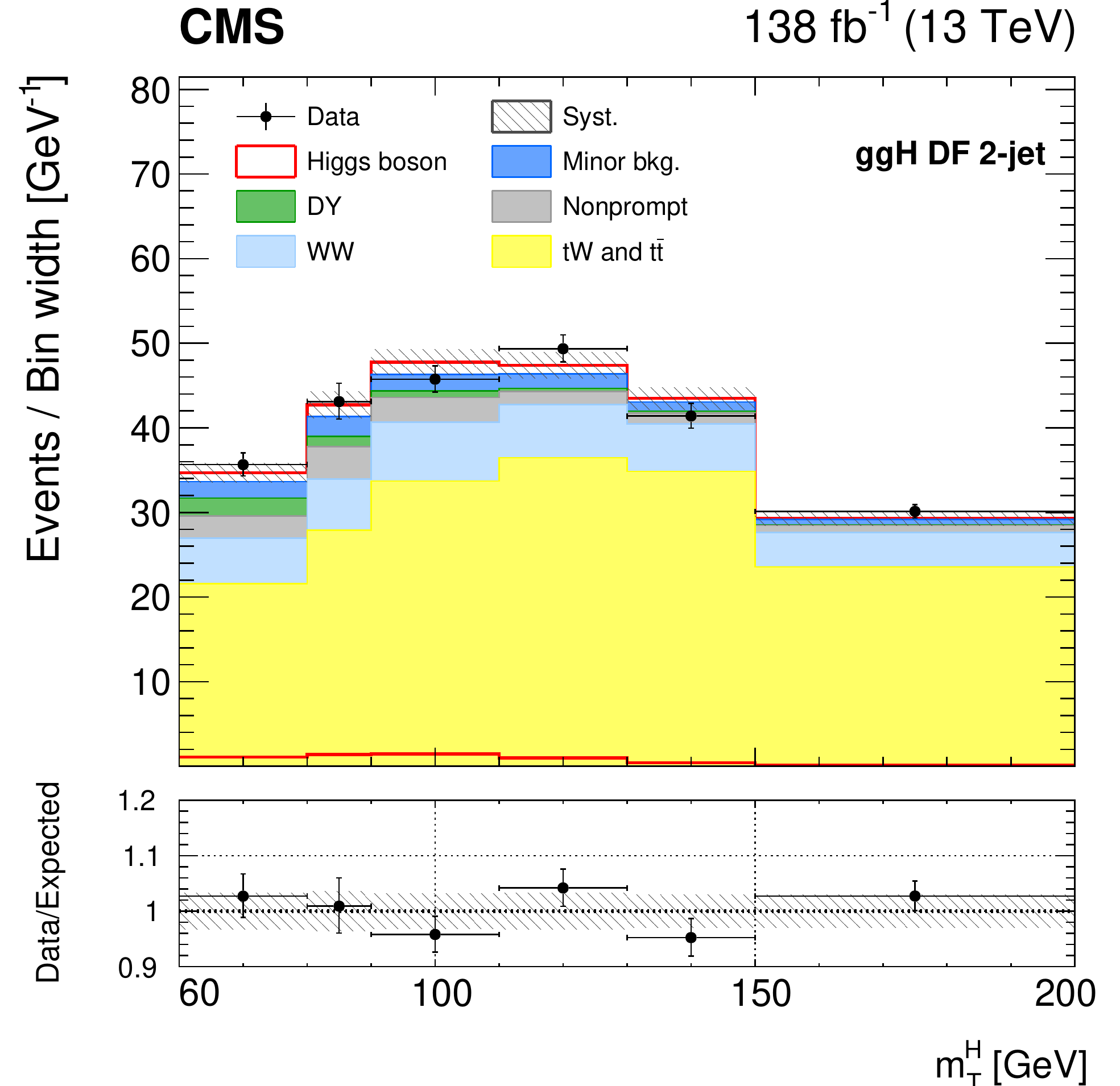}
    \caption{Observed distributions of the \mll (left) and \mTH (right) fit variables in the 2-jet \ggH DF category. A detailed description is given in the Fig.~\ref{fig:mll_mth_ggH_DF_0j} caption.}
    \label{fig:mll_mth_ggH_DF_2j}
\end{figure*}

\begin{figure*}[htbp]
    \centering
    \includegraphics[width=0.49\textwidth]{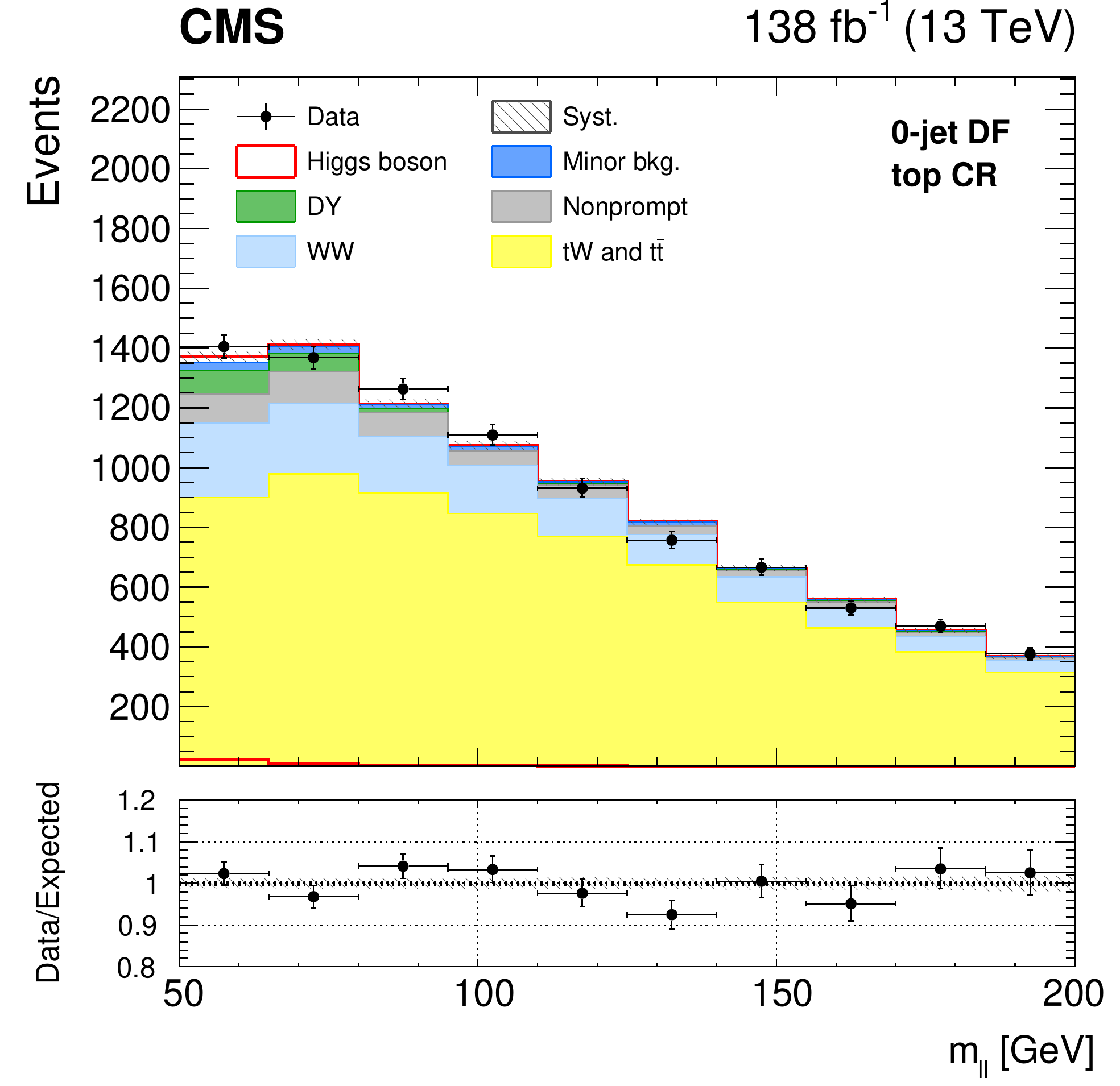}
    \includegraphics[width=0.49\textwidth]{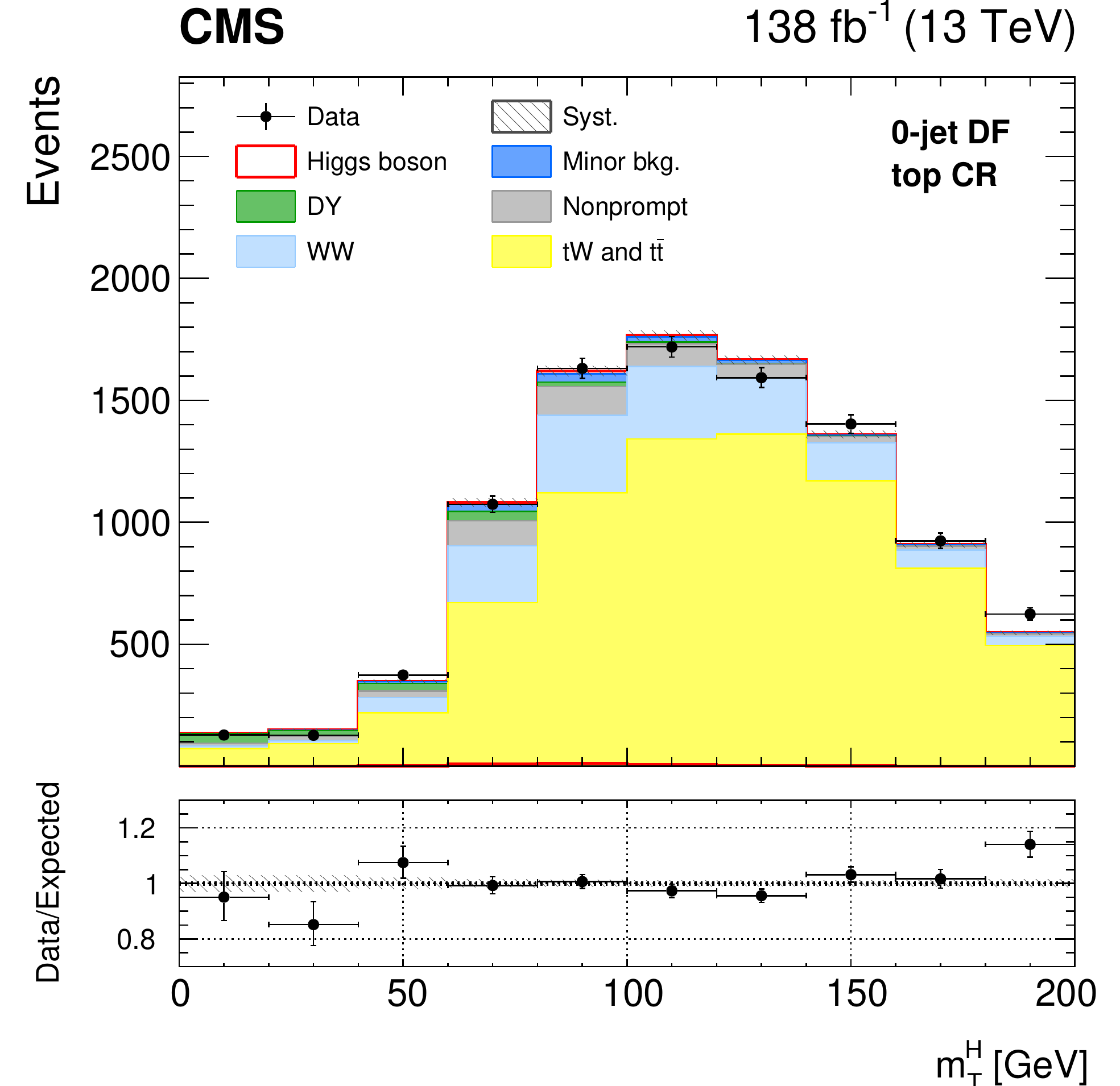}
    \caption{Observed distributions of the \mll (left) and \mTH (right) variables in the 0-jet DF top quark control region. A detailed description is given in the Fig.~\ref{fig:mll_mth_ggH_DF_0j} caption.}
    \label{fig:mll_mth_topCR_0j}
\end{figure*}

\begin{figure*}[htbp]
    \centering
    \includegraphics[width=0.49\textwidth]{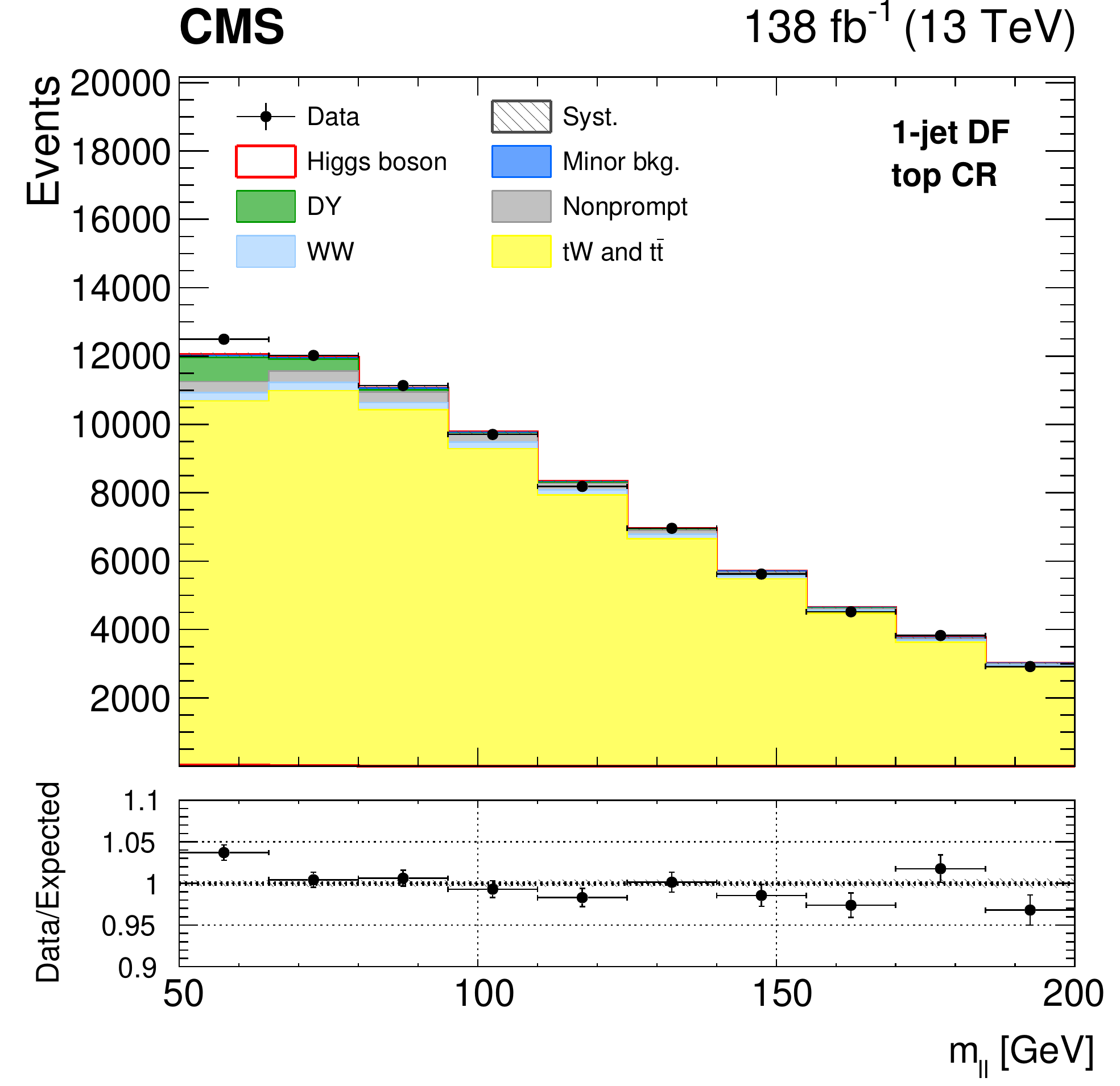}
    \includegraphics[width=0.49\textwidth]{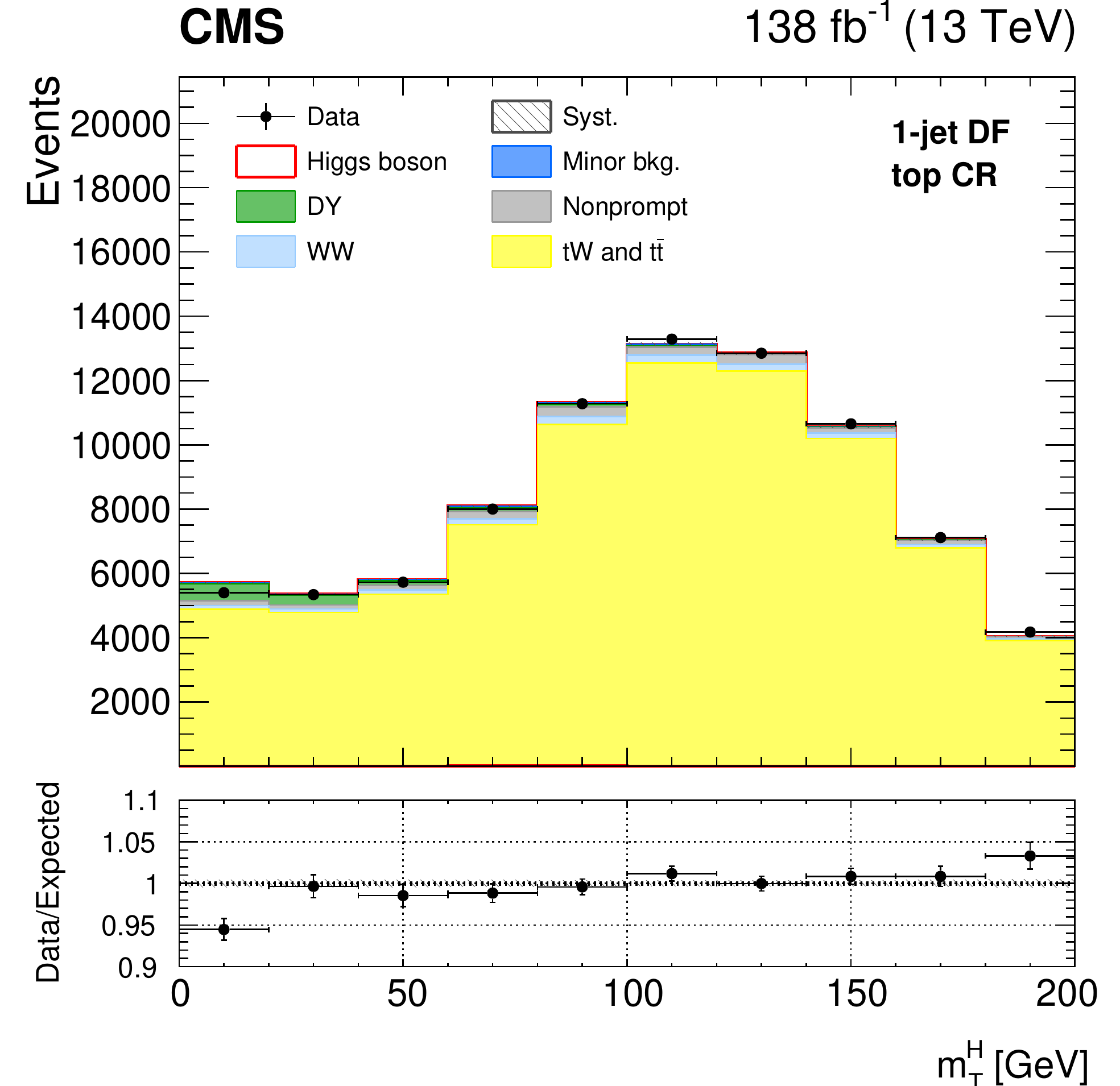}
    \caption{Observed distributions of the \mll (left) and \mTH (right) variables in the 1-jet DF top quark control region. A detailed description is given in the Fig.~\ref{fig:mll_mth_ggH_DF_0j} caption.}
    \label{fig:mll_mth_topCR_1j}
\end{figure*}

\begin{figure*}[htbp]
    \centering
    \includegraphics[width=0.49\textwidth]{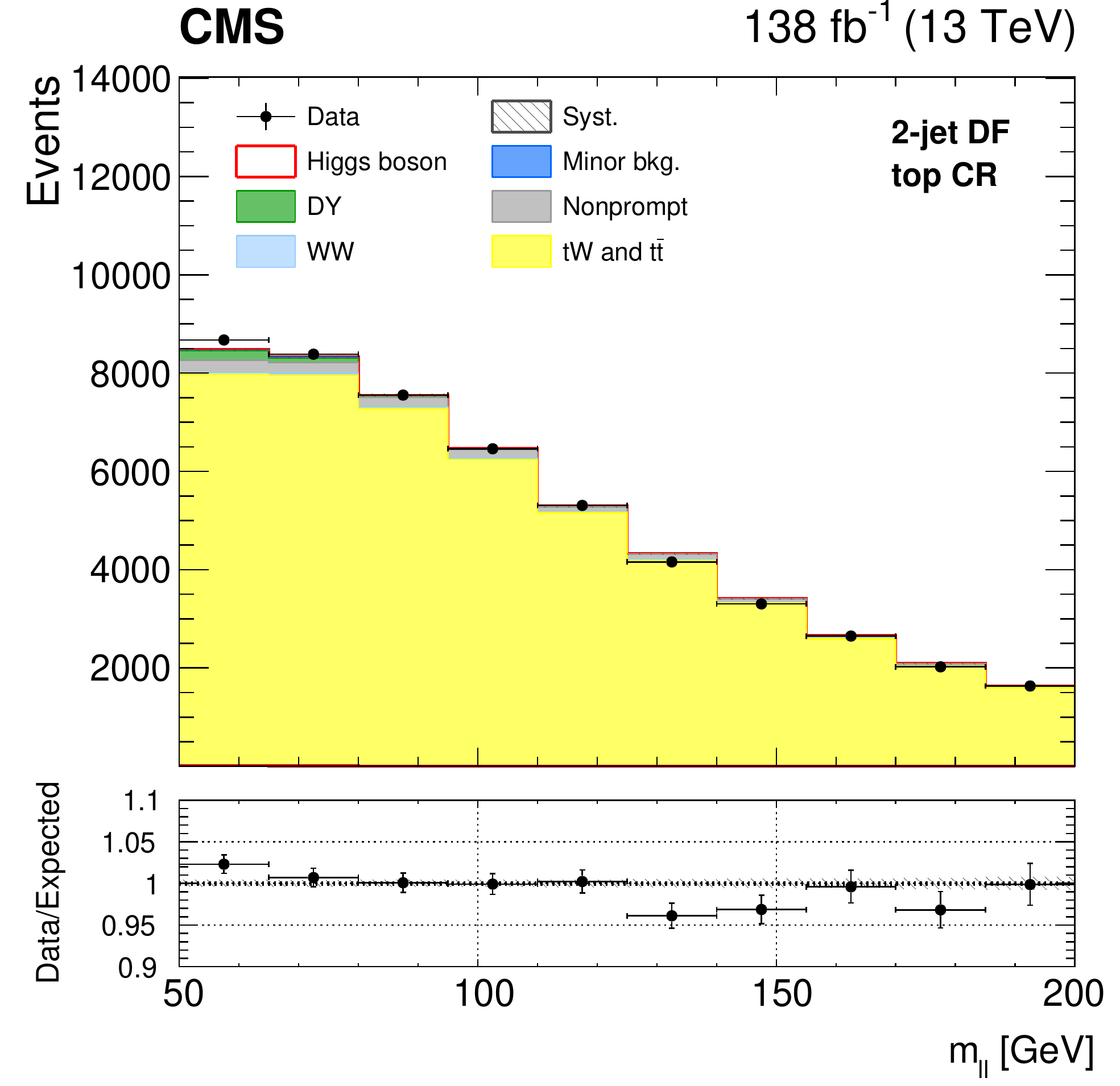}
    \includegraphics[width=0.49\textwidth]{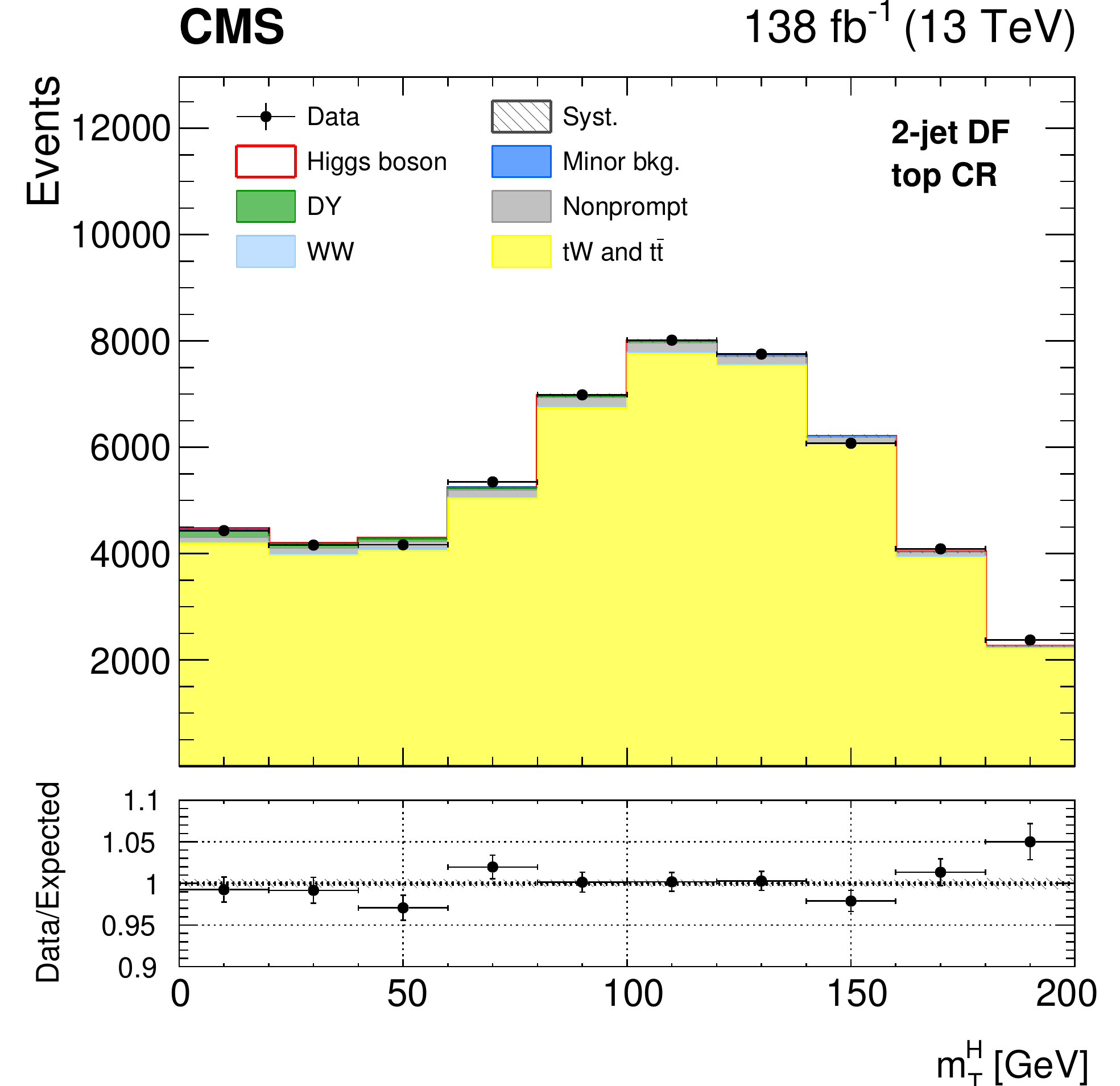}
    \caption{Observed distributions of the \mll (left) and \mTH (right) variables in the 2-jet DF top quark control region. A detailed description is given in the Fig.~\ref{fig:mll_mth_ggH_DF_0j} caption.}
    \label{fig:mll_mth_topCR_2j}
\end{figure*}

\begin{figure*}[htbp]
    \centering
    \includegraphics[width=0.49\textwidth]{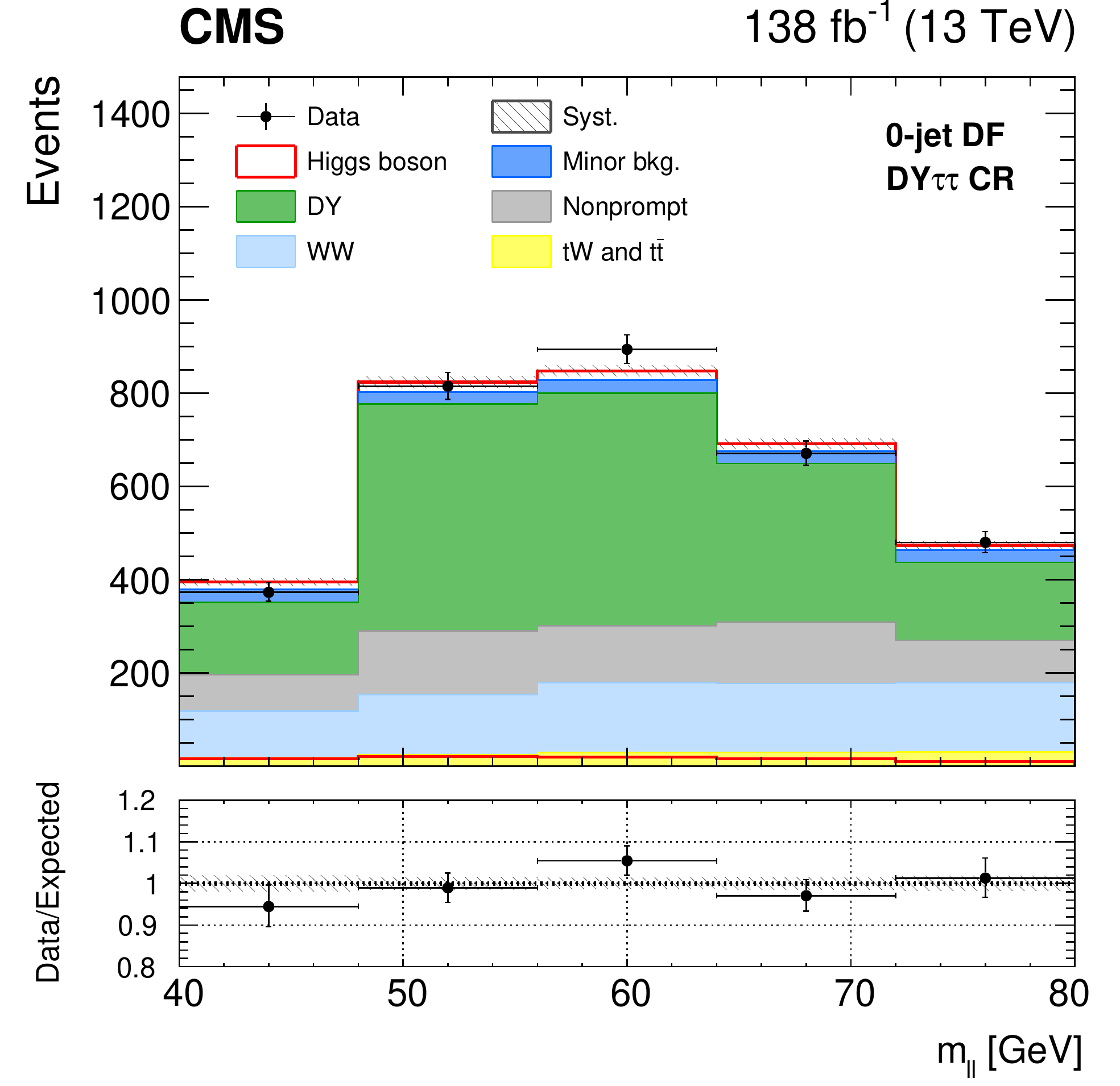}
    \includegraphics[width=0.49\textwidth]{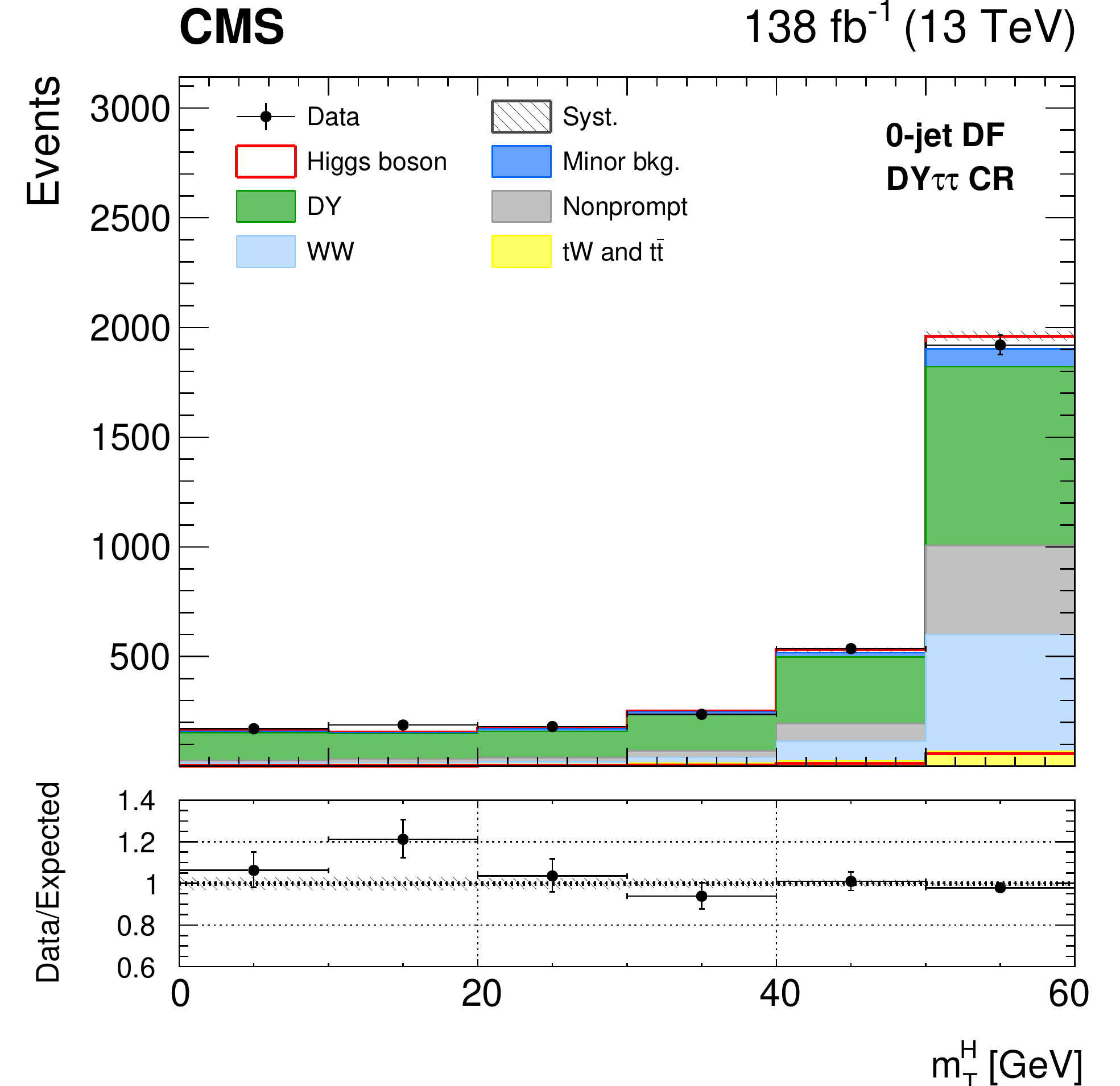}
    \caption{Observed distributions of the \mll (left) and \mTH (right) variables in the 0-jet DF \tautau control region. A detailed description is given in the Fig.~\ref{fig:mll_mth_ggH_DF_0j} caption.}
    \label{fig:mll_mth_dyttCR_0j}
\end{figure*}

\begin{figure*}[htbp]
    \centering
    \includegraphics[width=0.49\textwidth]{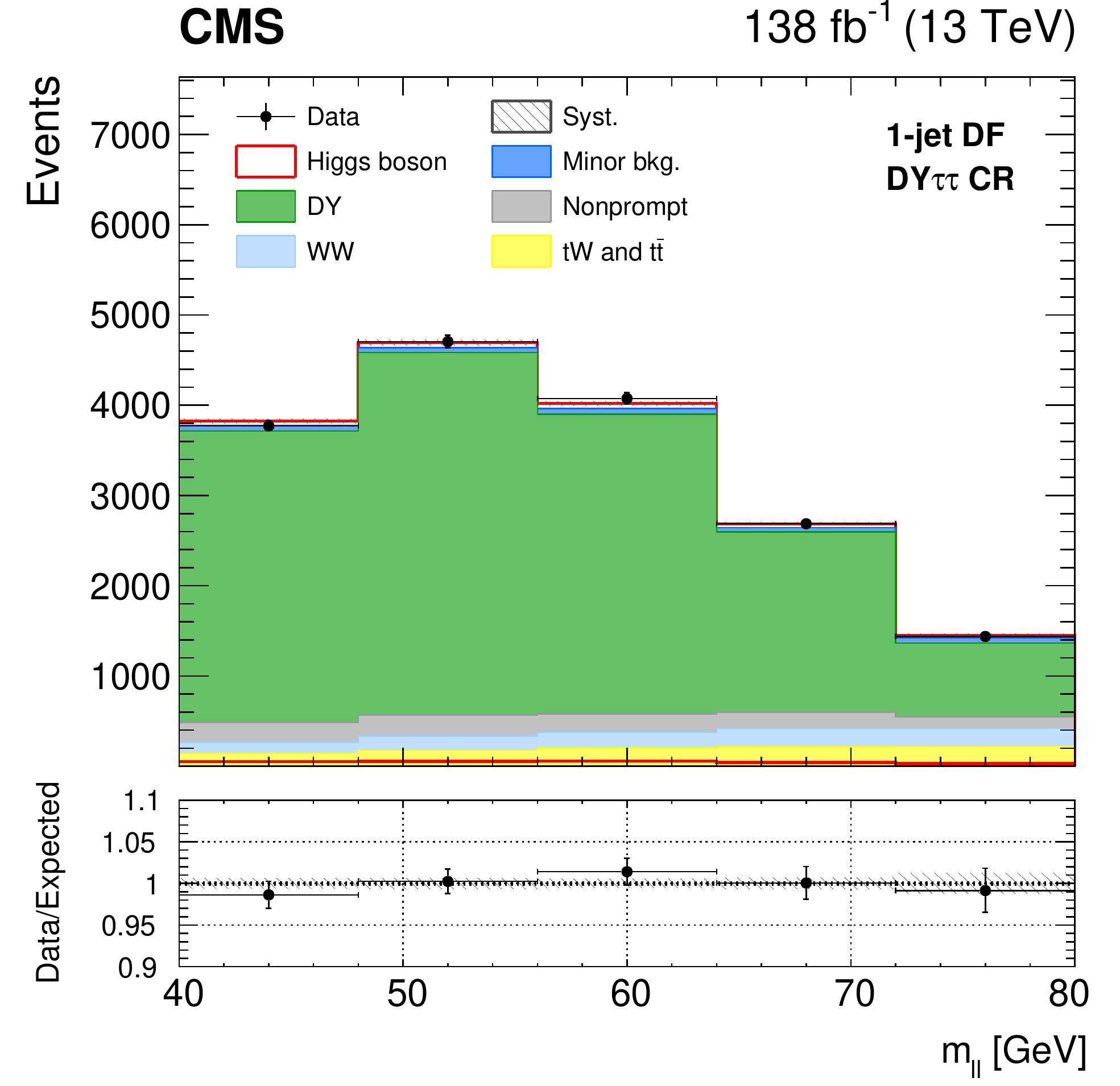}
    \includegraphics[width=0.49\textwidth]{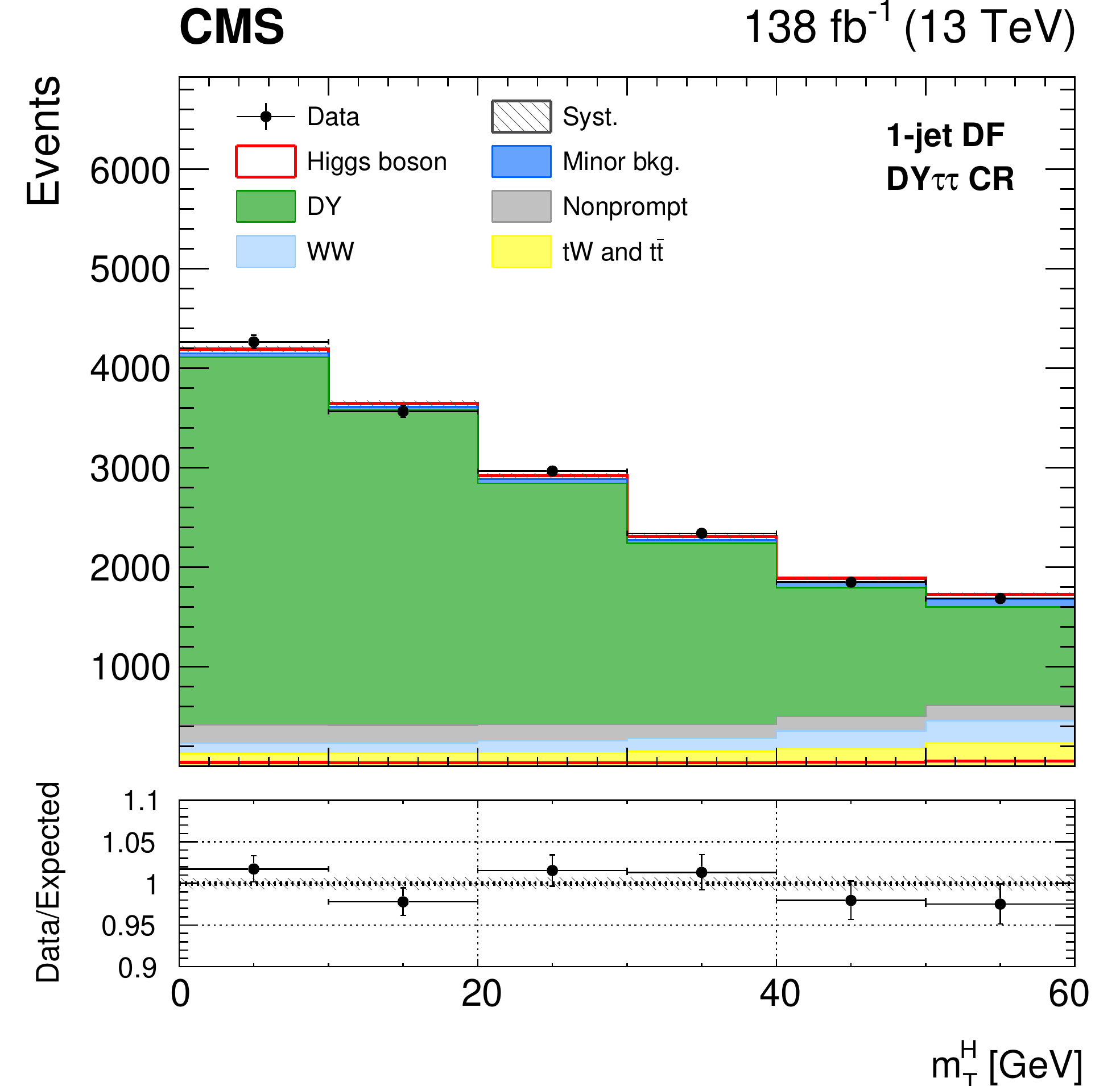}
    \caption{Observed distributions of the \mll (left) and \mTH (right) variables in the 1-jet DF \tautau control region. A detailed description is given in the Fig.~\ref{fig:mll_mth_ggH_DF_0j} caption.}
    \label{fig:mll_mth_dyttCR_1j}
\end{figure*}

\begin{figure*}[htbp]
    \centering
    \includegraphics[width=0.49\textwidth]{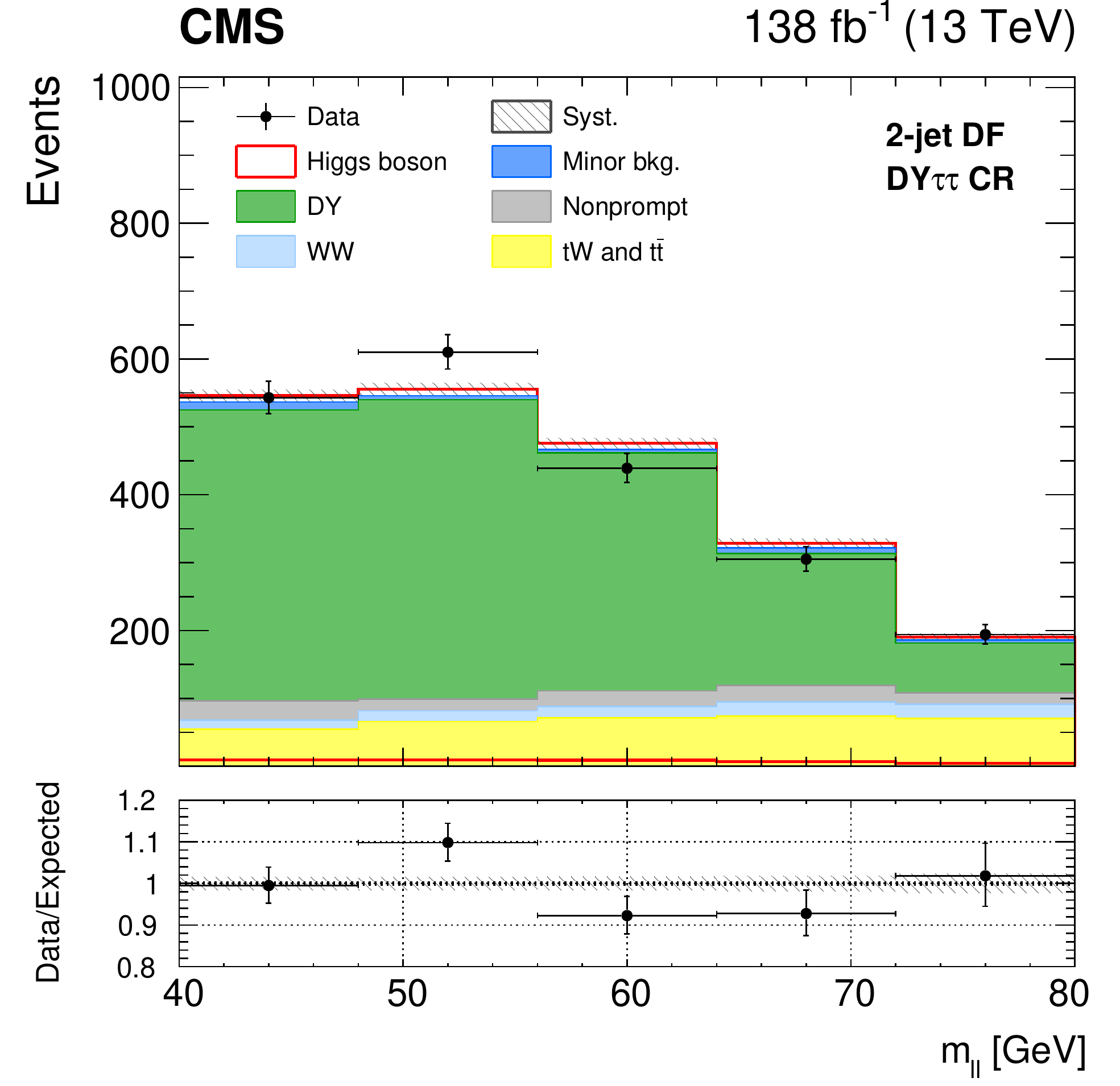}
    \includegraphics[width=0.49\textwidth]{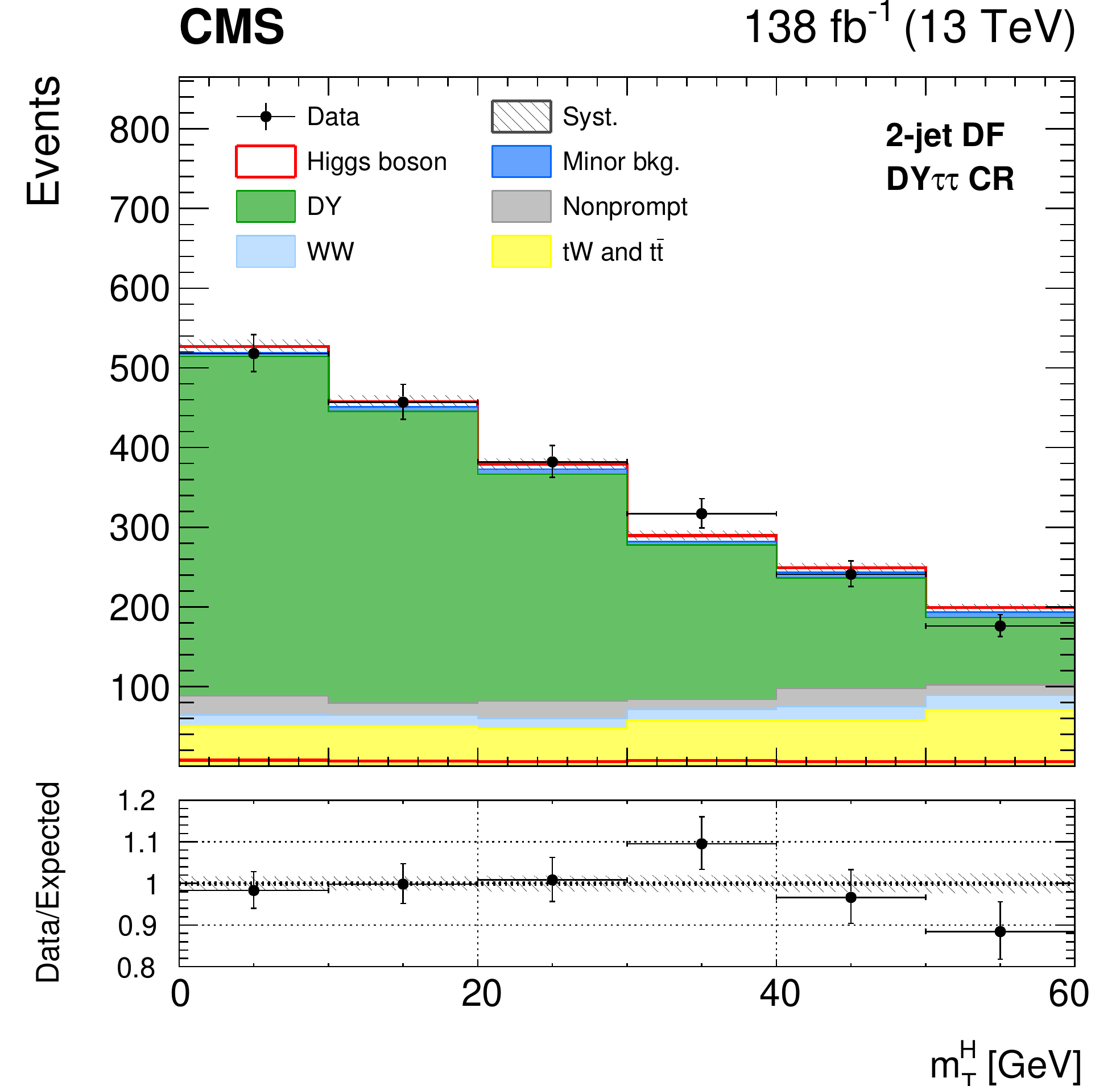}
    \caption{Observed distributions of the \mll (left) and \mTH (right) variables in the 2-jet DF \tautau control region. A detailed description is given in the Fig.~\ref{fig:mll_mth_ggH_DF_0j} caption.}
    \label{fig:mll_mth_dyttCR_2j}
\end{figure*}

In order to optimize background subtraction in the signal region (SR), two additional orthogonal selections are defined for each jet multiplicity category. These define two sets of control regions (CR), enriched in \tautau and top quark events, respectively. They are defined by the same selection as the SR, but inverting the \PQb jet veto for the top CR and the \mTH requirement for the \tautau CR. The full selection and categorization strategy is summarized in Table~\ref{tab:ggh_DF_selection}. Observed distributions for \mll and \mTH for the 0-, 1-, and 2-jet \ggH categories are shown in Figs.~\ref{fig:mll_mth_ggH_DF_0j},~\ref{fig:mll_mth_ggH_DF_1j}, and~\ref{fig:mll_mth_ggH_DF_2j}, respectively. The \WZ, \ZZ, $\PV\PGg$, $\PV\PGg^{\ast}$, and $\PV\PV\PV$ backgrounds are shown together as minor backgrounds. The observed \mll and \mTH distributions for the 0-, 1-, and 2-jet CRs enriched in top quark events are shown in Figs.~\ref{fig:mll_mth_topCR_0j},~\ref{fig:mll_mth_topCR_1j}, and~\ref{fig:mll_mth_topCR_2j}, and for the \tautau CRs in Figs.~\ref{fig:mll_mth_dyttCR_0j},~\ref{fig:mll_mth_dyttCR_1j}, and~\ref{fig:mll_mth_dyttCR_2j}.

\begin{table*}[h!]
    \centering
    \topcaption{Summary of the selection used in different-flavor \ggH categories.}
    \begin{tabular}{cc}
        \hline
        Subcategories & Selection \\
        \hline
        \underline{\textit{Global selection}} & \\
        \multirow{3}{*}{\NA} & $\ptone > 25\GeV$, $\pttwo > 10\GeV$ (2016) or 13\GeV \\
                             & $\ptmiss > 20\GeV$, $\ptll > 30\GeV$, $\mll > 12\GeV$ \\
                             & {\Pe}{\PGm} pair with opposite charge \\ [\cmsTabSkip]
        \underline{\textit{0-jet \ggH category}} & \\
        \multirow{4}{*}{$\Pell^{\pm}\Pell^{\mp}$, $\pttwo \lessgtr 20\GeV$} & $\mTH > 60\GeV$, $\mT({\Pell_2,\ptmiss}) > 30\GeV$ \\
                                                                           & $\pttwo \lessgtr 20\GeV$ \\
                                                                           & No jet with $\pt > 30\GeV$ \\
                                                                           & No \PQb-tagged jet with $\pt > 20\GeV$ \\ [\cmsTabSkip]
        \multirow{2}{*}{Top quark CR}                                      & As SR but with no \mTH requirement, $\mll>50\GeV$ \\
                                                                           & At least 1 \PQb-tagged jet with $20 < \pt < 30\GeV$ \\ [\cmsTabSkip]
        \multirow{2}{*}{\tautau CR}                                        & As SR but with $\mTH <60\GeV$ \\
                                                                           & $40 < \mll < 80\GeV$ \\ [\cmsTabSkip]
        \underline{\textit{1-jet \ggH category}} & \\
        \multirow{4}{*}{$\Pell^{\pm}\Pell^{\mp}$, $\pttwo \lessgtr 20\GeV$} & $\mTH > 60\GeV$, $\mT({\Pell_2,\ptmiss}) > 30\GeV$ \\
                                                                           & $\pttwo \lessgtr 20\GeV$ \\
                                                                           & 1 jet with $\pt > 30\GeV$ \\
                                                                           & No \PQb-tagged jet with $\pt > 20\GeV$ \\ [\cmsTabSkip]
        \multirow{2}{*}{Top quark CR}                                      & As SR but with no \mTH requirement, $\mll>50\GeV$ \\
                                                                           & At least 1 \PQb-tagged jet with $\pt > 30\GeV$ \\ [\cmsTabSkip]
        \multirow{2}{*}{\tautau CR}                                        & As SR but with $\mTH <60\GeV$ \\
                                                                           & $40 < \mll < 80\GeV$ \\ [\cmsTabSkip]
        \underline{\textit{2-jet \ggH category}} & \\
        \multirow{5}{*}{SR}                                                & $\mTH > 60\GeV$, $\mT({\Pell_2,\ptmiss}) > 30\GeV$ \\
                                                                           & $\pttwo \lessgtr 20\GeV$ \\
                                                                           & At least 2 jets with $\pt > 30\GeV$ \\
                                                                           & No \PQb-tagged jet with $\pt > 20\GeV$ \\
                                                                           & $\mjj<65\GeV$ or $105 < \mjj < 120\GeV$ \\ [\cmsTabSkip]
        \multirow{2}{*}{Top quark CR}                                      & As SR but with no \mTH requirement, $\mll>50\GeV$ \\
                                                                           & At least one \PQb-tagged jet with $\pt > 30\GeV$ \\ [\cmsTabSkip]
        \multirow{2}{*}{\tautau CR}                                        & As SR but with $\mTH <60\GeV$ \\
                                                                           & $40 < \mll < 80\GeV$ \\
        \hline
    \end{tabular}
    \label{tab:ggh_DF_selection}
\end{table*}




\subsection{Same-flavor ggH categories}
\label{subsec:ggH_SF}

The categories described in this section target the \ggH production mechanism in final states with either two electrons or two muons. The two leading leptons in the event are required to form an oppositely charged {\Pe}{\Pe} or {\PGm}{\PGm} pair. Events containing at least one \PQb-tagged jet with $\pt > 20\GeV$ are discarded. Low-mass resonances are suppressed by requiring $\mll>12\GeV$. The {\PW}+jets background is reduced by requiring the \pt of the dilepton system to exceed 30\GeV. Events are also required to have $\ptmiss>20\GeV$ to enrich the selection in processes with genuine missing transverse momentum. Finally, to reduce the DY background, which is dominant in this channel, a veto is placed on events in which \mll is within 15\GeV of the nominal mass of the \PZ boson ($m_\PZ$).

Events are divided in subcategories based on the number of hadronic jets, and further selections on \mTH, \mll, and the azimuthal angle between the two leading leptons (\dphill) are applied depending on the subcategory. A dedicated multivariate discriminant based on a DNN, called DYMVA in the following, is built and trained with the \textsc{TensorFlow} package~\cite{tensorflow2015-whitepaper} to distinguish signal events from DY events. The DNN is trained separately for each jet multiplicity subcategory. The architecture of the DNN is that of a feed-forward multilayer perceptron, taking 21, 22, and 27 input variables in the 0-, 1-, and 2-jet categories, respectively. These include kinematic information from the dilepton system, \ptvecmiss, and jets where present. To better constrain the top quark and \WW backgrounds, two CRs are defined in each jet multiplicity subcategory, enriched in the respective processes. The full selection is given in Table~\ref{tab:ggh_SF_selection}. The selection efficiency of the requirement on the DYMVA score in 0-jet categories is found to be approximately 50, 7, and 30\% for signal, DY, and total background events, respectively. In 1- and 2-jet categories the corresponding efficiencies are $\approx$50, 1, and 10\%.
\begin{table*}[htbp]
    \centering
    \topcaption{Summary of the selection used in same-flavor \ggH categories. The DYMVA threshold is optimized separately in each subcategory and data set.}
    \begin{tabular}{cc}
        \hline
        Subcategories & Selection \\
        \hline
        \underline{\textit{Global selection}}    & \\
        \multirow{4}{*}{\NA}                     & $\ptone > 25\GeV$, $\pttwo > 10\GeV$ (2016) or 13\GeV \\
                                                 & $\ptmiss > 20\GeV$, $\ptll > 30\GeV$ \\
                                                 & {\Pe}{\Pe} or {\PGm}{\PGm} pair with opposite charge \\
                                                 & $\mll>12\GeV$, $\abs{\mll-m_\PZ} > 15\GeV$ \\ [\cmsTabSkip]
        \underline{\textit{0-jet \ggH category}} & \\
        \multirow{3}{*}{$\Pe\Pe$, $\PGm\PGm$}    & $\mll < 60\GeV$, $\mTH > 90\GeV$, $\abs{\dphill}<2.3$ \\
                                                 & No \PQb-tagged jets with $\pt > 20\GeV$ \\
                                                 & DYMVA above threshold \\ [\cmsTabSkip]
        \multirow{2}{*}{\WW CR}                  & As SR but with $\mll > 100\GeV$ \\
                                                 & $\mTH > 60\GeV$, $\mT({\Pell_2,\ptmiss}) > 30\GeV$ \\  [\cmsTabSkip]
        \multirow{2}{*}{Top quark CR}            & As SR but with $\mll > 100\GeV$, $\mT({\Pell_2,\ptmiss}) > 30\GeV$ \\
                                                 & At least one \PQb-tagged jet with $20 < \pt < 30\GeV$ \\ [\cmsTabSkip]
        \underline{\textit{1-jet \ggH category}} & \\
        \multirow{3}{*}{$\Pe\Pe$, $\PGm\PGm$}    & $\mll < 60\GeV$, $\mTH > 80\GeV$, $\abs{\dphill}<2.3$ \\
                                                 & No \PQb-tagged jets with $\pt > 20\GeV$ \\
                                                 & DYMVA above threshold \\ [\cmsTabSkip]
        \multirow{2}{*}{\WW CR}                  & As SR but with $\mll > 100\GeV$ \\
                                                 & $\mTH > 60\GeV$, $\mT({\Pell_2,\ptmiss}) > 30\GeV$ \\ [\cmsTabSkip]
        \multirow{2}{*}{Top quark CR}            & As SR but with $\mll > 100\GeV$, $\mT({\Pell_2,\ptmiss}) > 30\GeV$ \\
                                                 & At least one \PQb-tagged jet with $\pt > 30\GeV$ \\ [\cmsTabSkip]
        \underline{\textit{2-jet \ggH category}} & \\
        \multirow{3}{*}{$\Pe\Pe$, $\PGm\PGm$}    & $\mll < 60\GeV$, $65 < \mTH < 150\GeV$ \\
                                                 & No \PQb-tagged jets with $\pt > 20\GeV$ \\
                                                 & DYMVA above threshold \\ [\cmsTabSkip]
        \multirow{2}{*}{\WW CR}                  & As SR but with $\mll > 100\GeV$ \\
                                                 & $\mTH > 60\GeV$, $\mT({\Pell_2,\ptmiss}) > 30\GeV$ \\ [\cmsTabSkip]
        \multirow{2}{*}{Top quark CR}            & As SR but with $\mll > 100\GeV$, $\mT({\Pell_2,\ptmiss}) > 30\GeV$ \\
                                                 & At least one \PQb-tagged jet with $\pt > 30\GeV$ \\
        \hline
    \end{tabular}
    \label{tab:ggh_SF_selection}
\end{table*}
Once the selection is performed, the signal is extracted via a simultaneous fit to the number of events in each category.

\section{Vector boson fusion categories}
\label{subsec:vbf}

This section describes the categories targeting the VBF production mechanism, both in DF and SF final states.
This mode involves the production of a Higgs boson in association with a pair of forward-backward jets. The dijet system is characterized by a large \mjj, large pseudorapidity separation \detajj, and low hadronic activity in the pseudorapidity region between the tagging jets. The fully leptonic final state in the VBF category therefore consists of two isolated leptons, large \ptmiss from the two undetectable neutrinos, and a pair of forward-backward jets.
The main background processes for the VBF categories are the same as for the \ggH categories.
An additional complication however arises in the entanglement of VBF and \ggH events, given the identical decay mode and the fact that the \ggH cross section is larger than the VBF one by one order of magnitude.

\subsection{Different-flavor VBF categories}

On top of the common global selection, the same requirements on leptons and \ptmiss used in the DF \ggH categories are applied.
In this case, however, there are no subcategories based on jet multiplicity.
Instead, exactly two jets with $\pt>30\GeV$ and $\mjj>120\GeV$ are required,
while still requiring the absence of \PQb-tagged jets with  $\pt>20\GeV$. In this category the \textsc{DeepFlavor} tagger~\cite{CMS-DP-2017-013} is used.
Finally, $60 < \mTH < 125\GeV$ is required.

\begin{figure*}
    \centering
    \includegraphics[width=0.49\textwidth]{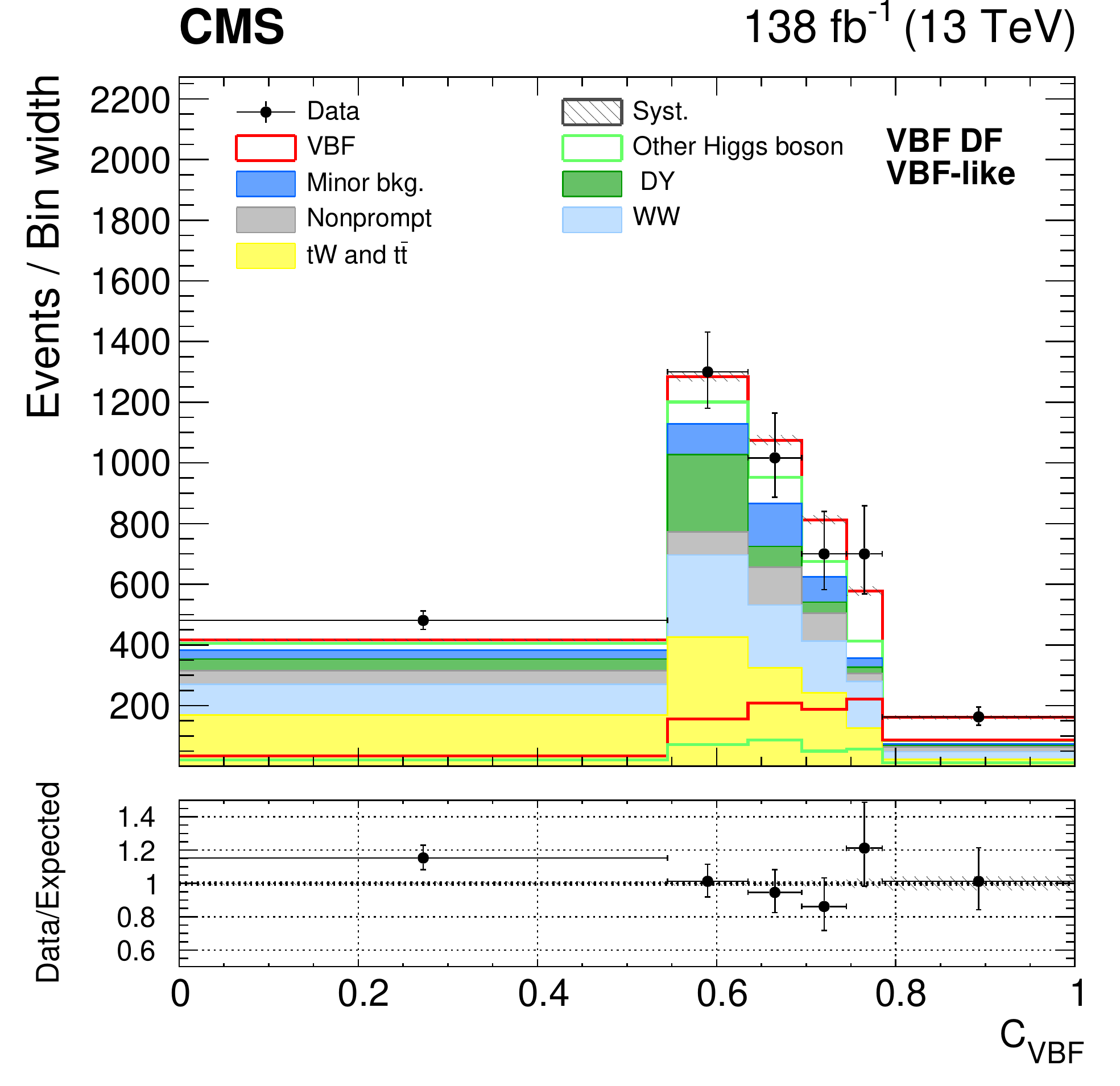}
    \includegraphics[width=0.49\textwidth]{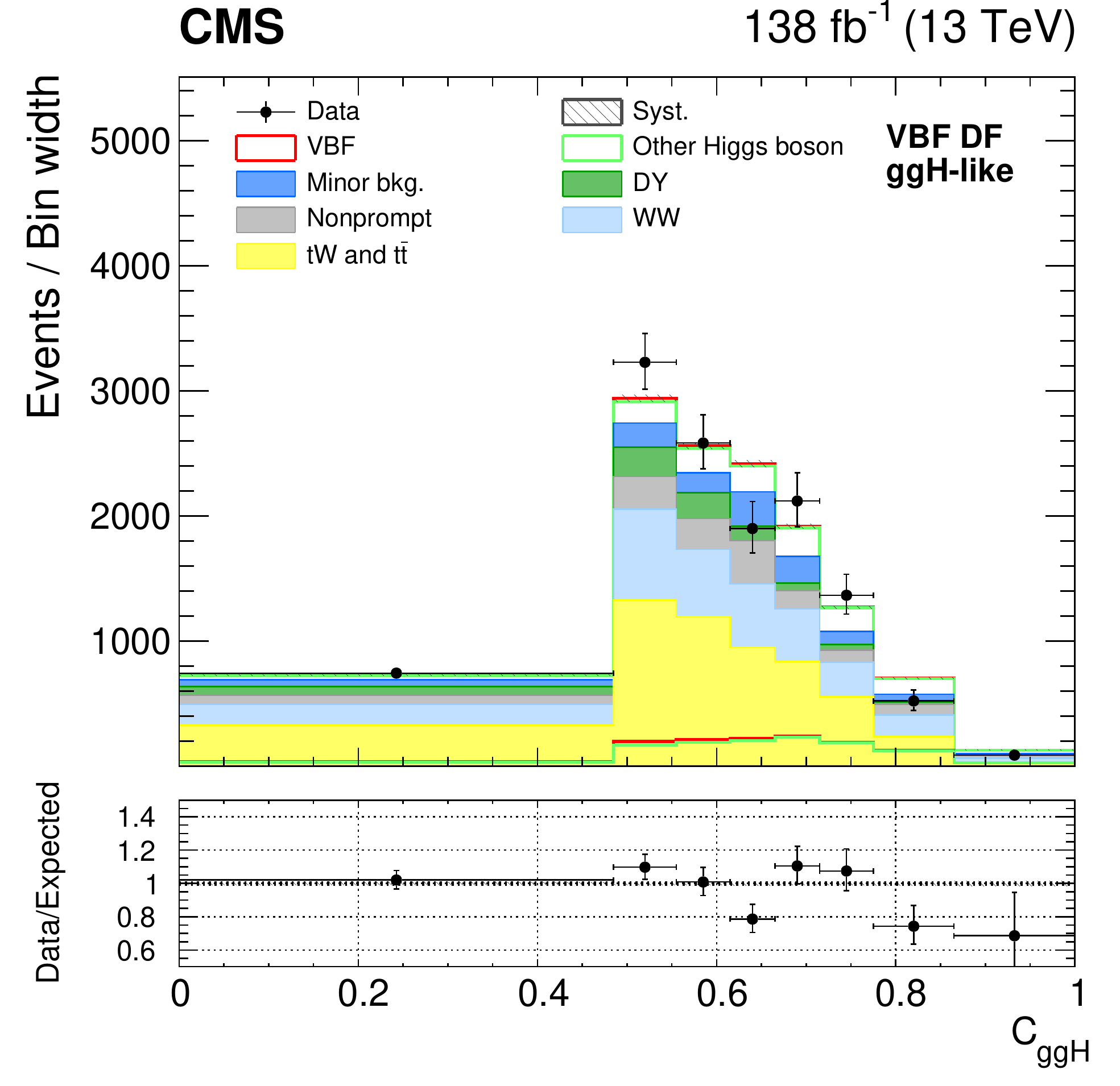}
    \caption{Distributions for the $C_{\mathrm{VBF}}$ (left) and $C_{\ggH}$ (right) classifiers in the \textit{VBF-like} and \textit{ggH-like} VBF DF categories, respectively. A detailed description is given in the Fig.~\ref{fig:mll_mth_ggH_DF_0j} caption.}
    \label{fig:vbf_df_disc}
\end{figure*}

In order to separate the signal from the background, a DNN approach has been followed.
The DNN is constructed to perform a multiclass classification of an event as either signal (VBF) or any of the three main background processes, namely:  \WW, top quark production, and \ggH.
As a result, a vector $\vec{\boldmath{o}}$ of four numbers is attributed to an event. Each number represents the degree of agreement of the event with the signal and the three background processes. 
Each of these outputs can be interpreted as a probability, since they are normalized to one.
Therefore, for a given event, the process $j$ with the highest output $o_j$ is interpreted as the most probable process. For this reason, the four outputs are referred to as classifiers: $C_{\mathrm{VBF}}$, $C_{\PQt}$, $C_{\WW}$, and $C_{\ggH}$.
In the SR four orthogonal categories are made using the classifiers.
If, for a given event, $C_j$ is higher than the other three, 
the event is classified in the \textit{j-like} category, and $C_j$ is used as the discriminating variable.
A shape-based analysis is hence performed in these categories. The DNN is trained on a set of 26 input variables, including kinematic information from the dilepton system, \ptvecmiss, and jets. The variables with the most discrimination power are found to be \mjj, \detajj and \mll.
As done in the DF \ggH categories, in order to optimize background subtraction in the SR, two CRs are defined, enriched in \tautau and top quark events, respectively. 
They are defined by the same selection as the SR, but inverting the \PQb jet veto for the top quark CR and the \mTH requirement for the \tautau CR. 
The full selection and categorization strategy is summarized in Table~\ref{tab:vbf_df_selection}. 
Observed distributions for the $C_{\mathrm{VBF}}$ and $C_{\ggH}$ classifiers in the \textit{VBF-like} and \textit{ggH-like} categories respectively are shown in Fig.~\ref{fig:vbf_df_disc}.

\begin{table*}[htbp]
    \centering
    \topcaption{Selection used in the different-flavor VBF categories.}
    \begin{tabular}{cc}
        \hline
        Subcategories & Selection \\
        \hline
        \underline{\textit{Global selection}} \\
        \multirow{3}{*}{\NA} & $\ptone > 25\GeV$, $\pttwo > 10\GeV$ (2016) or 13\GeV \\
                             & $\ptmiss > 20\GeV$, $\ptll > 30\GeV$, $\mll > 12\GeV$ \\
                             & {\Pe}{\PGm} pair with opposite charge \\ [\cmsTabSkip]
        \underline{\textit{2-jet VBF category}} \\
        \multirow{3}{*}{SR}            & $60 < \mTH < 125\GeV$, $m_\mathrm{T}({\Pell_2,\ptmiss}) > 30\GeV$ \\
                                       & 2 jets with $\pt > 30\GeV$, $ m_\mathrm{jj} > 120\GeV$ \\
                                       & No \PQb-tagged jet with $\pt > 20\GeV$ \\ [\cmsTabSkip]
        \multirow{2}{*}{Top quark CR}  & As SR but with no \mTH requirement, $\mll>50\GeV$ \\
                                       & At least one \PQb-tagged jet with $\pt > 30\GeV$ \\ [\cmsTabSkip]
        \multirow{2}{*}{\tautau CR}    & As SR but with $\mTH <60\GeV$ \\
                                       & $40 < \mll < 80\GeV$ \\
        \hline
    \end{tabular}
    \label{tab:vbf_df_selection}
\end{table*}

\begin{figure}
    \centering
    \includegraphics[width=\cmsFigureWidth]{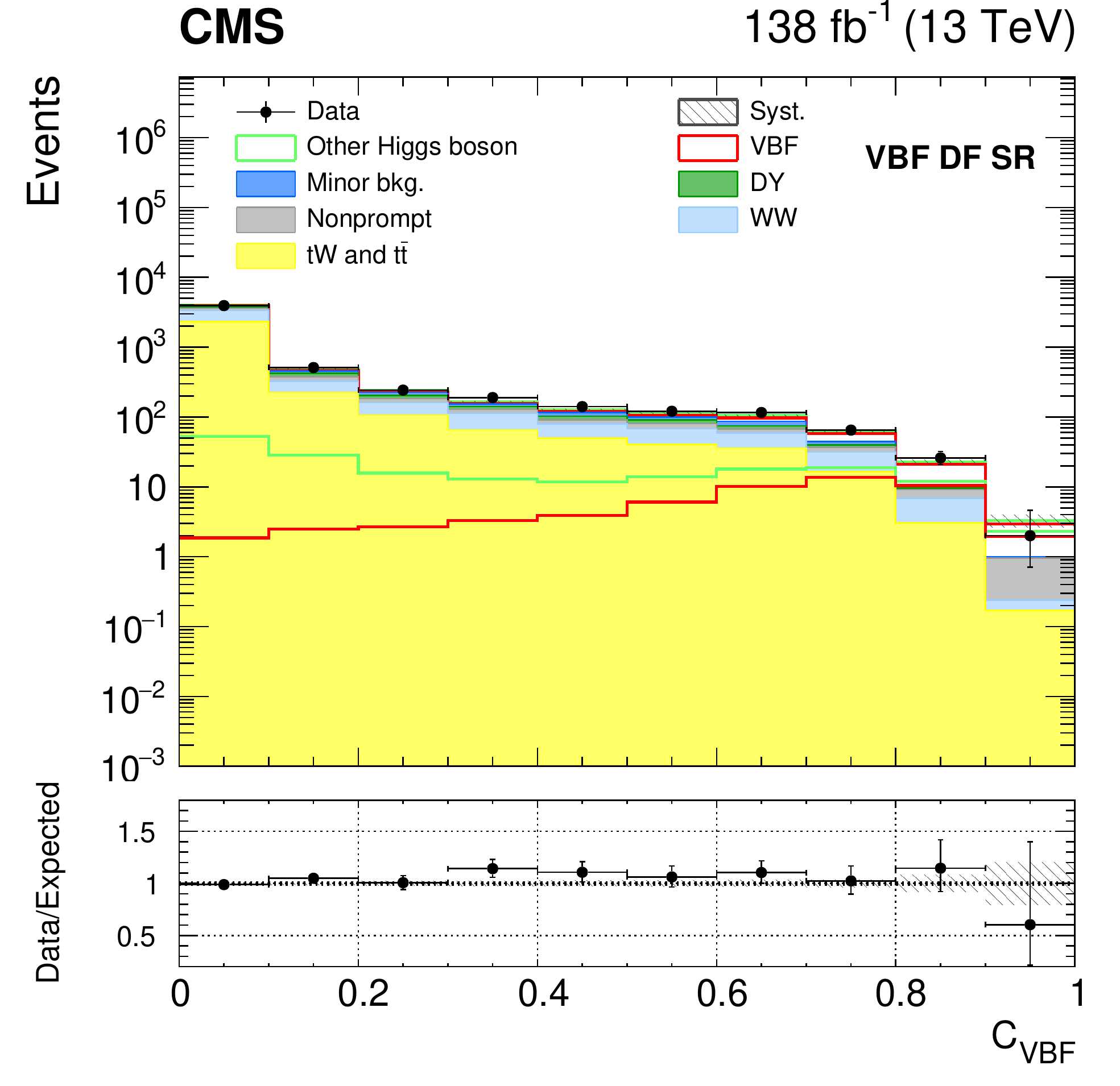}
    \caption{Distribution of the $C_{\mathrm{VBF}}$ classifier in the VBF DF SR, before the further event categorization based on the classifier outputs. A detailed description is given in the Fig.~\ref{fig:mll_mth_ggH_DF_0j} caption.}
    \label{fig:vbf_disc_presel}
\end{figure}

In order to verify that the simulated background processes agree with data in the DNN classifiers, the distributions are also checked at the level of the VBF SR global selection, \ie, before the further event categorization based on the classifier outputs. The 
$C_{\mathrm{VBF}}$ DNN output in the aforementioned global selection region is shown in Fig.~\ref{fig:vbf_disc_presel}.

\subsection{Same-flavor VBF categories}

On top of the common global selection, the same selection used in the SF \ggH categories is applied.
However, in this case, at least two jets with $\pt > 30\GeV$ are required, with $\mjj > 350\GeV$, 
while also rejecting events that contain any \PQb-tagged jets with $\pt>20\GeV$.
To define a Higgs-boson-enriched phase space, a selection on the DYMVA DNN is added. The DNN is trained and optimized separately in each category.
Two background CRs help in constraining the normalization of the top quark and \WW backgrounds. 
These CRs consist in regions of phase space orthogonal but as close as possible to the signal phase space.
This channel utilizes a simple counting experiment analysis, thus the event requirements are chosen to maximize the expected signal significance.
The full selection and categorization strategy is summarized in Table~\ref{tab:vbf_sf_selection}. 

\begin{table*}[htbp]
    \centering
    \topcaption{Selection used in the same-flavor VBF categories. The DYMVA threshold is optimized separately in each subcategory and data set.}
    \begin{tabular}{cc}
        \hline
        Subcategories & Selection \\
        \hline
        \underline{\textit{Global selection}} \\
        \multirow{4}{*}{\NA} & $\ptone > 25\GeV$, $\pttwo > 10\GeV$ (2016) or 13\GeV \\
                             & $\ptmiss > 20\GeV$, $\ptll > 30\GeV$ \\
                             & {\Pe}{\Pe} or {\PGm}{\PGm} pair with opposite charge \\
                             & $\mll > 12\GeV$, $\abs{\mll-m_\PZ}>15\GeV$ \\
        \underline{\textit{2-jet VBF category}} \\
        \multirow{5}{*}{$\Pe\Pe$, $\PGm\PGm$} & $\mll < 60\GeV$, $65 < \mTH < 150\GeV$ \\
                                              & At least 2 jets with $\pt > 30\GeV$ \\
                                              & $\abs{\dphill}<1.6$,  $\mjj > 350\GeV$ \\
                                              & No \PQb-tagged jets with $\pt > 20\GeV$ \\
                                              & DYMVA above threshold \\ [\cmsTabSkip]
        \multirow{2}{*}{\WW CR}               & As SR but with $\mll > 100\GeV$ \\
                                              & $\mTH > 60\GeV$, $\mT({\Pell_2,\ptmiss}) > 30\GeV$ \\ [\cmsTabSkip]
        \multirow{2}{*}{Top quark CR}         & As SR but with $\mll > 100\GeV$, $\mT({\Pell_2,\ptmiss}) > 30\GeV$ \\
                                              & At least one of the leading jets \PQb-tagged \\
        \hline
    \end{tabular}
    \label{tab:vbf_sf_selection}
\end{table*}

\section{Vector boson associated production categories}
\label{sec:vh}

This section describes categories targeting the \VH production mode. Four subcategories are defined (\WH{}SS, \WH{}3\Pell, \ZH{}3\Pell, and \ZH{}4\Pell) to target final states in which the vector boson \PV, produced in association with the Higgs boson, decays leptonically. Two more categories (\VH{}2j DF/SF) select events in which the V boson decays into two resolved jets. An additional selection is applied in each category to reduce the background, as well as an event categorization, defining phase spaces more sensitive to either signal or specific backgrounds. Details on the event selection and categorization are given below.

\subsection{WHSS categories}
\label{subsec:WHSS}

The \WH{}SS category targets the $\PW\PH\to 2\Pell 2\PGn \PQq\PQq$ final state, where the two charged leptons are required to have same sign to reduce DY background. Therefore, the final state contains two same-sign leptons, \ptmiss, and at least one jet. The analysis requires the leading (subleading) lepton to have $\pt>25\,(20)\GeV$. To remove contributions from low-mass resonances, \mll is required to be greater than 12\GeV. The two leptons must have a pseudorapidity separation ($\Delta\eta_{\Pell\Pell}$) of less than two. Events are also required to have $\ptmiss > 30\GeV$, as well as no \PQb-tagged jet with $\pt > 20\GeV$.

Signal region events are further categorized based on the number of jets and the lepton flavor composition. Events in the 1-jet category are required to contain exactly one jet with $\pt > 30\GeV$ and $\abs{\eta}<4.7$. Events in the 2-jet category are required to contain at least two jets with the same kinematic constraints. For events containing more than two jets, only the two jets with highest \pt are considered for the analysis. These jets must have $\mjj <100\GeV$. The SRs are further divided into {\Pe}{\PGm} and {\PGm}{\PGm} categories. Events with two electrons are not considered, as this flavor category is less sensitive to signal.

To improve discrimination between signal and background, the variable \mHtil is defined, which serves as a proxy for $m_\PH$. It is computed as the invariant mass of the dijet pair four-momentum $P_\mathrm{jj}=(E_\mathrm{jj}, \vec{p}_\mathrm{jj})$ and twice the four-momentum of the lepton closest to the dijet pair $P_{\Pell} = (p_{\Pell},\vec{p}_{\Pell})$:
\begin{linenomath}
\begin{equation}
    \mHtil=\sqrt{(P_{jj}+2P_{\Pell})^2}.
\end{equation}
\end{linenomath}
The second lepton four-momentum serves as a proxy for the neutrino. If an event in the 1-jet category contains a second jet with $20 < \pt < 30\GeV$, this jet is included in the computation of this variable; otherwise the four-momentum of the single jet is used. Events in all categories are required to have $\mHtil > 50\GeV$. A summary of the event selection is given in Table~\ref{tab:WHSS_selection}.

\begin{table*}
    \centering
    \topcaption{Event selection and categorization in the \WH{}SS category.}
    \begin{tabular}{cc}
        \hline
        Subcategories & Selection \\
        \hline
        \underline{\textit{Global selection}} \\
        \multirow{3}{*}{\NA}           & $\ptone > 25\GeV$, $\pttwo > 20\GeV$ \\
                                       & $\mll > 12\GeV$, $\abs{\Delta\eta_{\Pell\Pell}} < 2$, $\ptmiss > 30\GeV$ \\
                                       & $\mHtil > 50\GeV$, no \PQb-tagged jet with $\pt > 20\GeV$ \\ [\cmsTabSkip]
        \underline{\textit{Signal region}} \\
        \multirow{2}{*}{1-jet $\Pe\PGm (\PGm\PGm)$} & One jet with $\pt > 30\GeV$ \\
                                                    & $\Pe\PGm (\PGm\PGm)$ pair with same charge \\ [\cmsTabSkip]
        \multirow{2}{*}{2-jet $\Pe\PGm (\PGm\PGm)$} & At least two jets with $\pt > 30\GeV$, $\mjj <100\GeV$ \\
                                                    & $\Pe\PGm (\PGm\PGm)$ pair with same charge \\ [\cmsTabSkip]
        \underline{\textit{Control region}} \\
        WZ                                          & Shared with \ZH{}3\Pell \\
        \hline
    \end{tabular}
    \label{tab:WHSS_selection}
\end{table*}

The main backgrounds in the \WH{}SS category are \WZ, {\PW}+jets, $\PV\PGg$, and $\PV\PGg^{\ast}$ production. Additional backgrounds are top quark, triboson, \WW, and \ZZ production. The {\PW}+jets events pass the selection when a nonprompt lepton passes the lepton selection. This nonprompt background is estimated from data, as described in Section~\ref{sec:background}. The remaining backgrounds are estimated using MC simulation. The \WZ background normalization is estimated in the 1- and 2-jet CRs shared with the \ZH{}3\Pell category, described in Section~\ref{subsec:ZH3l}.

\begin{figure*}
    \centering
    \includegraphics[width=0.49\textwidth]{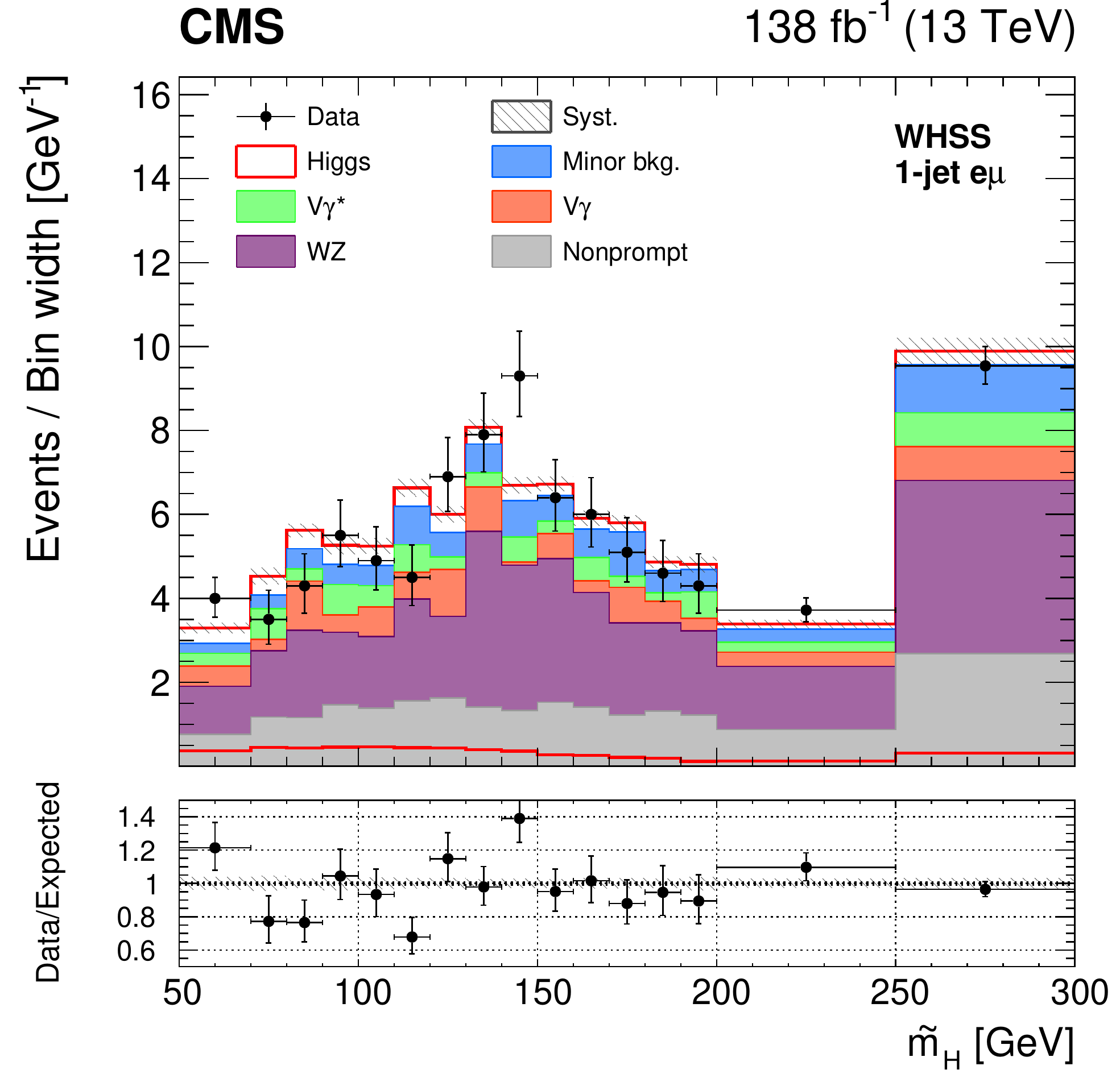}
    \includegraphics[width=0.49\textwidth]{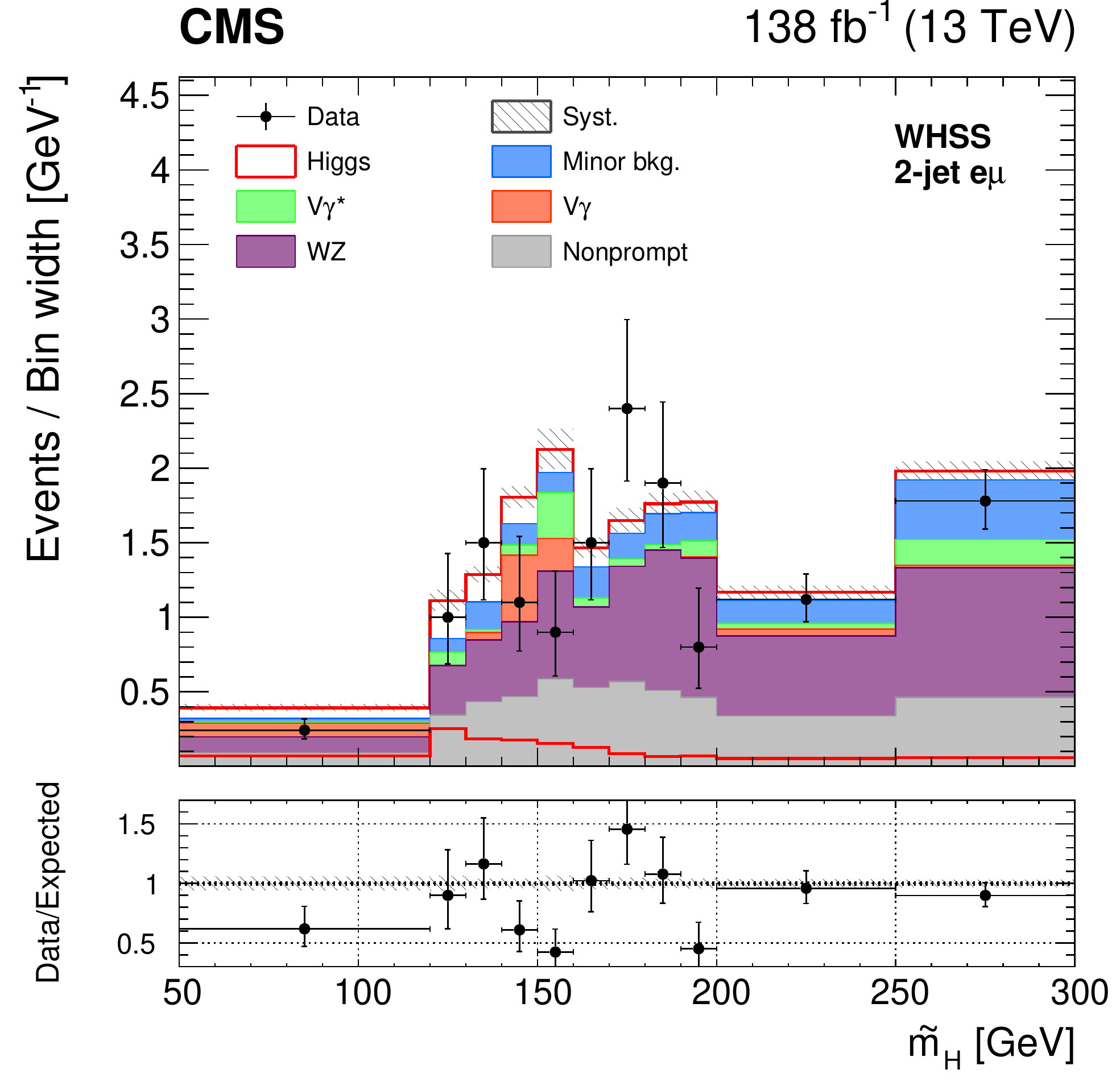}
    \\
    \includegraphics[width=0.49\textwidth]{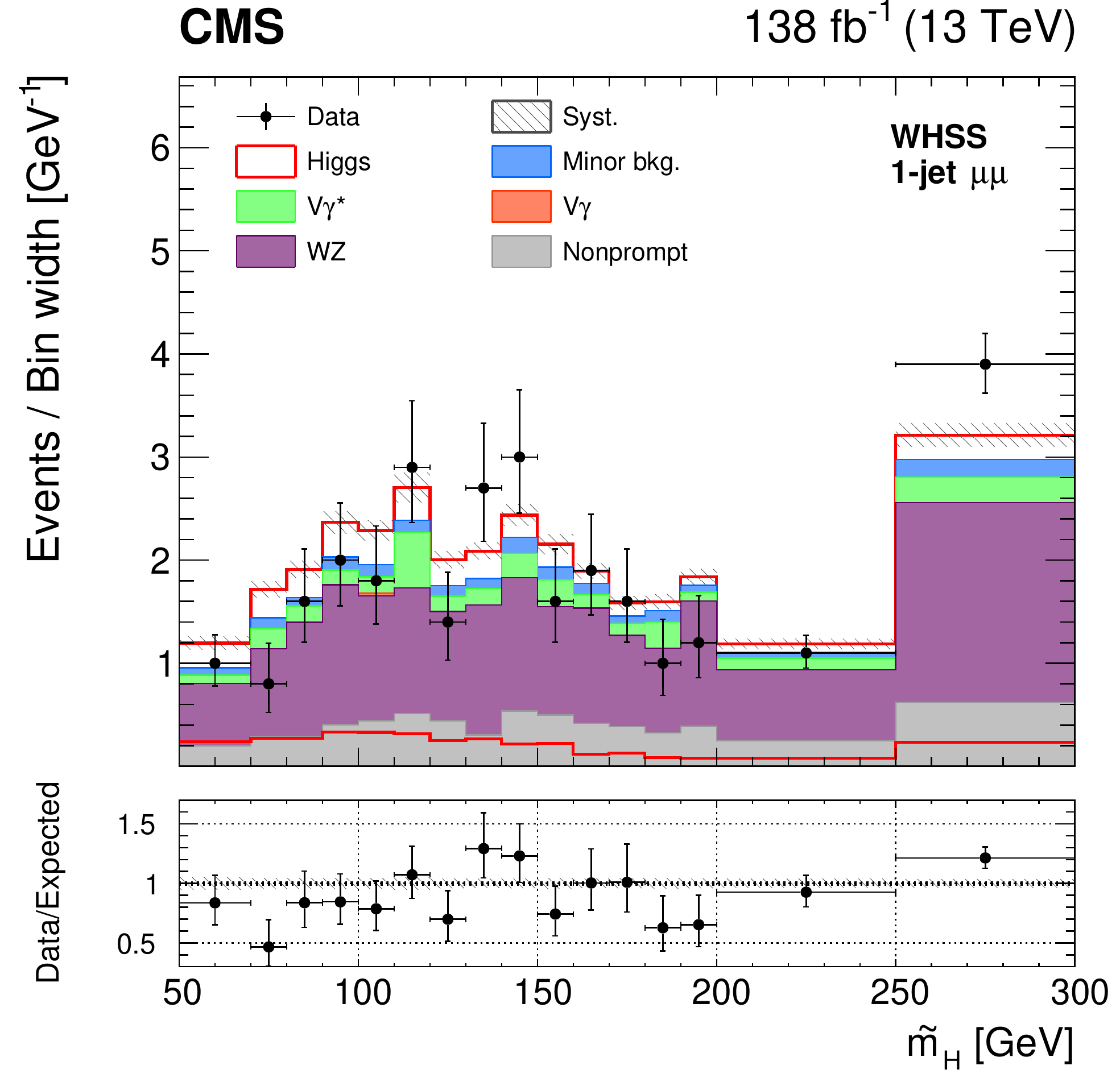}
    \includegraphics[width=0.49\textwidth]{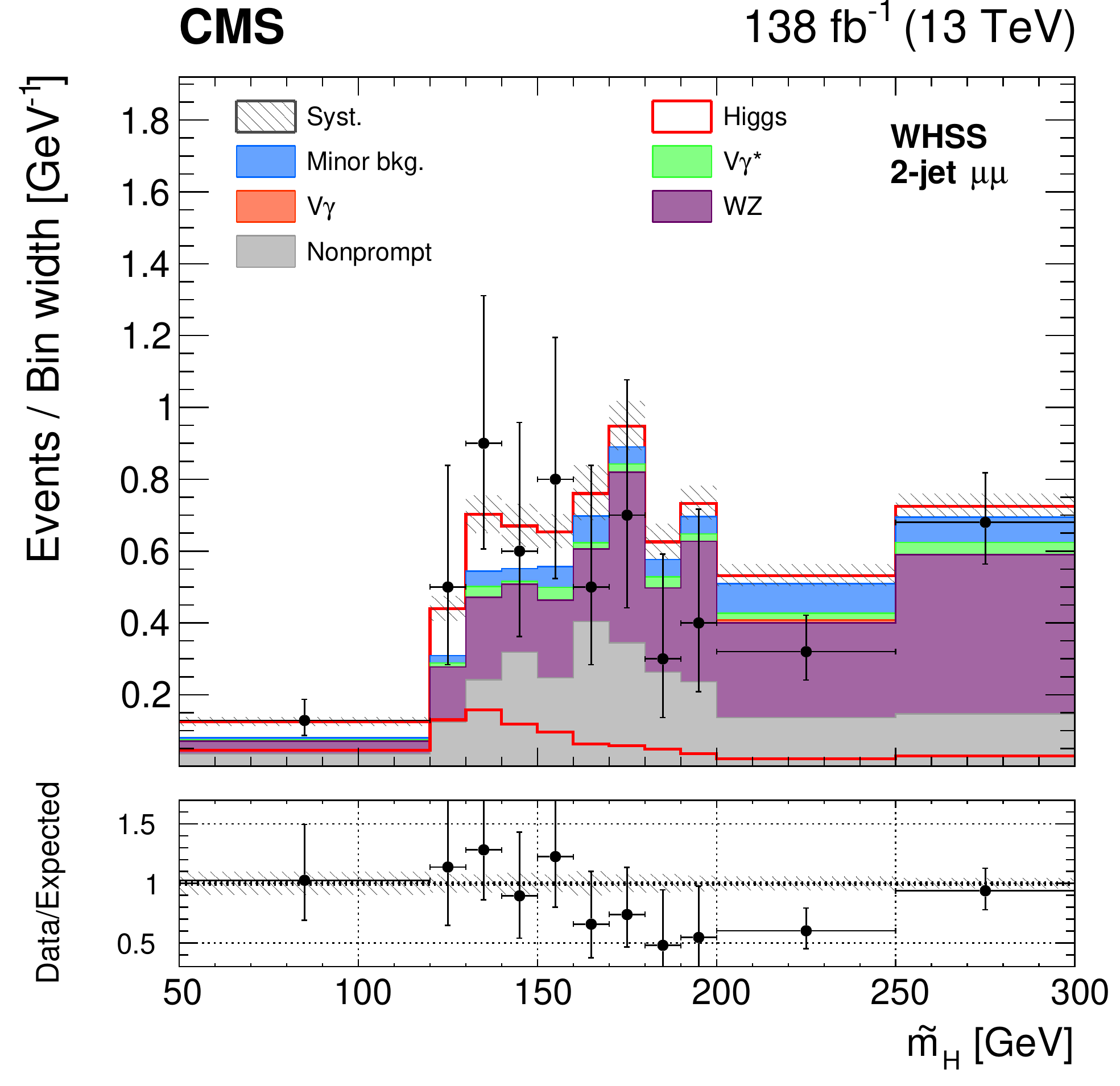}
    \caption{Observed distributions of the \mHtil fit variable in the \WH{}SS 1-jet {\Pe}{\PGm} (upper left), 2-jet {\Pe}{\PGm} (upper right), 1-jet {\PGm}{\PGm} (lower left), and 2-jet {\PGm}{\PGm} (lower right) SRs. A detailed description is given in the Fig.~\ref{fig:mll_mth_ggH_DF_0j} caption.}
    \label{fig:mlljj20_whss}
\end{figure*}

To extract the Higgs boson production cross section, a binned fit is performed to the \mHtil variable. Figure~\ref{fig:mlljj20_whss} shows the \mHtil distribution after the fit to the data.

\subsection{WH3\texorpdfstring{\Pell}{l} categories}
\label{subsec:WH3l}

The \WH{}3\Pell category targets the $\PW\PH\to 3\Pell 3\PGn$ decay. The final state therefore contains three leptons and \ptmiss. The analysis selects events containing three leptons with $\pt>25$, 20, and 15\GeV, respectively and total charge ($\text{Q}_{3\Pell}$) $\pm$1. The invariant mass of any dilepton pair is required to be greater than 12\GeV to remove low-mass resonances. Events are rejected if they contain a jet with $\pt > 30\GeV$, or any \PQb-tagged jet with $\pt > 20\GeV$.

Events in the SR are categorized based on the flavor composition of the lepton pairs. Events with at least one opposite-sign SF (OSSF) lepton pair are placed in the OSSF category, while all other events are placed in the same-sign SF (SSSF) category. To reject backgrounds containing \PZ bosons, events in the OSSF SR must pass a \PZ boson veto, where all lepton pairs must satisfy $\abs{\mll - m_\PZ} > 20\GeV$, as well as $\ptmiss > 40\GeV$.

The main backgrounds in the \WH{}3\Pell category are \WZ, \ZZ, $\PV\PGg$, and $\PV\PGg^{\ast}$ production, as well as backgrounds containing nonprompt leptons. Nonprompt backgrounds are estimated from data, as described in Section~\ref{sec:background}. The remaining backgrounds are estimated from simulated samples. The \WZ and {\PZ}{\PGg} backgrounds are normalized using dedicated CRs, matching the OSSF SR with the exception of an inverted \PZ boson veto, a differing \ptmiss requirement, and an additional selection on the invariant mass of the full lepton system ($m_{3\Pell}$). A summary of the event selection and categorization is given in Table~\ref{tab:WH3l_selection}.

\begin{table*}
    \centering
    \topcaption{Event selection and categorization in the \WH{}3\Pell category.}
    \begin{tabular}{cc}
        \hline
        Subcategories & Selection \\
        \hline
        \underline{\textit{Global selection}} \\
        \multirow{4}{*}{\NA}           & $\ptone > 25\GeV$, $\pttwo > 20\GeV$, $\ptthr > 15\GeV$ \\
                                       & $\text{Q}_{3\Pell}=\pm 1$, $\min(\mll) > 12\GeV$, $\Delta\eta_{\Pell\Pell}>2.0$ \\
                                       & $\ptmiss > 30\GeV$, $\mHtil>50\GeV$\\
                                       & No jets with $\pt > 30\GeV$, no \PQb-tagged jet with $\pt > 20\GeV$ \\ [\cmsTabSkip]
        \underline{\textit{Signal region}} \\
        OSSF                           & OSSF lepton pair, $\abs{\mll - m_\PZ} > 20\GeV$, $\ptmiss>40\GeV$ \\ [\cmsTabSkip]
        SSSF                           & No OSSF lepton pair \\ [\cmsTabSkip]
        \underline{\textit{Control region}} \\
        \multirow{2}{*}{\WZ}           & OSSF lepton pair, $\abs{\mll - m_\PZ} < 20\GeV$ \\
                                       & $\ptmiss>45\GeV$, $m_{3\Pell}>100\GeV$\\ [\cmsTabSkip]
        \multirow{2}{*}{$\PZ\PGg$}     & OSSF lepton pair, $\abs{\mll - m_\PZ} < 20\GeV$ \\
                                       & $\ptmiss<40\GeV$, $80<m_{3\Pell}<100\GeV$\\   
        \hline
    \end{tabular}
    \label{tab:WH3l_selection}
\end{table*}

\begin{figure*}
    \centering
    \includegraphics[width=0.49\textwidth]{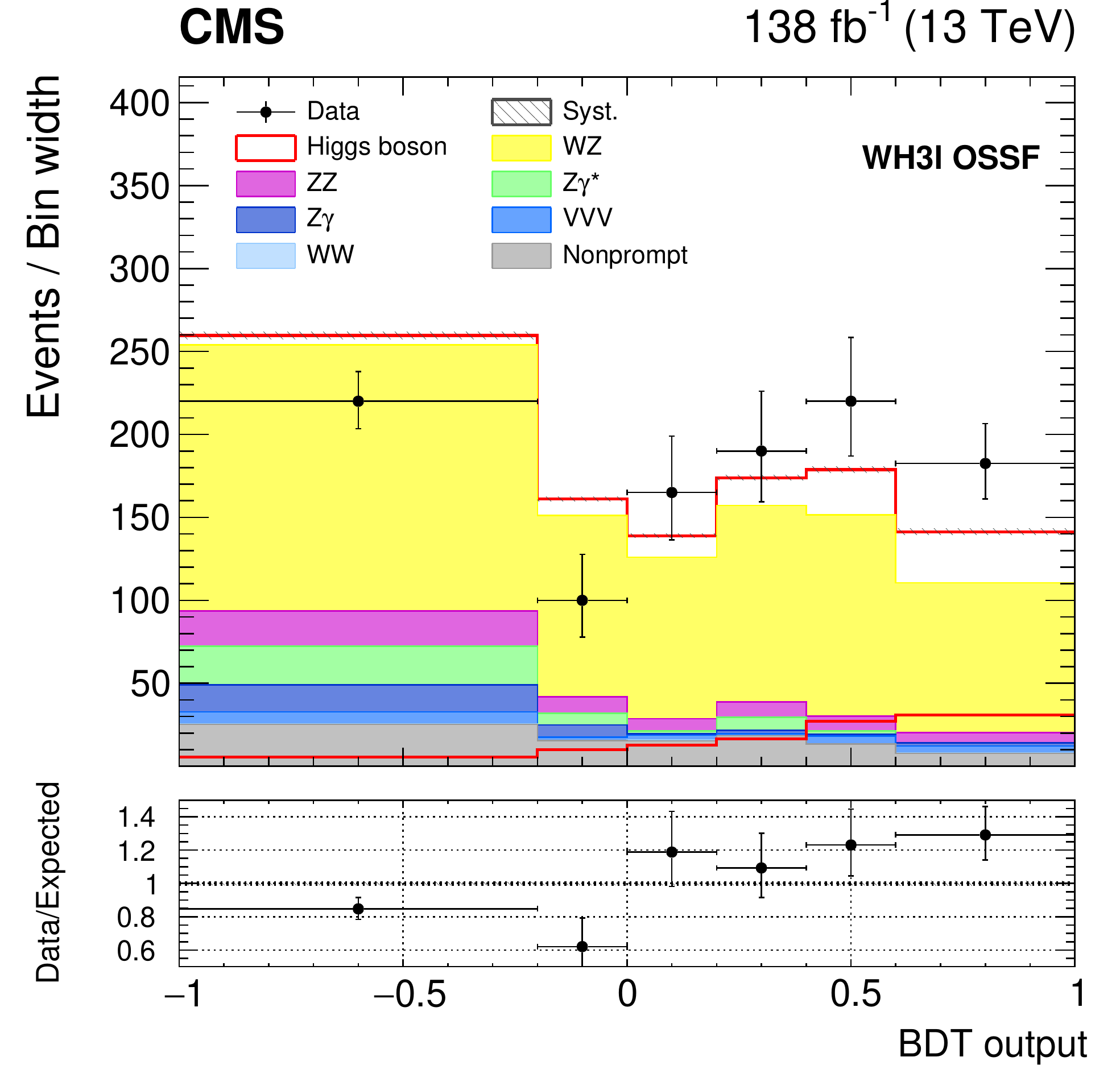}
    \includegraphics[width=0.49\textwidth]{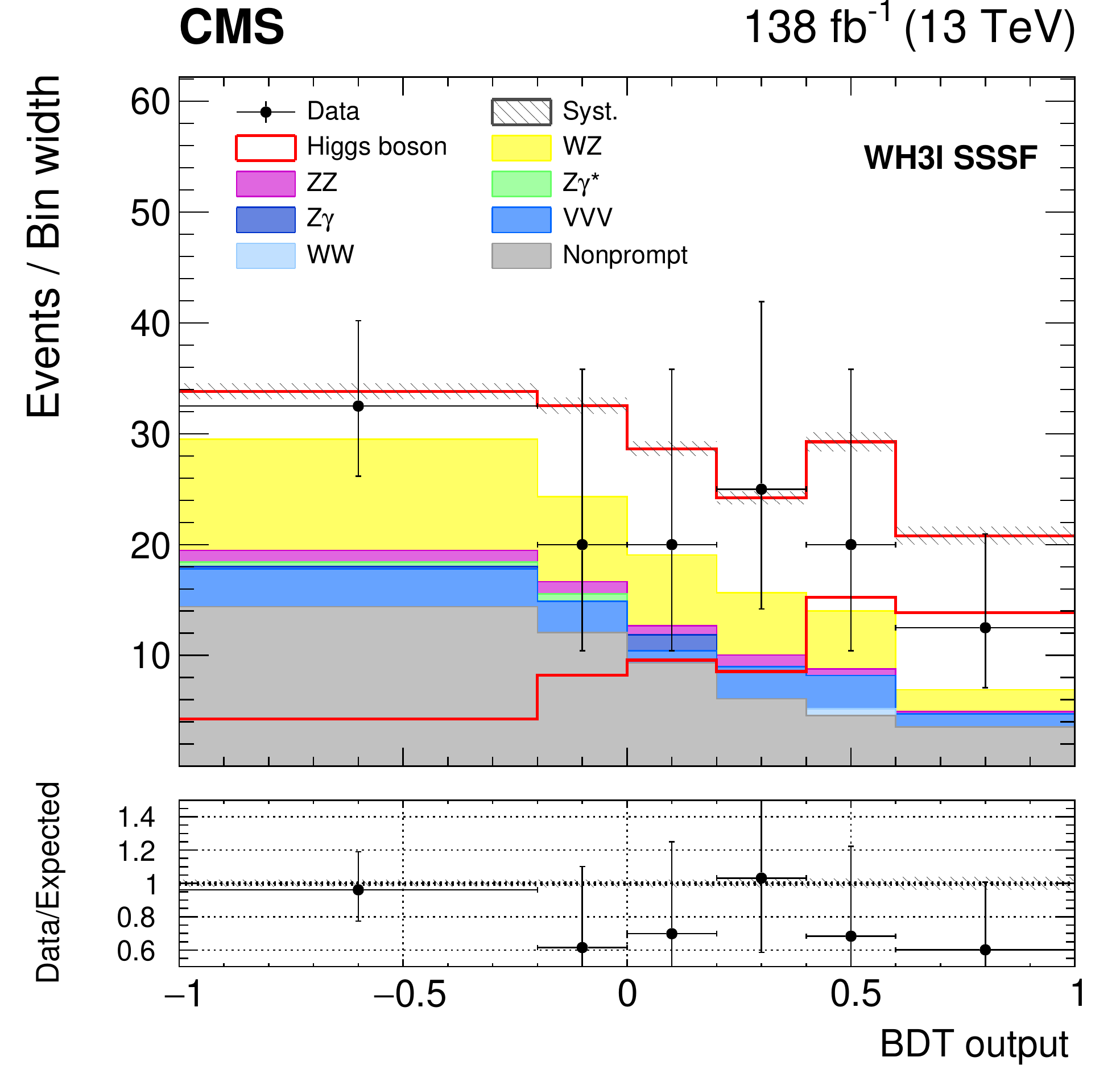}
    \caption{Observed distributions of the BDT score in the \WH{}3\Pell OSSF (left) and SSSF (right) SRs. A detailed description is given in the Fig.~\ref{fig:mll_mth_ggH_DF_0j} caption.}
    \label{fig:BDT_wh3l}
\end{figure*}

To discriminate between signal and background, two BDTs, trained separately for the OSSF and SSSF categories, are used. The BDTs are built using the TMVA~\cite{TMVA_arxiv} package and trained on events passing the OSSF and SSSF SR selections without the $\abs{\mll - m_\PZ}$ requirement. The number of input variables used in the BDT training is 19 and 15 in the OSSF and SSSF regions, respectively. They include kinematic information on the leptons, \ptvecmiss, \PQb tagging scores for the leading jets, and various invariant masses built from leptons and \ptvecmiss, with the minimum invariant mass and $\Delta R$ separation of the opposite sign lepton pairs giving the most discrimination power. To extract the Higgs boson production cross section, a binned fit is performed to the BDT score. Figure~\ref{fig:BDT_wh3l} shows the BDT discriminant distributions after the fit to the data.

\subsection{ZH3\texorpdfstring{\Pell}{l} categories}
\label{subsec:ZH3l}

The \ZH{}3\Pell category targets the $\PZ\PH\to 3\Pell\PGn \PQq\PQq$ decay. The final state therefore contains three leptons with total charge $\pm$1. The invariant mass of any dilepton pair is required to be greater than 12\GeV to reject low-mass resonances. The event must contain an OSSF lepton pair with invariant mass $\abs{\mll - m_\PZ} < 25\GeV$. Events are rejected if any \PQb-tagged jet with $\pt > 20\GeV$ passing the medium WP of the tagging algorithm is found.

Events are categorized based on the number of jets. Events in the 1-jet category contain exactly one jet with $\pt > 30\GeV$ and $\abs{\eta} < 4.7$, while events in the 2-jet category contain at least two jets passing these requirements. Signal region events must also have an azimuthal separation between the two \PW bosons ($\Delta\phi(\Pell\ptmiss, j(j))$), represented by the {\Pell}+{\ptmiss} and (di)jet systems respectively, below $\pi/2$, and pass a \PZ boson internal conversion veto $\abs{m_{3\Pell} - m_{\PZ}} > 20\GeV$.

The main backgrounds in the \ZH{}3\Pell analysis are \WZ, \ZZ, and {\PZ}+jets events. The {\PZ}{\PGg}/$\PGg^{\ast}$, $\PV\PV\PV$, and {\ttbar}+jets processes also contribute. The {\PZ}+jets events pass the selection when a nonprompt lepton passes the lepton selection. This background is estimated from data as described in Section~\ref{sec:background}. The remaining backgrounds are modeled using MC simulation. The \WZ normalization as a function of the number of jets is extracted from dedicated CRs, which are categorized by the number of jets in the same way as the SRs. The \WZ CRs are also used to constrain the \WZ background in the \WH{}SS category. A summary of the event selection and categorization is shown in Table~\ref{tab:ZH3l_selection}.

\begin{table*}
    \centering
    \topcaption{Event selection and categorization in the \ZH{}3\Pell category.}
    \begin{tabular}{cc}
        \hline
        Subcategories & Selection \\
        \hline
        \underline{\textit{Global selection}} \\
        \multirow{4}{*}{\NA}           & $\ptone > 25\GeV$, $\pttwo > 20\GeV$, $\ptthr > 15\GeV$ \\
                                       & $\text{Q}_{3\Pell}=\pm 1$, $\min(m_{\Pell\Pell}) > 12\GeV$ \\
                                       & $\abs{m_{\Pell\Pell} - m_{\PZ}}<25\GeV$, $\abs{m_{3\Pell} - m_{\PZ}}>20\GeV$\\
                                       & No \PQb-tagged jet with $\pt > 20\GeV$ \\ [\cmsTabSkip]
        \underline{\textit{Signal region}} \\
        1-jet                                 & $=$1 jet with $\pt>30\GeV$, $\Delta\phi(\Pell\ptmiss,j(j))< \pi/2$ \\ [\cmsTabSkip]
        2-jet                                 & $\geq$2 jets with $\pt>30\GeV$, $\Delta\phi(\Pell\ptmiss,j(j))< \pi/2$ \\ [\cmsTabSkip]
        \underline{\textit{Control region}} \\
        1-jet \WZ                            & $=$1 jet with $\pt>30\GeV$, $\Delta\phi(\Pell\ptmiss,j(j))> \pi/2$ \\ [\cmsTabSkip]
        2-jet \WZ                            & $\geq$2 jets with $\pt>30\GeV$, $\Delta\phi(\Pell\ptmiss,j(j))> \pi/2$ \\
        \hline
    \end{tabular}
    \label{tab:ZH3l_selection}
\end{table*}

\begin{figure*}
    \centering
    \includegraphics[width=0.49\textwidth]{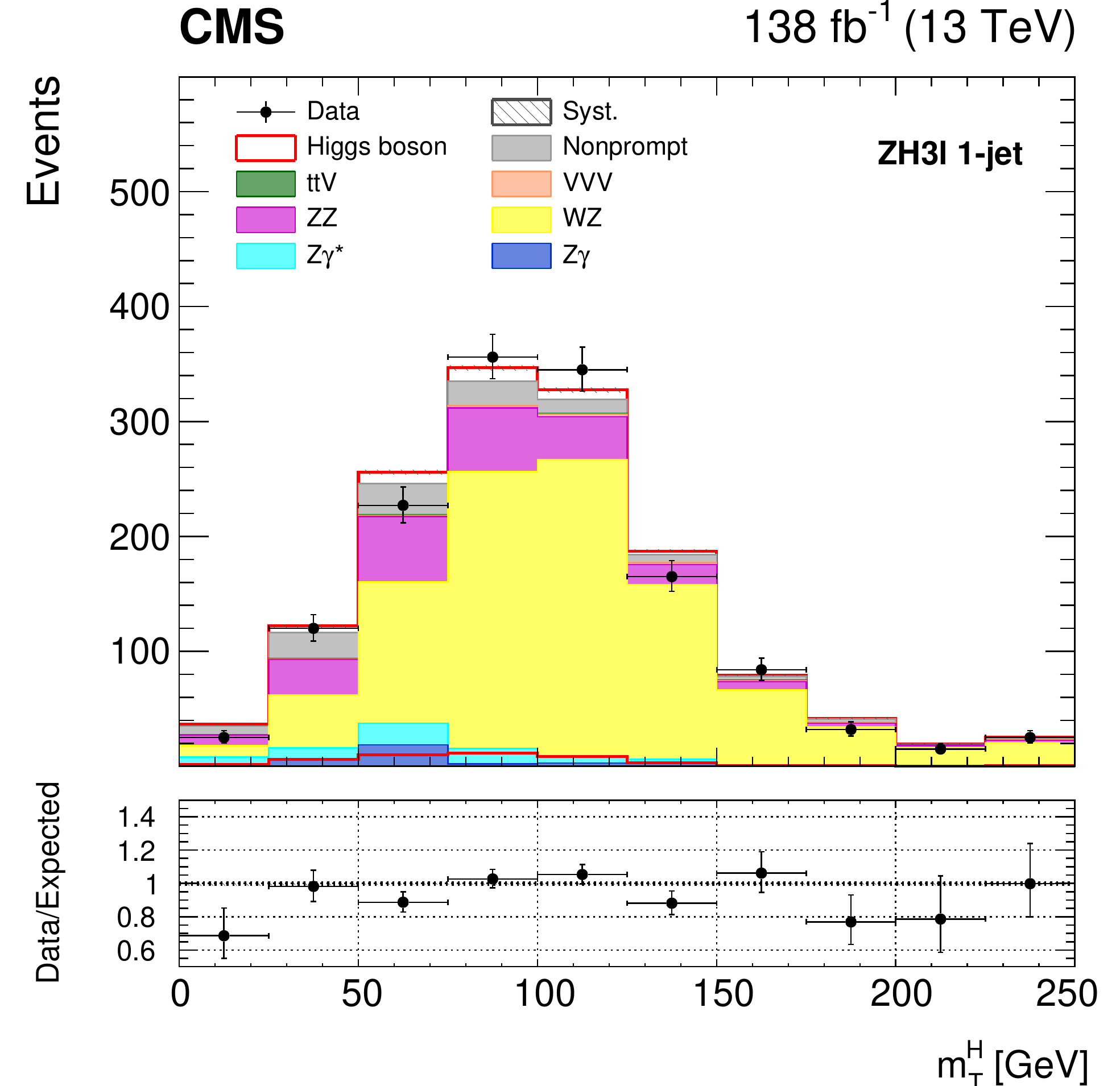}
    \includegraphics[width=0.49\textwidth]{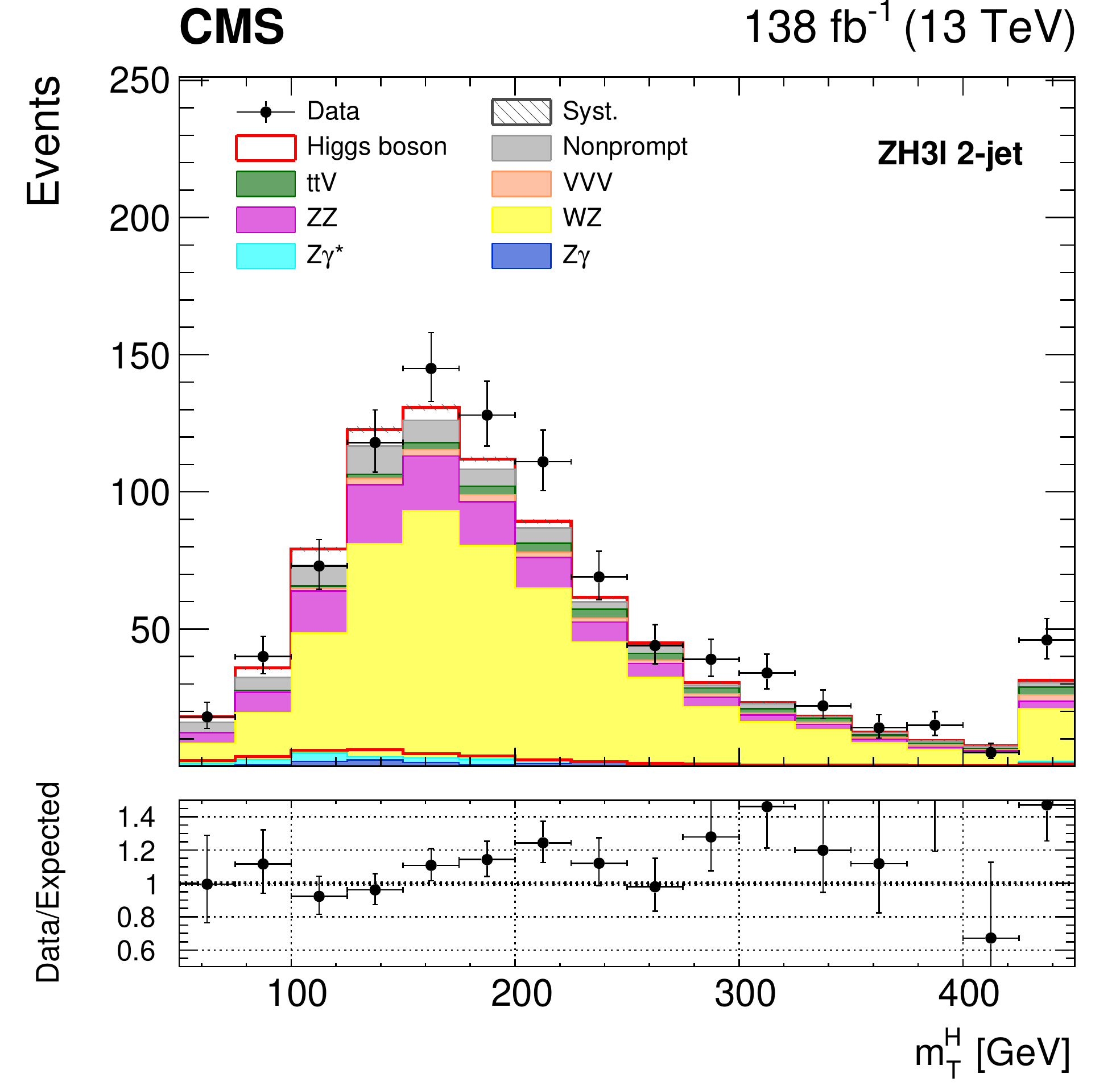}
    \caption{Observed distributions of the \mTH fit variable in the \ZH{}3\Pell 1-jet (left) and 2-jet (right) SRs. A detailed description is given in the Fig.~\ref{fig:mll_mth_ggH_DF_0j} caption.}
    \label{fig:mth_zh3l}
\end{figure*}

To extract the Higgs boson production cross section, a binned fit is performed to the $\mTH = \mT(\Pell\ptmiss, j(j))$ variable, defined in Eq.~(\ref{eq:mth}). Figure~\ref{fig:mth_zh3l} shows the \mTH distributions after the fit to the data.

\subsection{ZH4\texorpdfstring{\Pell}{l} categories}
\label{subsec:ZH4l}

The \ZH{}4\Pell category targets the $\PZ\PH\to 4\Pell 2\PGn$ decay. The final state therefore contains four leptons and \ptmiss. The analysis selects events containing four leptons with $\pt > 25$, 15, 10, and 10\GeV, respectively, and null total charge ($\text{Q}_{4\Pell}$). The invariant mass of any dilepton pair is required to be greater than 12\GeV to reject low-mass resonances. The opposite-sign SF lepton pair with \mll closest to $m_\PZ$ is designated as the \PZ boson candidate, while the remaining lepton pair is referred to as the \PX candidate. The \PZ boson candidate mass is required to be within 15\GeV of $m_\PZ$. Events are rejected if they contain any b-tagged jet with $\pt > 20\GeV$.

Events are categorized based on the flavor of the lepton pair forming the \PX candidate. Events in the {\PX}SF category have an SF \PX lepton pair, while events in the {\PX}DF category have a DF \PX lepton pair. In the {\PX}SF category, events are required to satisfy $m_{4\Pell} > 140\GeV$, $10 < m_{\Pell\Pell}^\PX < 60\GeV$, and $\ptmiss > 35\GeV$. Events in the {\PX}DF category must have $10 < m_{\Pell\Pell}^\PX < 70\GeV$ and $\ptmiss > 20\GeV$.

Production of \ZZ pairs is the main background in this category. Additional contributions arise from {\ttbar}{\PZ}, $\PV\PV\PV$, and V{\PGg} processes. These backgrounds are all modeled with MC simulation. The \ZZ normalization is extracted from data in a dedicated CR defined by the requirements $75 < m_{\Pell\Pell}^\PX < 105\GeV$ and $\ptmiss < 35\GeV$. The event selection and categorization in the \ZH{}4\Pell category is summarized in Table~\ref{tab:ZH4l_selection}.

\begin{table*}
    \centering
    \topcaption{Event selection and categorization in the \ZH{}4\Pell category.}
    \begin{tabular}{cc}
        \hline
        Subcategories & Selection \\
        \hline
        \underline{\textit{Global selection}} \\
        \multirow{3}{*}{\NA}               & $\ptone > 25\GeV$, $\pttwo > 15\GeV$, $\ptthr > 10\GeV$, $\pt{}_4 > 10\GeV$ \\
                                           & $\text{Q}_{4\Pell}=0$, $\min(\mll) > 12\GeV$, $\abs{\mll - m_\PZ}<15\GeV$ \\
                                           & No \PQb-tagged jet with $\pt > 20\GeV$ \\ [\cmsTabSkip]
        \underline{\textit{Signal region}} \\
        \multirow{2}{*}{{\PX}SF}           & Same-flavor \PX pair, $m_{4\Pell}>140\GeV$  \\
                                           & $10 < m_{\Pell\Pell}^\PX < 60\GeV$, $\ptmiss>35\GeV$ \\ [\cmsTabSkip]
        \multirow{2}{*}{{\PX}DF}           & Different-flavor \PX pair, $10 < m_{\Pell\Pell}^\PX < 70\GeV$ \\
                                           & $\ptmiss>20\GeV$ \\ [\cmsTabSkip]
        \underline{\textit{Control region}} \\
        \ZZ                                & $75 < m_{\Pell\Pell}^\PX < 105\GeV$, $\ptmiss<35\GeV$ \\
        \hline
    \end{tabular}
    \label{tab:ZH4l_selection}
\end{table*}

\begin{figure*}
    \centering
    \includegraphics[width=0.49\textwidth]{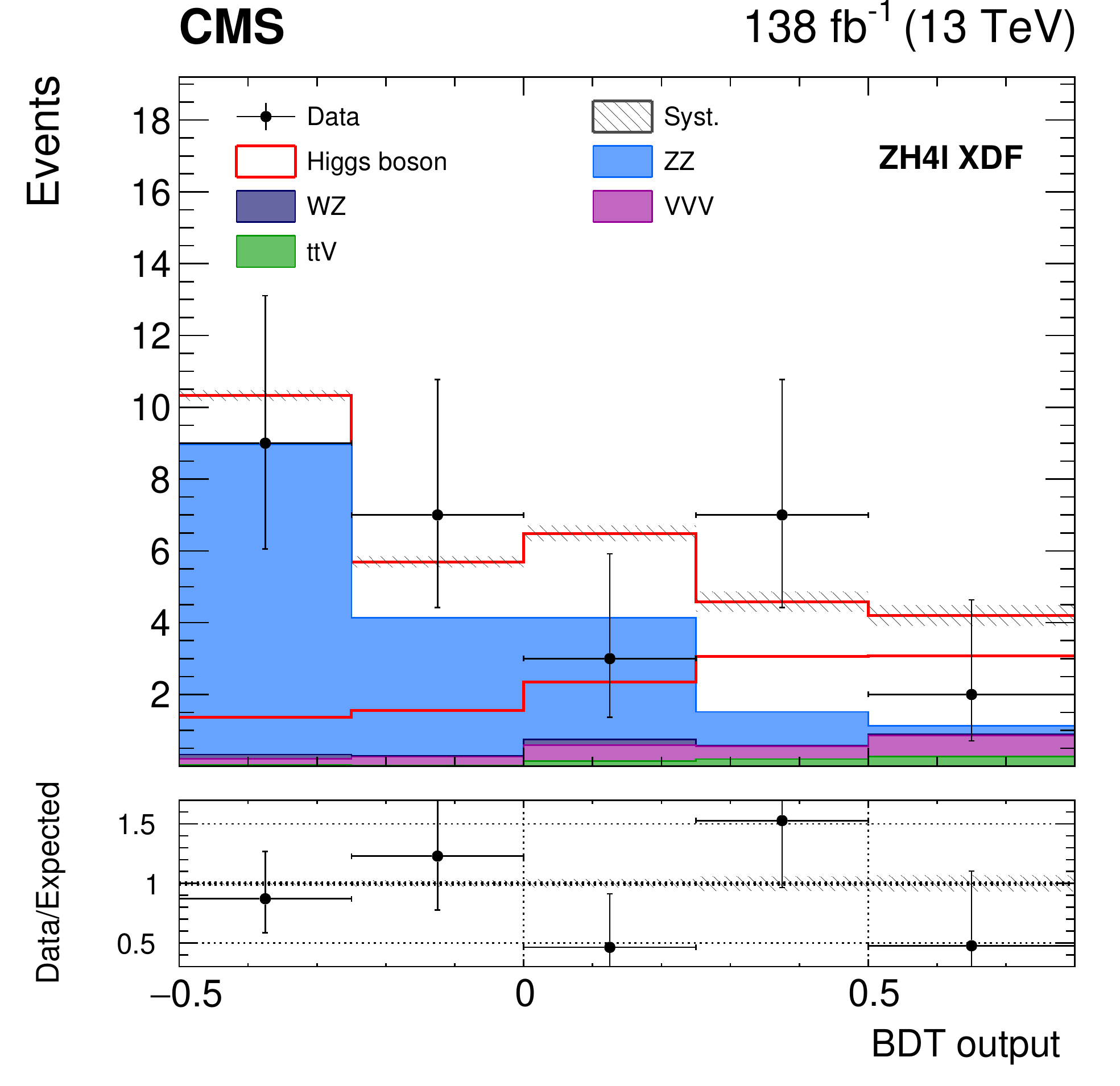}
    \includegraphics[width=0.49\textwidth]{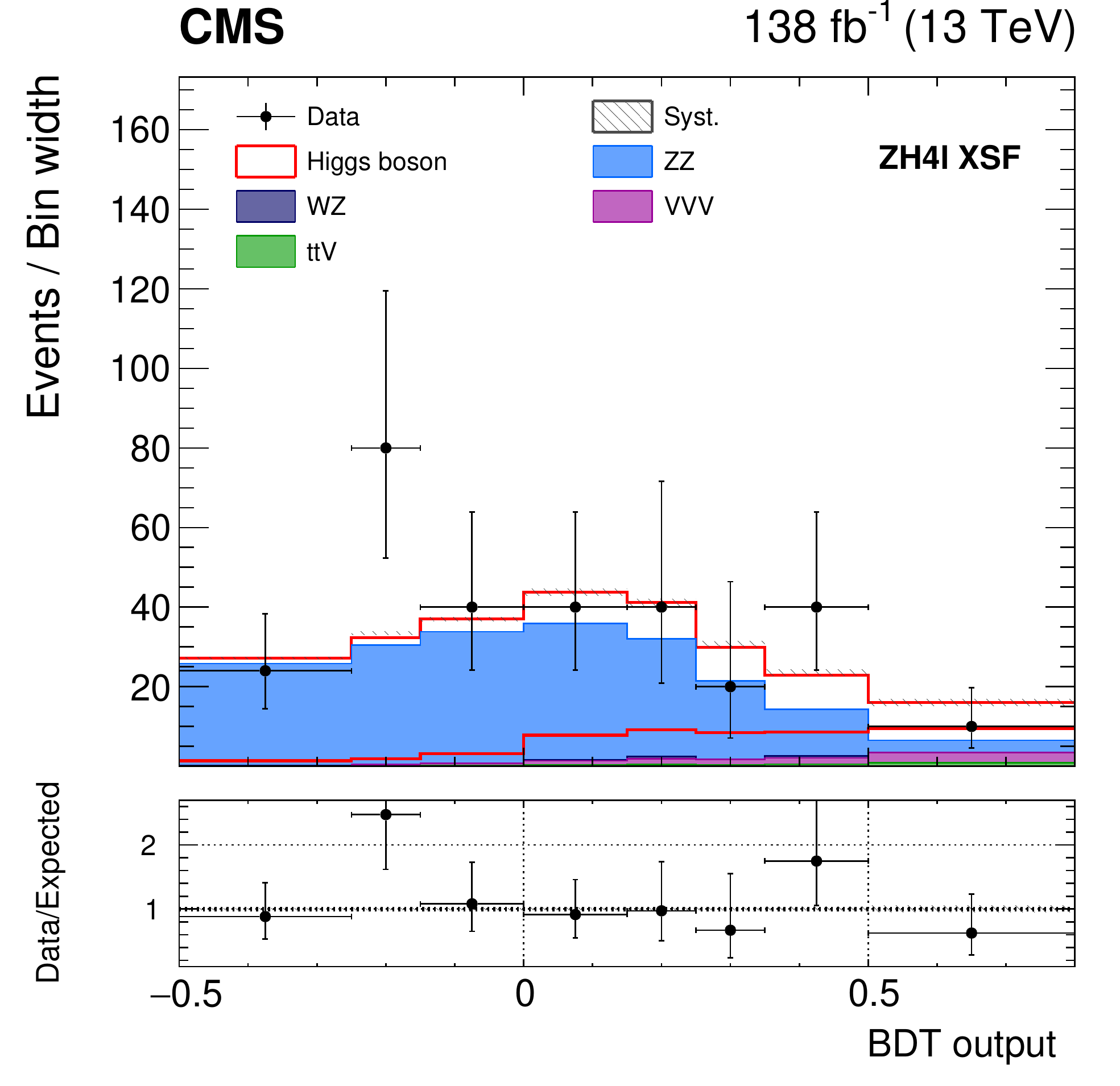}
    \caption{Observed distributions of the BDT score in the \ZH{}4\Pell {\PX}DF (left) and {\PX}SF (right) SRs. A detailed description is given in the Fig.~\ref{fig:mll_mth_ggH_DF_0j} caption.}
    \label{fig:BDT_zh4l}
\end{figure*}

A BDT approach is used to discriminate between signal and background. The BDT is trained on events passing the global selection, with $\ptmiss > 20\GeV$ and $10 < m_{\Pell\Pell}^\PX < 70\GeV$. The number of inputs used in the BDT is eight, and these include separation in the $\eta$-$\phi$ plane between the leptons in each dilepton pair, transverse masses of combinations of leptons and \ptvecmiss, as well as \ptmiss itself. The kinematic variables of the \PX candidate give the most discriminating power, along with \ptmiss. To extract the Higgs boson cross section, a binned fit is performed on the BDT score. Figure~\ref{fig:BDT_zh4l} shows the BDT score distributions after the fit to the data.

\subsection{Different-flavor \texorpdfstring{{\VH}2j}{VH2j} categories}
\label{subsec:VH2j_DF}

This category targets \VH events in which the vector boson decays into two resolved jets and the Higgs boson decays to an {\Pe}{\PGm} pair and neutrinos. The final state, and therefore the selection, is analogous to that of the \ggH DF 2-jet category, with the added requirement that the dijet invariant mass be close to that of the \PW and \PZ bosons.

The main backgrounds in this category are top quark and nonresonant \WW pair production, as well as \tautau pair production. The top quark and \tautau backgrounds are normalized to the data in dedicated CRs. The full selection is summarized in Table~\ref{tab:vh2j_DF_selection}. The VH production is found to contribute about 30\% of the total signal in the {\VH}2j DF SR.

\begin{table*}
    \centering
    \topcaption{Summary of the selection applied to different-flavor {\VH}2j categories.}
    \begin{tabular}{cc}
        \hline
        Subcategory & Selection \\
        \hline
        \underline{\textit{Global selection}} \\
        \multirow{3}{*}{\NA}                      & $\ptone > 25\GeV$, $\pttwo > 10\GeV$ (2016) or 13\GeV \\
                                                  & $\ptmiss > 20\GeV$, $\ptll > 30\GeV$, $m_{\Pell\Pell} > 12\GeV$ \\
                                                  & {\Pe}{\PGm} pair with opposite charge \\ [\cmsTabSkip]
        \underline{\textit{Signal region}} \\
        \multirow{4}{*}{\NA}                      & At least 2 jets with $\pt > 30\GeV$, $\abs{\eta_{j1}},\abs{\eta_{j2}} < 2.5$ \\
                                                  & $\detajj< 3.5$, $65 < \mjj < 105\GeV$ \\
                                                  & $60\GeV < \mTH < 125\GeV$, $\dR_{\Pell\Pell}<2$ \\
                                                  & No \PQb-tagged jet with $\pt > 20\GeV$ \\ [\cmsTabSkip]
        \underline{\textit{Control region}} \\
        \multirow{2}{*}{Top quark CR}             & As SR but with no \mTH requirement, $\mll>50\GeV$ \\
                                                  & At least 1 \PQb-tagged jet with $\pt > 30\GeV$ \\ [\cmsTabSkip]
        \multirow{2}{*}{\tautau CR}               & As signal region but with $\mTH <60\GeV$ \\
                                                  & $40 < \mll < 80\GeV$ \\
        \hline
    \end{tabular}
    \label{tab:vh2j_DF_selection}
\end{table*}

\begin{figure}[htbp]
    \centering
    \includegraphics[width=\cmsFigureWidth]{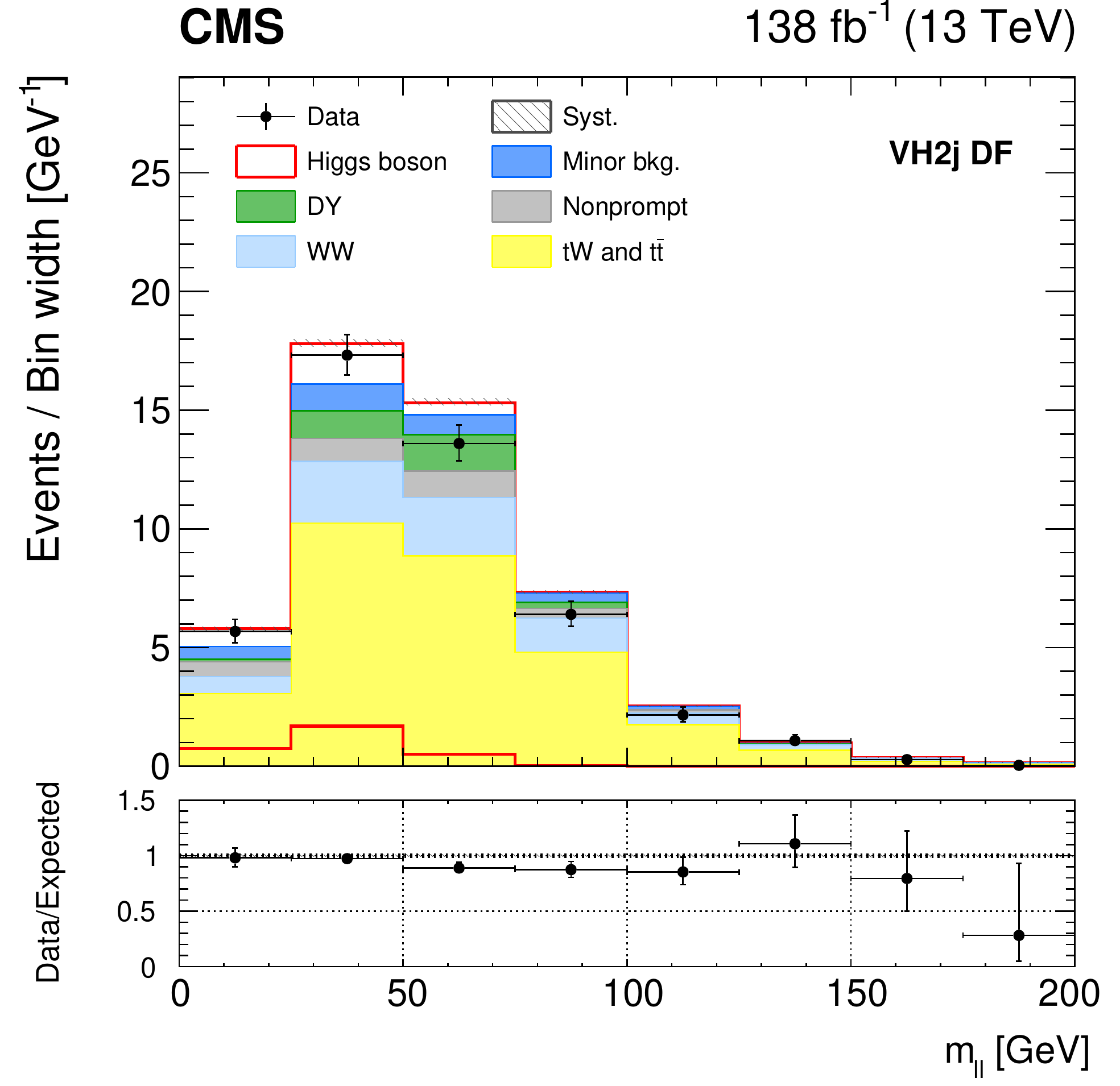}
    \caption{Observed distribution of the \mll fit variable in the {\VH}2j DF SR. A detailed description is given in the Fig.~\ref{fig:mll_mth_ggH_DF_0j} caption.}
    \label{fig:vh2j_fit_variables}
\end{figure}

The signal extraction fit is performed on a binned template shape of \mll, which has a different profile for the signal and the nonresonant \WW background. The distribution of \mll after the fit to the data is shown in Fig.~\ref{fig:vh2j_fit_variables}.

\subsection{Same-flavor \texorpdfstring{{\VH}2j}{VH2j} categories}
\label{subsec:VH2j_SF}

This category targets \VH events in which the vector boson decays into two jets and the Higgs boson decays to either an {\Pe}{\Pe} or a {\PGm}{\PGm} pair and neutrinos. The selection is identical to the 2-jet \ggH SF categories described in Section~\ref{subsec:ggH_SF} and Table~\ref{tab:ggh_SF_selection}, with the following modifications: the additional requirement $65 < \mjj < 105\GeV$ is imposed, the \mll threshold is moved to 70\GeV, a selection on $\mTH < 150\GeV$ is added, and the angle between the two leptons in the transverse plane ($\dphill$) is required to be less than 1.6. The threshold on the DYMVA is tuned to achieve the highest signal-to-background ratio. The signal is extracted via a simultaneous fit to the number of events in each category.

\section{The STXS measurement}
\label{sec:STXS}

\begin{figure*}
    \centering
    \includegraphics[width=0.49\textwidth]{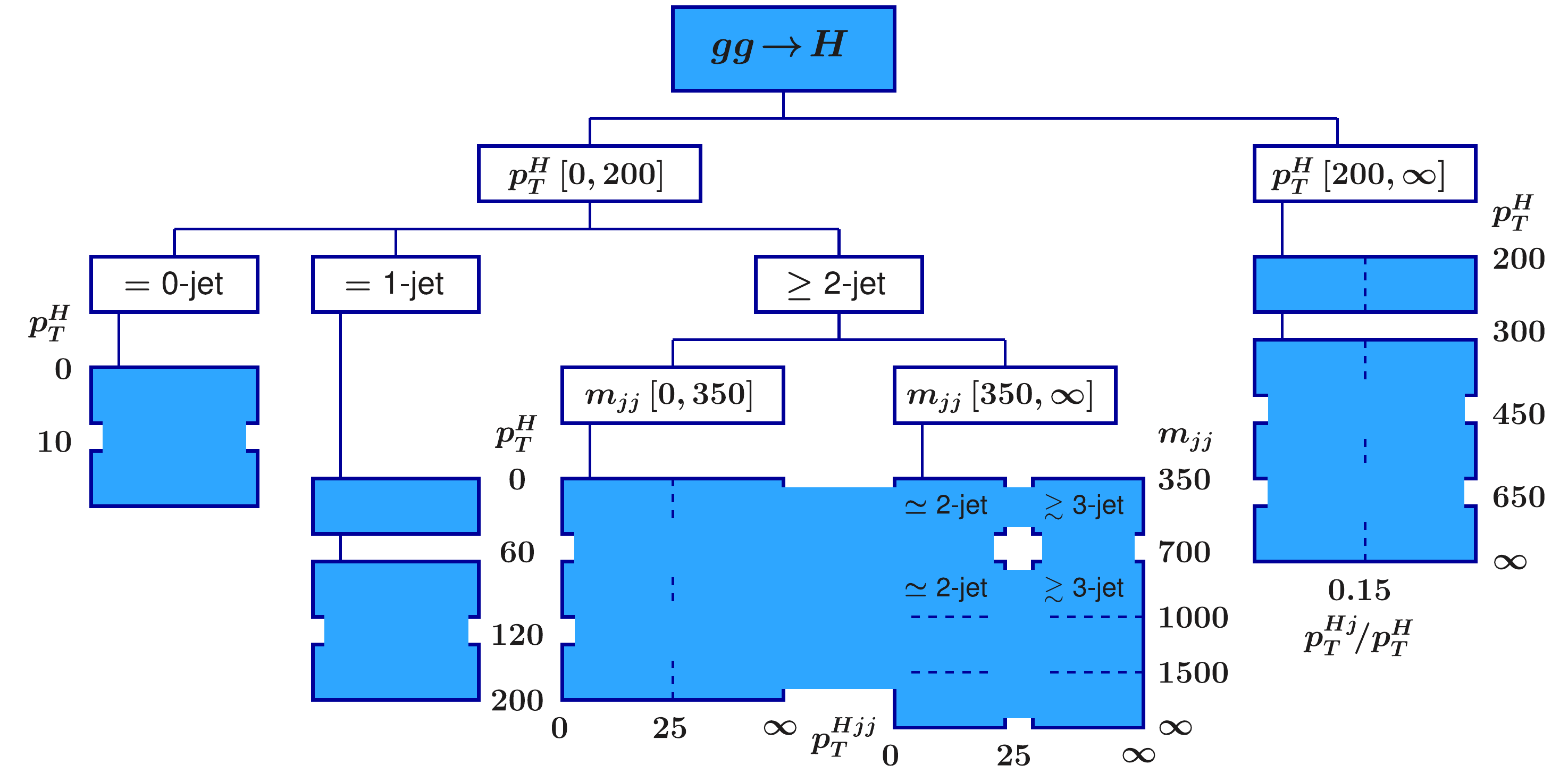}
    \includegraphics[width=0.49\textwidth]{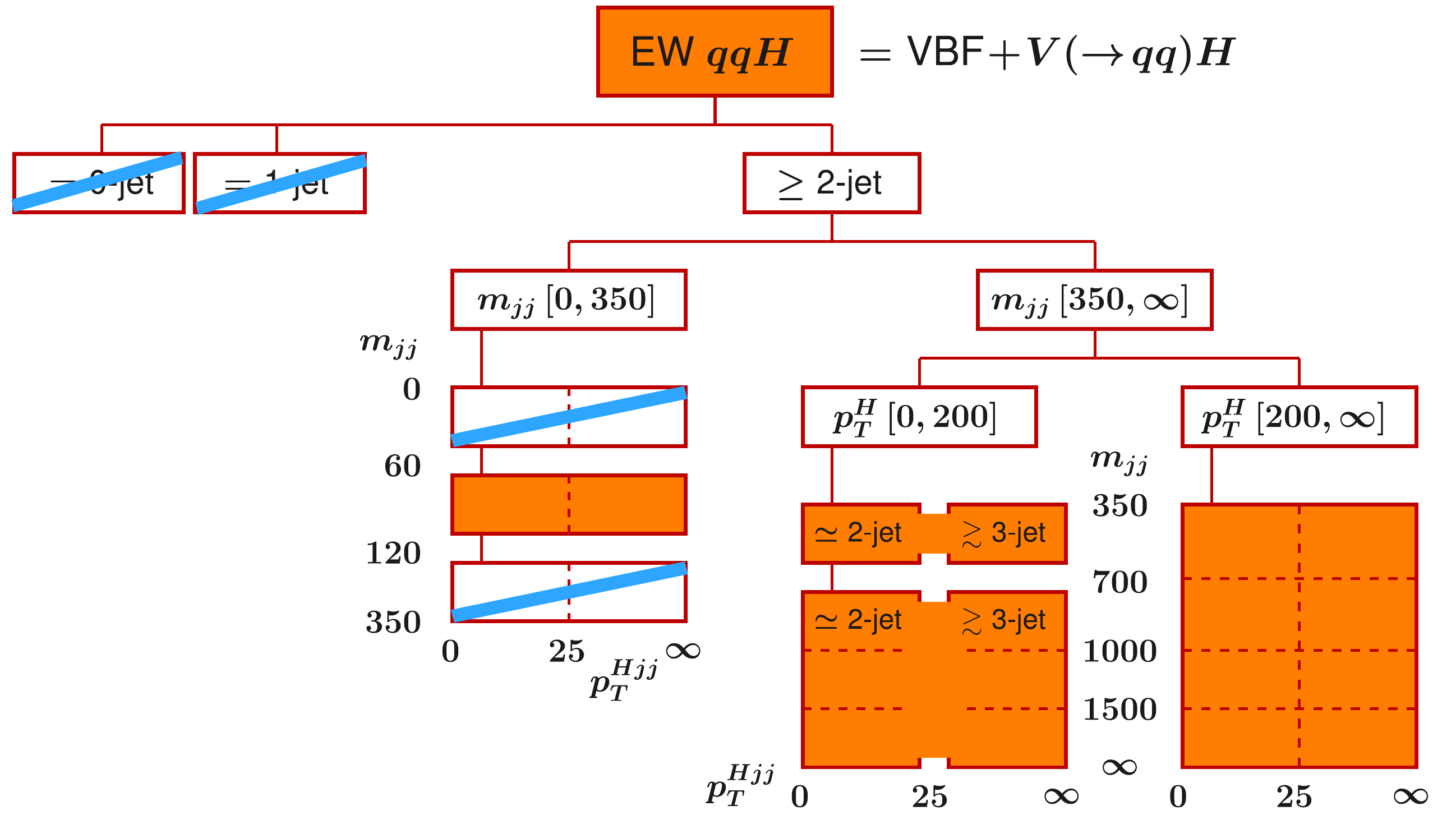}
    \includegraphics[width=0.49\textwidth]{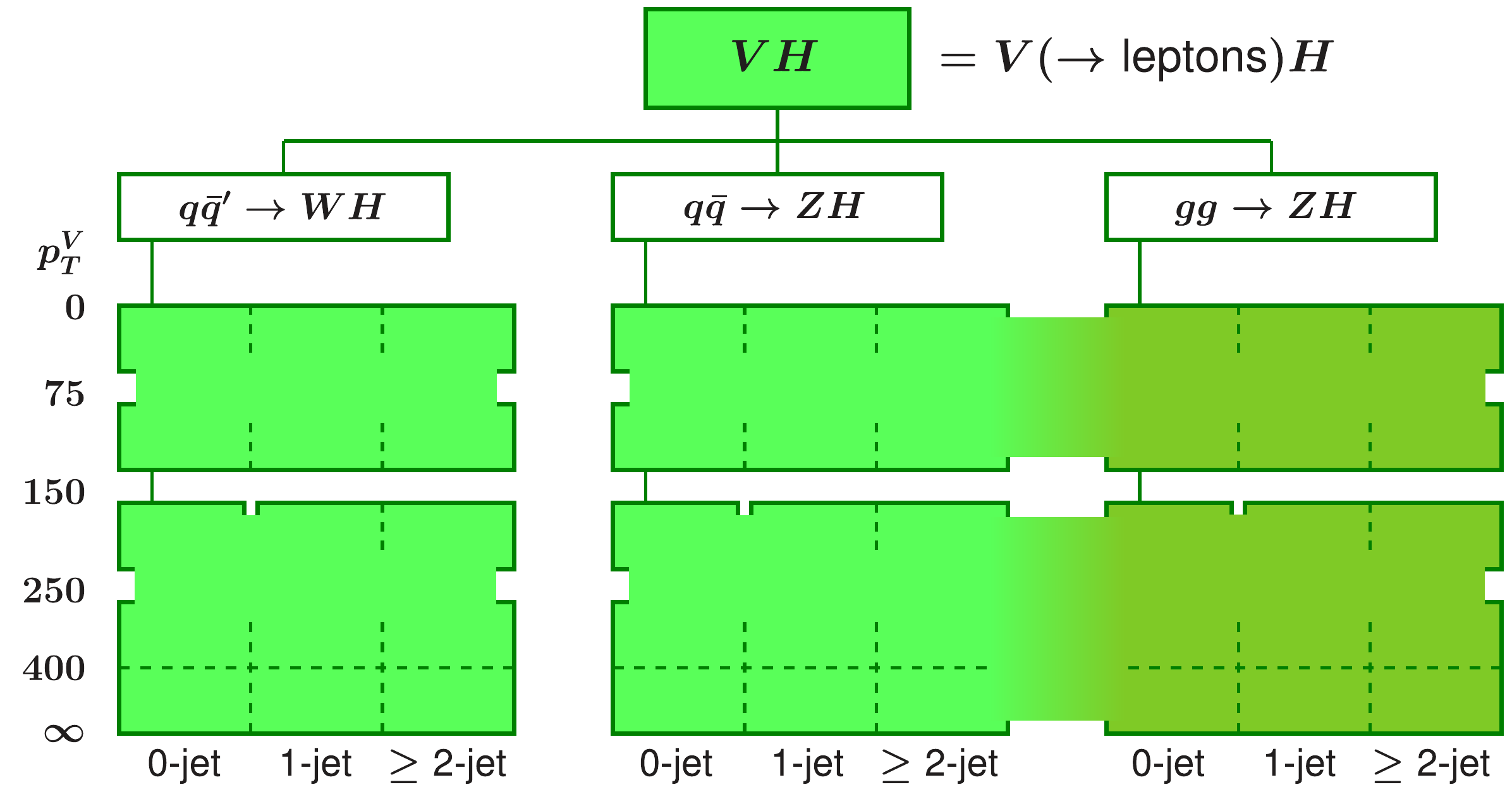}
    \caption{The STXS Stage 1.2 binning scheme. Each rectangle corresponds to one of the STXS Stage 1.2 bins. Dashed lines indicate a possible finer splitting of some of the bins (not used in this analysis). Bins fused together with solid colors are merged in the analysis, \ie, they are measured as a single bin. Crossed-out bins are not measured.}
    \label{fig:STXS_maximal}
\end{figure*}

\begin{figure*}
    \centering
    \includegraphics[width=\textwidth]{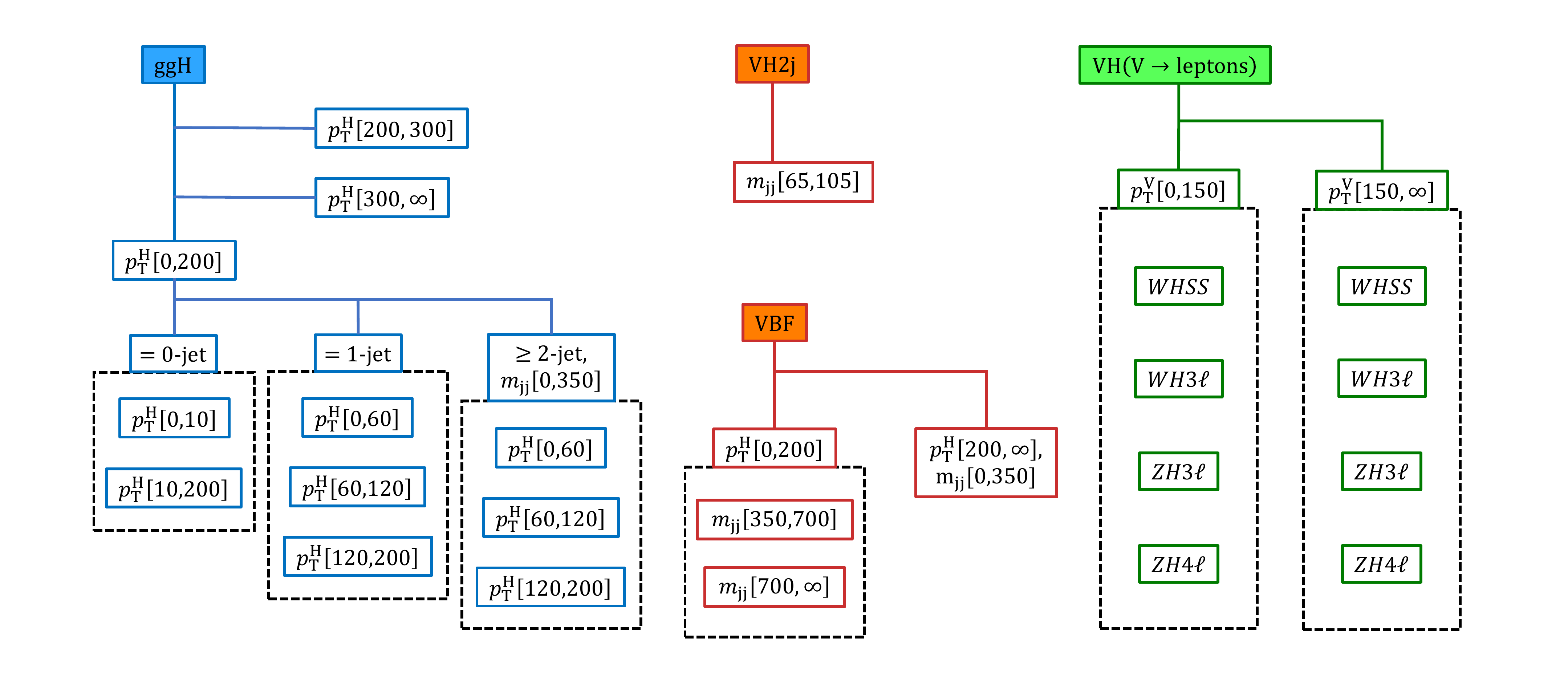}
    \caption{Analysis categories for the STXS measurement. The baseline \ggH, VBF, and VH selections are identical to what was described in Sections~\ref{sec:ggH}--\ref{sec:vh}. All dimensional quantities are measured in \GeVns.}
    \label{fig:STXS_ggh_fit_setup}
\end{figure*}

\begin{figure*}
    \centering
    \includegraphics[width=\textwidth]{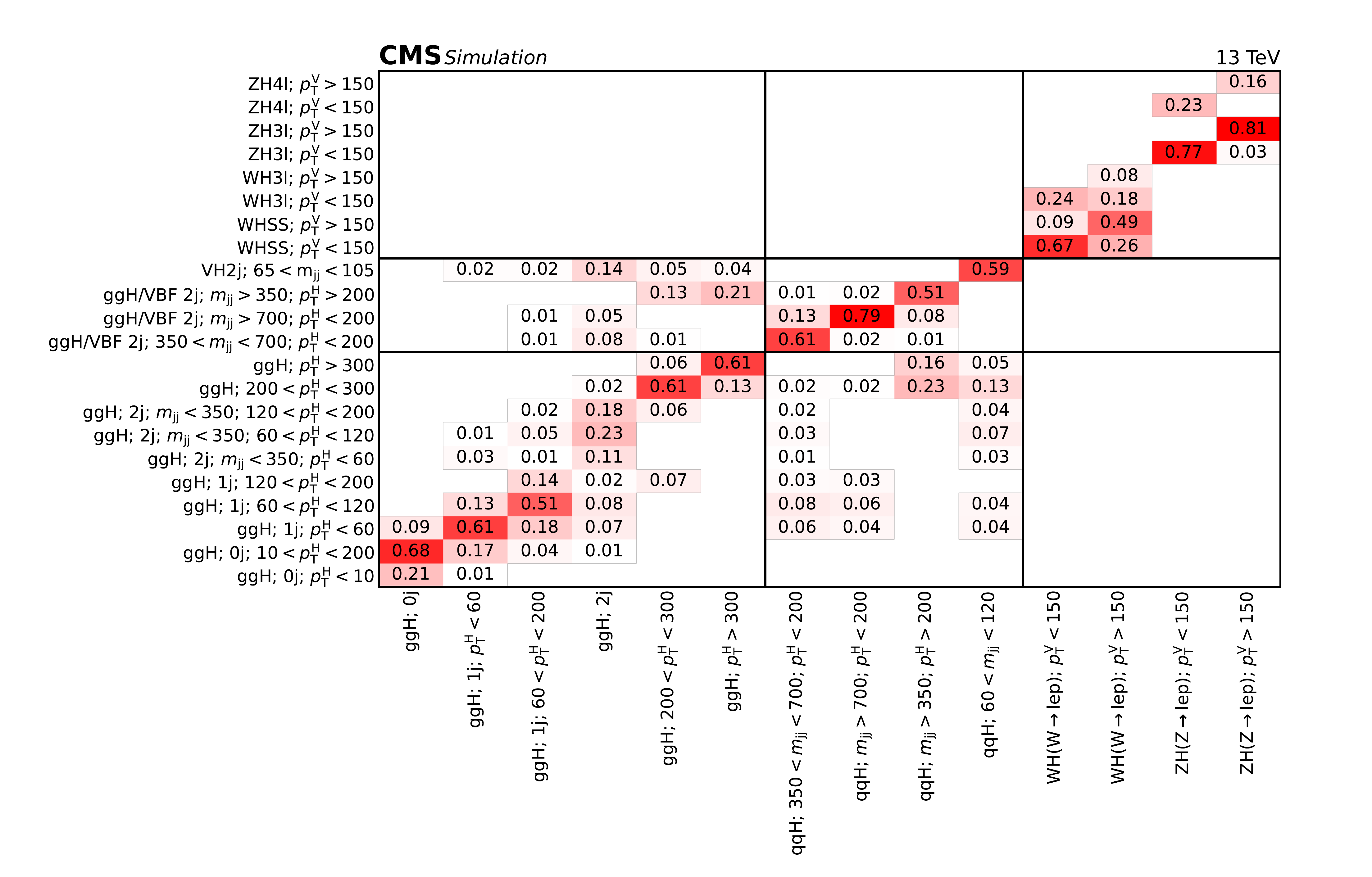}
    \caption{Expected signal composition in each STXS bin. Generator-level bins are reported in the horizontal axis, and the corresponding analysis categories on the vertical axis. All quantities in the definitions of bins are measured in \GeVns.}
    \label{fig:STXS_migration_matrix}
\end{figure*}

Together with inclusive production cross sections, differential cross section measurements are also presented. These are performed within the STXS framework, using Stage 1.2 definitions~\cite{deFlorian:2227475}. In the STXS framework, the cross sections of different Higgs boson production mechanisms are measured in mutually exclusive regions of generator-level phase space, referred to as STXS bins, designed to enhance sensitivity to possible deviations from the SM. The full set of Stage 1.2 STXS bins is given in Fig.~\ref{fig:STXS_maximal}. The selections used in the STXS measurement match the ones described in the previous section, and the measurement is carried out by defining a set of analysis categories that target each STXS bin, as summarized in Fig.~\ref{fig:STXS_ggh_fit_setup}. The same CR setup as described in the previous section is maintained, and each CR is then subdivided to match the STXS categorization shown in Fig.~\ref{fig:STXS_ggh_fit_setup}. In all cases, the number of events is used as a fit variable in CRs. Results are then unfolded to the generator level, with the contribution from each STXS bin to each analysis category estimated from MC simulation, as shown in Fig.~\ref{fig:STXS_migration_matrix}. Given the statistical power of the present data set, sensitivity to some of the Stage 1.2 bins is limited. Some bins are therefore measured together, by fixing the corresponding cross section ratios to the value predicted by the SM. We refer to this procedure as bin merging. Some STXS bins have been excluded, given the very low sensitivity. Groups of STXS bins merged with this procedure are highlighted in Fig.~\ref{fig:STXS_maximal}.

In the DF \ggH and VBF categories, the discriminants of the same DNN explained in Section~\ref{subsec:vbf} are used for the categories which are common between VBF and \ggH ($\mjj>350\GeV$ and $\ptH<200\GeV$), and in the category exclusive to the VBF ($\mjj>350\GeV$ and $\pt>200\GeV$).
The signal extraction fit is performed on the two-dimensional (\mll, \mjj) template in the VH2j DF category ($60<\mjj<120\GeV$), while either \mll or (\mll, \mTH) templates are used in the remaining DF categories, depending on the number of expected events in each.
In the same flavor categories a similar approach is followed, but only the number of events is used for the fit.

In the \VH categories with a leptonic decay of the \PV boson, to extract the cross section as a function of the vector boson \pt, events are categorized into corresponding regions 
of reconstructed vector boson \pt. The reconstructed vector boson \pt is defined differently depending on the 
vector boson type and decay channel. Because in the \WH{}SS and \WH{}3\Pell categories the \PW boson \pt ($\ptvec^{\PW}$) cannot be fully reconstructed 
due to the unobserved neutrino, proxies are defined in both cases. In the \WH{}SS category, the four-momenta of the lepton and neutrino from the associated \PW boson decay can be designated $\Pell_{\PW}$ and $\PGn_{\PW}$, 
while the four-momenta of the lepton and neutrino from the Higgs boson decay can be designated $\Pell_{\PH}$ and $\PGn_{\PH}$. The lepton from the \PW boson
decay is identified as the one with the largest azimuthal separation from the jet or dijet. The transverse momentum of the \PW boson is defined 
as $\vec{\Pell}_{\PW,\mathrm{T}}+\vec{\PGn}_{\PW,\mathrm{T}}$, where $\vec{\PGn}_{\PW}$ is defined as:
\begin{linenomath}
\begin{equation}
\vec{\PGn}_{\PW,\mathrm{T}} = \ptvecmiss - \vec{\PGn}_{\PH,\mathrm{T}} = \ptvecmiss - \vec{\Pell}_{\PH,\mathrm{T}} \left( \frac{125\GeV}{ \abs{\vec{\Pell}_{\PH} + \vec{jj}} } - 1 \right)
\end{equation}
\end{linenomath}
for events with two jets, or $\ptvecmiss - \vec{\Pell}_{\PH,\mathrm{T}}$ for events with fewer than two jets. Here $\vec{jj}$ indicates the dijet momentum.
In the \WH{}3\Pell category, $\ptvec^{\PW}$ is difficult to resolve given the ambiguities from the three neutrinos in the final state. Instead,
$\pt(\Pell_{\PW})$ is used as a proxy for the \PW boson \pt in the \WH{}3\Pell category. Here, $\Pell_{\PW}$ is defined as the lepton pointing 
away from the opposite-sign dilepton pair with smallest angular separation \dR.
In the \ZH{}3\Pell and \ZH{}4\Pell categories, the reconstructed \PZ boson \pt ($\ptvec^{\PZ}$) is defined as the \pt of the OSSF dilepton 
pair with \mll closest to $m_\PZ$. The variables used in the fit are the same as described in Section~\ref{sec:vh}.

\begin{figure*}
    \centering
    \includegraphics[width=\textwidth]{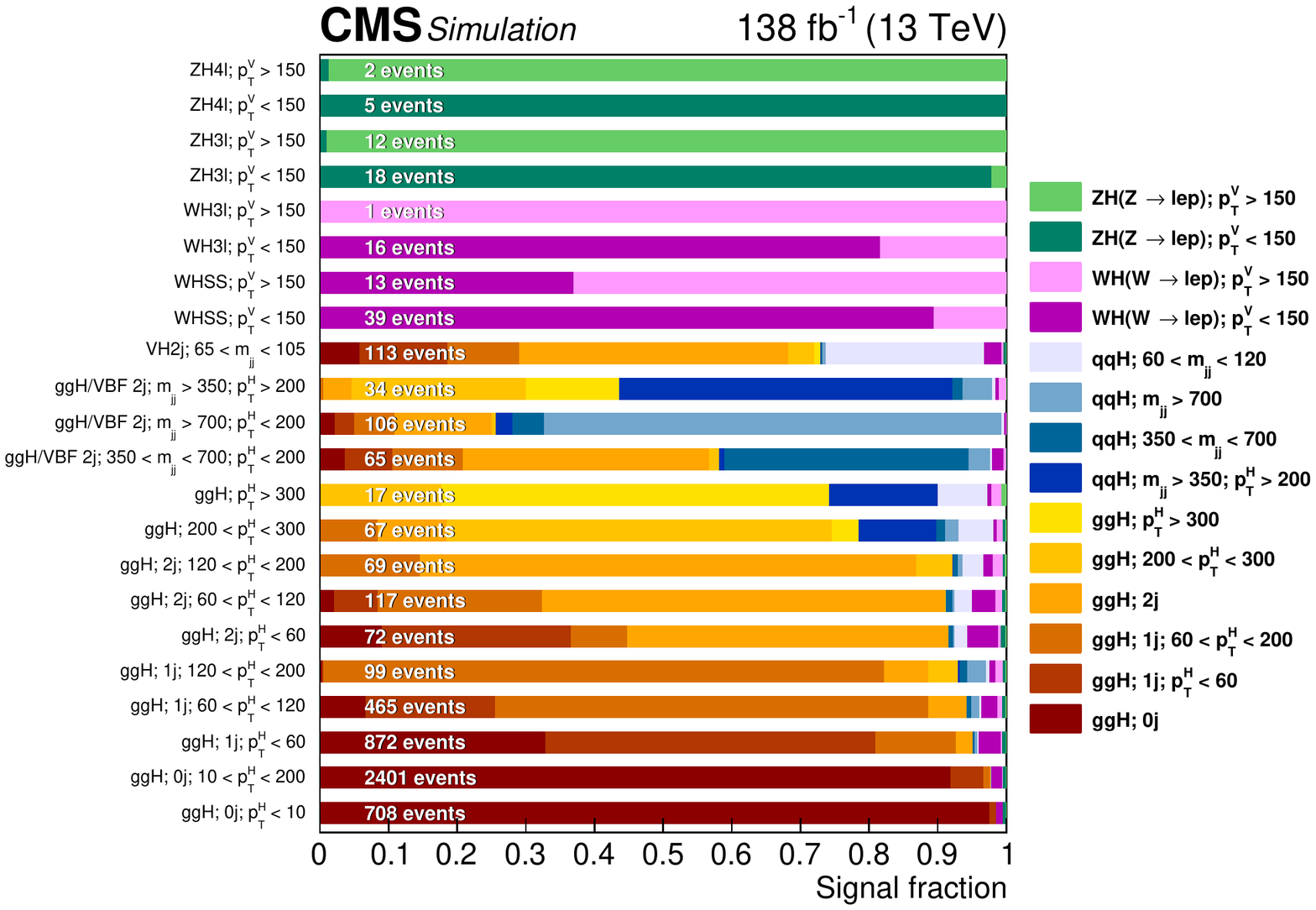}
    \caption{Expected relative fractions of different STXS signal processes in each category. The total number of expected \hww signal events in each category is also shown. All dimensional quantities in the definitions of bins are measured in \GeVns.}
    \label{fig:stxs_signal_fraction}
\end{figure*}

A summary of the expected signal fraction of the considered STXS signal processes in each category is shown in Fig.~\ref{fig:stxs_signal_fraction}, together with the total number of expected \hww signal events.

\section{Background estimation}
\label{sec:background}

\subsection{Nonprompt lepton background}
The nonprompt lepton backgrounds originating from leptonic decays of heavy quarks, hadrons misidentified as leptons, and electrons from photon conversions are suppressed by the identification and isolation requirements imposed on electrons and muons. The nonprompt lepton background in the two-lepton final state primarily originates from {\PW}+jets events, while the nonprompt lepton background in the three-lepton final state primarily comes from {\PZ}+jets events. Top quark production with a jet misidentified as a lepton also contributes to the three-lepton final state. The nonprompt lepton background gives a negligible contribution in the four-lepton final state.
This background is estimated from data, as described in detail in Ref.~\cite{Sirunyan:2018egh}. The rate at which a nonprompt lepton passing a loose selection further passes a tight selection (misidentification rate) is measured in a data sample enriched in events composed uniquely of jets produced through the strong interaction, referred to as QCD multijet events. The corresponding rate for a prompt lepton to pass this selection (prompt rate) is measured using a tag-and-probe method~\cite{CMS:2010svw} in a data sample enriched in DY events. The misidentification and prompt rates are used to construct a relation between the number of leptons passing the loose selection, the number of leptons passing the tight selection, and the number of true prompt leptons in an event. This relation is applied as a transfer function to a data sample containing leptons passing the loose selection, weighting the events by the probability for $N$ leptons to pass the tight selection while fewer than $N$ leptons are truly prompt.
The nonprompt background with two leptons is validated with data in a CR enriched with {\PW}+jets events, in which a pair of same-sign leptons is required, while the nonprompt background with three leptons is validated in a CR enriched with top quark events or DY events.
The systematic uncertainty in the misidentification rate determination, which arises mainly from the different jet flavor composition between the events entering the QCD multijet and the analysis phase space, is estimated with a twofold approach. First, a validation check in the aforementioned CRs yields a normalization uncertainty of about 30\% that fully covers any differences with respect to data in all the kinematic distributions of interest in this analysis. Second, a shape uncertainty is estimated by varying the jet \pt threshold used in the calculation of the misidentification rate in the 15--25\GeV range, in bins of the lepton $\eta$ and \pt. For each threshold variation, the fake rate is recomputed and the difference with respect to the nominal fake rate is taken as a systematic uncertainty.

\subsection{Top quark background}
The background contributions from top quark processes are estimated using a combination of MC simulations and dedicated regions in data. A reweighting of the top quark and antiquark \pt spectra at parton level is performed for the \ttbar simulation in order to match the NNLO and next-to-next-to-leading logarithmic (NNLL) QCD predictions, including also the NLO EW contribution~\cite{Czakon:2017wor}. A shape uncertainty based on renormalization  ($\mu_\text{R}$) and factorization ($\mu_\text{F}$) scale variations is taken into account. For the \ggH, VBF, and {\VH}2j categories, in which the contribution of top quark backgrounds is dominant, the normalization of the simulated templates is left unconstrained in the fit separately for 0-, 1-, 2-jet \ggH, \VH, and VBF categories. The normalizations in these phase spaces are therefore measured from the data, by constraining the free-floating normalization parameters in top quark enriched CRs.

\subsection{Nonresonant \texorpdfstring{\WW}{WW} background}
The nonresonant \WW background is estimated using a combination of MC simulations and dedicated regions in data, and the quark-induced \WW simulated events are reweighted to match the diboson \pt spectrum computed at NNLO+NNLL QCD accuracy~\cite{Meade:2014fca,Jaiswal:2014yba}. The shape uncertainties related to the missing higher-order corrections are estimated by varying the $\mu_\text{R}$ and $\mu_\text{F}$ scales, as well as considering the independent variation of the resummation scale from its nominal value, taken as the mass of the \PW boson. For the \ggH, VBF, and {\VH}2j categories, the normalizations of the quark-induced and gluon-induced \WW backgrounds are left unconstrained in the fit (the ratio between the two is kept fixed within the uncertainty), keeping a different parameter for each signal phase space as done for the top quark background. In the DF final states the normalization parameters are constrained directly in the SRs without the need of defining CRs, as the SRs span the high-\mll phase space enriched in \WW events with a negligible Higgs boson signal contribution. Since in SF final states a counting analysis is performed, dedicated CRs enriched in \WW events are defined selecting events with high \mll.
The normalizations of the EW and QCD {\WW}+2 jets backgrounds are instead fixed to the respective SM cross sections provided by the MC simulation, taking into account the theoretical uncertainties arising from the variation of the $\mu_\text{R}$ and $\mu_\text{F}$ scales.

\subsection{\texorpdfstring{Drell--Yan background}{Drell-Yan background}}
The backgrounds arising from DY+jets processes are estimated using a different approach depending on the signal category.

In the \ggH, VBF, and {\VH}2j DF categories, the only source of DY background arises from \tautau production with subsequent leptonic decays of the \PGt leptons. This background process is estimated with a data-embedding technique~\cite{tau_embedding}, in which $\PGmp\PGmm$ events with well-identified muons are selected in a data sample. In each event, the selected muons are removed and replaced with simulated \PGt leptons, keeping the same four-momentum of the initial muons. The embedded sample is then corrected using scale factors related to the simulation of \PGt leptons. The usage of the embedded sample allows for a better modeling of the observables that are sensitive to the detector response and calibration, such as \ptvecmiss and other variables related to the hadronic activity in the event. Since the embedded sample takes into account all processes with a \tautau pair decaying to either electrons or muons, simulated \ttbar, single top, and diboson background events that contain a \tautau pair are not considered in the analysis to avoid any double counting. To correct for any additional discrepancy associated with the different acceptance of the \hww signal phase space, the normalization of the embedded samples is left unconstrained in the fit as done for top quark and \WW backgrounds. An orthogonal \tautau enriched CR is defined for the 0-, 1-, 2-jet \ggH{}-like, 2-jet VH-like, and 2-jet VBF-like phase spaces to help in constraining the free normalization parameters.
The embedded samples cover the events that pass the \Pe{}\PGm triggers, which represent the vast majority of the events selected in the DF final state. The contribution of the remaining \tautau events that enter the analysis phase space thanks to the single-lepton triggers ($\approx$5\% of the total) is estimated using MC simulation.

In the \ggH, VBF, and {\VH}2j SF categories, the dominant background contribution arises from DY production of $\Pell\Pell$ pairs and is estimated using a data-driven technique described in Ref.~\cite{Sirunyan:2018egh}. The $\Pell\Pell$ background contribution for events with $\abs{\mll-m_\PZ}>7.5\GeV$ is estimated by counting the number of events in data passing a selection with an inverted \mll requirement (\ie, under the \PZ boson mass peak), subtracting the non-\PZ-boson contribution from it, and scaling the obtained yield by the fraction of events outside and inside the \PZ boson mass region in MC simulation. The contribution of processes such as top quark and \WW production in the \PZ boson mass peak region, which have the same probability to decay into the \Pe{}\Pe, \Pe{}\PGm, \PGm{}\Pe, and \PGm{}\PGm final states, is estimated by counting the number of $\Pepm\PGm^{\mp}$ events in data, and applying a correction factor that accounts for the differences in the detection efficiency between electrons and muons. Other minor processes in the \PZ boson mass peak region (mainly \ZZ and {\PZ}{\PW}) are subtracted based on MC simulations. 
The yield obtained with this approach outside the \PZ boson mass peak is further corrected with a scale factor that takes into account the different acceptances between the estimation and SRs. 
The method is validated in orthogonal CRs enriched in DY events with a negligible signal contribution. The residual mismodeling between data and the estimated DY contribution arising from this validation is taken into account as a systematic uncertainty.
The same procedure is repeated separately for estimating and validating the DY contribution in the $\Pep\Pem$ and $\PGmp\PGmm$ final states.

In the leptonic \VH categories DY represents a minor background and is estimated using MC simulations.

\subsection{Multiboson background}
In categories with two charged leptons, the production of \WZ and \wgs contributes to the SRs whenever one of the three leptons is not identified. This background contribution is simulated as described in Section~\ref{sec:datasets}, and a data-to-simulation scale factor is derived in a three-lepton CR, orthogonal to the three-lepton SRs, as described in Ref.~\cite{Sirunyan:2018egh}. A normalization uncertainty of about 25\% is associated to the scale factor determination.
A different CR containing events with one pair of same-sign muons is also used as an additional validation of the \wgs simulation.
The contribution of the \wg process may also be a background in two-lepton SRs due to photon conversions in the detector material when one of the three leptons is not identified. This process is estimated using MC simulation and validated using data in a two-lepton CR requesting events with a leading \PGm and a trailing \Pe with same sign and a separation in \dR smaller than 0.5. These requirements mainly select events arising from \wg production where the \PW boson decays to \PGm{}\Pgngm and the photon is produced as final-state radiation from the muon.
The theoretical uncertainties in \wg and \wgs processes estimated using $\mu_\text{R}$ and $\mu_\text{F}$ scale variations are taken into account.

The \WZ process represents one of the main backgrounds in the leptonic VH categories and its normalization is left as a free parameter in the fit, separately for different jet multiplicity categories. Dedicated 0-, 1- and 2-jet CRs are included in the fit to help constraining the \WZ normalization parameters.

The production of a \PZ boson pair is the main background in the \ZH{}4\Pell category and is estimated using MC simulation. The normalization of this background is left free to float and constrained using data in a \ZZ-enriched CR.

Triple vector boson production is a minor background in all the considered categories and is estimated using MC simulation.

\section{Statistical procedure and systematic uncertainties}
\label{sec:uncertainties}

The statistical approach used to interpret the selected data sets for this analysis and to combine the results from the independent categories has been developed by the ATLAS and CMS Collaborations in the context of the LHC Higgs Combination Group~\cite{LHC-HCG}. All selections have been optimized entirely on MC simulation and have been frozen before comparing the templates to data, in order to minimize possible biases.
In all the categories considered, the signal extraction is performed using binned templates based on variables that allow for a good discrimination between signal and background, as summarized in Table~\ref{tab:fit_overview}. Therefore, the effect of each source of systematic uncertainty is either a change of the normalization of a given signal or background process, or a change of its template shape. The signal extraction is performed by a binned maximum likelihood fit, and each such change is modeled as a constrained nuisance parameter distributed according to a log-normal probability distribution function with standard deviation set to the size of the corresponding change. Where the change in shape of a template caused by a nuisance parameter is found to be negligible (\ie, its effect on the expected uncertainty on signal strength modifiers is well below 1\%), only its effect on the normalization is considered.

\begin{table*}[h]
    \centering
    \topcaption{Overview of the fit variables and CRs used in each analysis category. In all CRs, the number of events is used. The number of subcategories shown in the last column includes both SRs and CRs.}
    \begin{tabular}{lcccc}
        \hline
        Category & SR subcategorization & SR fit variable & Contributing CRs & $N_\mathrm{subcategories}$ \\
        \hline
        \multirow{2}{*}{\ggH DF} & (0j, 1j) $\times$ ($\pttwo \lessgtr 20\GeV$) $\times$ & \multirow{2}{*}{(\mll, \mTH)} & \multirow{2}{*}{Top quark, \tautau} & \multirow{2}{*}{15} \\
                                 & $\times$ ($\Pell^{\pm}\Pell^{\mp}$), ($\geq$2j)       &                                                                     &                     \\
        \ggH SF & (0j, 1j, $\geq$2j) $\times$ ({\Pe}{\Pe}, {\PGm}{\PGm}) & $N_{\text{events}}$ & Top quark, \WW & 12 \\
        VBF DF & $\max_{j} C_j$ & DNN output & Top quark, \tautau & 6 \\
        VBF SF & ({\Pe}{\Pe}, {\PGm}{\PGm}) & $N_{\text{events}}$ & Top quark, \WW & 4 \\
        \WH{}SS & (DF, SF) $\times$ (1j, 2j) & \mHtil & \WZ & 4 \\
        \multirow{2}{*}{\WH{}3\Pell} & SF lepton pair with   & \multirow{2}{*}{BDT output} & \multirow{2}{*}{\WZ, {\PZ}{\PGg}} & \multirow{2}{*}{4} \\
                                     & opposite or same sign &                             &                                   &                    \\
        \ZH{}3\Pell & (1j, 2j) & \mTH & \WZ & 4 \\
        \ZH{}4\Pell & (DF, SF) & BDT output & \ZZ & 3 \\
        {\VH}2j DF & \NA & \mll & Top quark, \tautau & 3 \\
        {\VH}2j SF & ({\Pe}{\Pe}, {\PGm}{\PGm}) & $N_{\text{events}}$ & Top quark, \WW & 4 \\
        \hline
    \end{tabular}
    \label{tab:fit_overview}
\end{table*}

The systematic uncertainties in this analysis arise either from an experimental or a theoretical source. The experimental uncertainties in the signal and background processes, as well as the theoretical uncertainties in the background processes, are taken into account for all the results discussed in Section~\ref{sec:results}. The treatment of the theoretical uncertainties in the signal processes is instead dependent on the measurement and interpretation being made. As an example, when measuring production cross sections for the STXS measurements, the theoretical uncertainties affecting the signal cross section in a given STXS bin are dropped and only the shape component is kept.

The following experimental uncertainties are included in the signal extraction fit.
\begin{itemize}
\item The integrated luminosities for the 2016, 2017, and 2018 data-taking years have 1.2--2.5\% individual uncertainties, while the overall uncertainty for the 2016--2018 period is 1.6\%~\cite{CMS-LUM-17-003,CMS-PAS-LUM-17-004,CMS-PAS-LUM-18-002}. This uncertainty is partially correlated among the three data sets, and is applied to all samples that are purely based on simulation.

\item The uncertainties in the trigger efficiency and lepton reconstruction and identification efficiencies are modeled in bins of the lepton \pt and $\eta$, independently for electrons and muons. These uncertainties cause both a normalization and a shape change of the signal and background templates and are kept uncorrelated among the three data sets. Their effect is of $\approx$2\% for electrons and $\approx$1\% for muons.

\item The uncertainties in the determination of the lepton momentum scale, jet energy scale, and unclustered energy scale cause the migration of the simulated events inside or outside the analysis acceptance, as well as migrations across the bins of the signal and background templates. The impact of these sources in the template normalizations is 0.6--1.0\% for the electron momentum scale, 0.2\% for the muon momentum scale, and 1--10\% for \ptvecmiss. The main contribution to these uncertainties arises from the limited data sample used for their estimation, and they are therefore treated as uncorrelated nuisance parameters among the three years. The jet energy scale uncertainty is modeled by implementing eleven independent nuisance parameters corresponding to different jet energy correction sources, six of which are correlated among the three data sets. Their effects vary in the range of 1--10\%, according mainly to the jet multiplicity in the analysis phase space.

\item The uncertainty in the jet energy resolution smearing applied to simulated samples to match the \pt resolution measured in data causes both a normalization and a shape change of the templates. This uncertainty has a minor impact on all the analyzed categories (effect below $\approx$1\%) and is uncorrelated among the three data sets.

\item The uncertainty in the pileup jet identification efficiency is modeled in bins of the jet \pt and $\eta$. It is considered for jets with $\pt<50\GeV$, since pileup jet identification techniques are only used for low-\pt jets. This uncertainty produces a change in both normalization and shape of the signal and background templates and is kept uncorrelated among the three data sets. The effect of this uncertainty on the measured quantities is found to be below 1\%.

\item The uncertainty in the \PQb tagging efficiency is modeled by implementing seventeen nuisance parameters, five of which are related to the theoretical uncertainties involved in the measurements and are therefore correlated among the three data sets. The remaining four parameters per data set, which arise from the statistical accuracy of the efficiency measurement, are kept uncorrelated~\cite{Sirunyan:2017ezt}. These uncertainties have an impact on both the shape of the templates and their normalization for all the simulated samples.

\item The uncertainties in the nonprompt lepton background estimation affect both the normalization and shape of the templates of this process. They arise from the limited size of the data set used for the misidentification rate measurement and the difference in the flavor composition of jets mismeasured as leptons between the measurement region and the signal phase space. Both sources are implemented as uncorrelated nuisance parameters between electrons and muons, given the different mismeasurement probabilities for the two flavors, and are uncorrelated among the three data sets. Their effects vary between few percent to $\approx$10\% depending on the SR. A further normalization uncertainty of 30\% is assigned to cover any additional mismodeling of the jet flavor composition using data in control samples, as described in Section~\ref{sec:background}. The latter uncertainty is correlated among the data sets, but uncorrelated among SRs containing different lepton flavor combinations, for which the main mechanism of nonprompt lepton production arises from different processes.

\item The statistical uncertainty due to the limited number of simulated events is associated with each bin of the simulated signal and background templates~\cite{BARLOW1993219}.

\end{itemize}

The theoretical uncertainties relevant to the simulated MC samples have different sources: the choice of the PDF set and the strong coupling constant \alpS, missing higher-order corrections in the perturbative expansion of the simulated matrix elements, and modeling of the pileup. Template variations, both in shape and normalization, associated with the aforementioned sources are treated as correlated nuisance parameters for the three data sets.

The uncertainties in the PDF set and \alpS choice are found to have a negligible effect on the simulated templates (the effect of the shape variation on the expected uncertainties was found to be below 1\%), therefore only the normalization change is considered, taking into account the effect due to the cross section and acceptance variation. These uncertainties are not considered for backgrounds with normalization constrained through data in dedicated CRs. For the Higgs boson signal processes, these theoretical uncertainties are computed by the LHC Higgs Cross Section Working Group~\cite{deFlorian:2227475} for each production mechanism.

The effect of missing higher-order corrections for the background processes is estimated by reweighting the MC simulation events with alternative event weights, where the $\mu_\text{R}$ and $\mu_\text{F}$ scales are varied by a factor of 0.5 or 2, and the envelopes of the varied templates are taken as the one standard deviation variation. All the combinations of the $\mu_\text{R}$ and $\mu_\text{F}$ scale variations are considered for computing the envelope, except for the extreme case where $\mu_\text{R}$ is varied by 0.5 and $\mu_\text{F}$ by 2, or vice versa. For backgrounds with normalization constrained using data in dedicated CRs, only the shape variation of the simulated templates arising from this procedure is considered. 
For the \WW background, an uncertainty in the higher-order reweighting described in Section~\ref{sec:background} is derived by shifting $\mu_\text{R}$, $\mu_\text{F}$, and the resummation scale.
For the \ggH signal sample, the uncertainties are decomposed into several sources according to Ref.~\cite{deFlorian:2227475}, to account for the overall cross section, migrations of events among jet multiplicity and \ptH bins, choice of the resummation scale, and finite top quark mass effects. For the VBF signal sample, different sources of uncertainty are also decoupled to account for the overall normalization, migrations of events among Higgs boson \pt, \njet, and \mjj bins, and EW corrections to the production cross section. 
The uncertainties due to missing higher-order corrections for the other signal samples are taken from Ref.~\cite{deFlorian:2227475}.
For both PDF and missing higher-order uncertainties, the nuisance parameters are correlated for the \WH and \ZH processes and uncorrelated for the other ones.

In order to assess the uncertainty in the pileup modeling, the total inelastic $\Pp\Pp$ cross section of 69.2\unit{mb}~\cite{ATLAS:2016pu,Sirunyan:2018nqx} is varied within a 5\% uncertainty, which includes the uncertainty in the inelastic cross section measurement, as well as the difference in the primary vertex reconstruction efficiency between data and simulation.

A theoretical uncertainty due to the modeling of the PS and UE is taken into account for all the simulated samples. The uncertainty in the PS modeling is evaluated by varying the PS weights computed by \PYTHIA 8.212 on an event-by-event basis, keeping the variations of the weights related to initial- and final-state radiation contributions uncorrelated. The uncertainty in the UE modeling is evaluated by shifting the nominal templates according to alternative MC simulations generated with a variation of the UE tune within its uncertainty. The corresponding nuisance parameter is correlated among all samples and between 2017 and 2018 data sets. An uncorrelated nuisance parameter is used for the 2016 data set, as the corresponding simulations are based on a different UE tune. The PS uncertainty affects the shape of the templates mainly through the migration of events across jet multiplicity bins, while the UE uncertainty is found to have a negligible impact on the shape of the templates and a normalization effect of $\approx$1.5\%.

Additional theoretical uncertainties in specific background processes are also taken into account. A 15\% uncertainty is assigned to the relative fraction of the gluon-induced component in the \WW background process~\cite{Caola:2016trd}. An uncertainty of 8\% is assigned to the relative fraction of single top quark and \ttbar processes. A 30\% uncertainty is assigned to the \wgs process associated with the measurement of the scale factor in the trilepton CR.

For the measurement of the signal cross sections in the STXS framework, the effect of theoretical uncertainties in the template normalizations is removed for signal processes in each STXS bin being measured. In cases where two or more STXS bins are measured together because of the lack of statistical accuracy in measuring single bin cross sections, the shape effect of theoretical uncertainties causing event migrations among the merged bins is kept.
In addition, residual theoretical uncertainties arising from $\mu_\text{R}$ and $\mu_\text{F}$ variations are taken into account to describe the acceptance effects that cause a shape variation of the signal templates within each STXS bin. The latter uncertainties are correlated among STXS bins that share a similar phase space definition, for example, \ggH 0-jet bins, \ggH 1-jet bins, \ggH high-\pt bins, and \ggH in VBF topology bins. A similar approach is used for the VBF STXS bins. For the measurement of leptonic VH cross sections in STXS bins, the aforementioned theoretical uncertainties are found to have a marginal impact with respect to the measurement statistical accuracy and have been neglected.

The contributions of different sources of systematic uncertainty in the signal strength measurement are summarized in Table~\ref{tab:unc_breakdown}.

\begin{table*}
    \centering
    \topcaption{Contributions of different sources of uncertainty in the signal strength measurement. The systematic component includes the combined effect from all sources besides background normalization and the size of the dataset, which make up the statistical part.}
    \begin{tabular}{lccccc}
        \hline
        Uncertainty source & $\Delta\PGm /\mu$ & $\Delta\mu_{\ggH}/\mu_{\ggH}$ & $\Delta\mu_{\mathrm{VBF}}/\mu_{\mathrm{VBF}}$ & $\Delta\mu_{\WH}/\mu_{\WH}$ & $\Delta\mu_{\ZH}/\mu_{\ZH}$ \\
        \hline
        Theory (signal) & 4\% & 5\% & 13\% & 2\% & $<$1\% \\
        Theory (background) & 3\% & 3\% & 2\% & 4\% & 5\% \\
        Lepton misidentification & 2\% & 2\% & 9\% & 15\% & 4\% \\
        Integrated luminosity & 2\% & 2\% & 2\% & 2\% & 3\% \\
        \PQb tagging & 2\% & 2\% & 3\% & $<$1\% & 2\% \\
        Lepton efficiency & 3\% & 4\% & 2\% & 1\% & 4\% \\
        Jet energy scale & 1\% & $<$1\% & 2\% & $<$1\% & 3\% \\
        Jet energy resolution & $<$1\% & 1\% & $<$1\% & $<$1\% & 3\% \\
        \ptmiss scale & $<$1\% & 1\% & $<$1\% & 2\% & 2\% \\
        PDF & 1\% & 2\% & $<$1\% & $<$1\% & 2\% \\
        Parton shower & $<$1\% & 2\% & $<$1\% & 1\% & 1\%\\
        Backg. norm. & 3\% & 4\% & 6\% & 4\% & 6\% \\ [\cmsTabSkip]
        Stat. uncertainty & 5\% & 6\% & 28\% & 21\% & 31\% \\
        Syst. uncertainty & 9\% & 10\% & 23\% & 19\% & 11\% \\ [\cmsTabSkip]
        Total uncertainty & 10\% & 11\% & 36\% & 29\% & 33\% \\
        \hline
    \end{tabular}
    \label{tab:unc_breakdown}
\end{table*}

\section{Results}
\label{sec:results}

\begin{figure}
    \centering
    \includegraphics[width=0.49\textwidth]{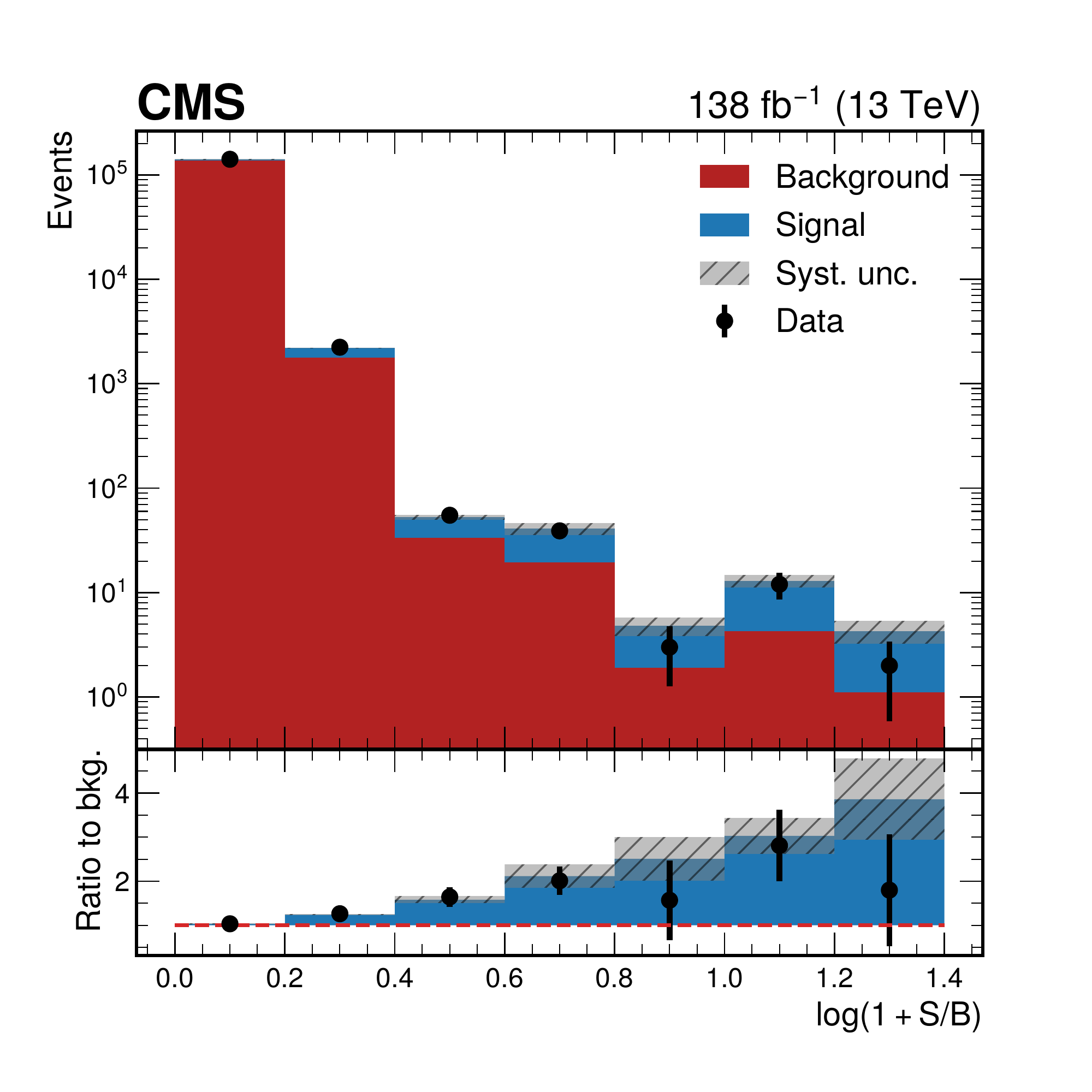}
    \caption{Distribution of events as a function of the statistical significance of their corresponding bin in the analysis template, including all categories. Signal and background contributions are shown after the fit to the data.}
    \label{fig:significance}
\end{figure}

Results are presented in terms of signal strength modifiers, STXS cross sections, and coupling modifiers. In all cases they are extracted via a simultaneous maximum likelihood fit to all the analysis categories, as explained in Section~\ref{sec:uncertainties}. The mass of the Higgs boson is assumed to be 125.38\GeV, as measured by the CMS Collaboration~\cite{HggMass}. The effect on event yields of varying $m_\PH$ within its uncertainty is found to be below 1\%. The number of expected and measured events for signal and background processes, as well as the number of observed events in each category, are reported in Tables~\ref{tab:yields_1}--\ref{tab:yields_4}. The normalization factors of the background contributions are found to be consistent with unity within their uncertainties. Figure~\ref{fig:significance} summarizes the full analysis template by showing the distribution of events as a function of the observed significance of the corresponding bins.

\begin{table*}
    \centering
    \topcaption{Number of events by process in the \ggH DF categories after the fit to the data, scaling the \ggH, VBF, \WH, and \ZH production modes separately. The \ttH contribution is fixed to its SM expectation. Numbers in parenthesis indicate expected yields.}
    \begin{tabular}{llll}
        \hline
                Process &           0-jets \ggH DF &            1-jet \ggH DF &         2-jets \ggH DF \\
        \hline
                   \ggH &     1875 $\pm$ 45 (2157) &       881 $\pm$ 28 (942) &        67 $\pm$ 5 (71) \\
                    VBF &          15 $\pm$ 2 (23) &          62 $\pm$ 7 (92) &          4 $\pm$ 1 (6) \\
                    \WH &         103 $\pm$ 7 (51) &        124 $\pm$ 10 (60) &         18 $\pm$ 2 (9) \\
                    \ZH &          38 $\pm$ 3 (19) &          33 $\pm$ 3 (17) &          7 $\pm$ 1 (4) \\
                   \ttH &                      \NA &            1 $\pm$ 1 (1) &          1 $\pm$ 1 (1) \\
  \textit{Total signal} &     2032 $\pm$ 51 (2250) &     1101 $\pm$ 31 (1111) &        99 $\pm$ 6 (90) \\ [\cmsTabSkip]

                     \WW &  37297 $\pm$ 285 (34781) &  12703 $\pm$ 307 (14932) &   748 $\pm$ 121 (1101) \\
              Top quark &  10165 $\pm$ 179 (10204) &  19711 $\pm$ 298 (19766) &  3989 $\pm$ 123 (3868) \\
              Nonprompt &    4407 $\pm$ 225 (5888) &    1999 $\pm$ 141 (2769) &     252 $\pm$ 42 (262) \\
                     DY &       495 $\pm$ 24 (563) &       822 $\pm$ 12 (792) &        87 $\pm$ 4 (86) \\
         $\PV\PZ$/$\PV\PGg^*$ &     1464 $\pm$ 45 (1776) &     1297 $\pm$ 44 (1531) &      123 $\pm$ 7 (140) \\
              $\PV\PGg$ &     1181 $\pm$ 19 (1273) &       723 $\pm$ 18 (777) &        57 $\pm$ 3 (56) \\
               Triboson &          38 $\pm$ 1 (39) &          66 $\pm$ 1 (72) &        13 $\pm$ 1 (14) \\
\textit{Total background} &  55045 $\pm$ 409 (54524) &  37321 $\pm$ 453 (40639) &  5269 $\pm$ 178 (5526) \\ [\cmsTabSkip]
\textit{Total prediction} &  57077 $\pm$ 412 (56773) &  38422 $\pm$ 454 (41750) &  5368 $\pm$ 178 (5616) \\
      \textit{Data}     &                    57024 &                    38373 &                   5380 \\
        \hline
    \end{tabular}
    \label{tab:yields_1}
\end{table*}

\begin{table*}
    \centering
    \topcaption{Number of events by process in the \ggH SF categories after the fit to the data, scaling the \ggH, VBF, \WH, and \ZH production modes separately. The \ttH contribution is fixed to its SM expectation. Numbers in parenthesis indicate expected yields.}
    \begin{tabular}{llll}
        \hline
                Process &         0-jets \ggH SF &          1-jet \ggH SF &        2-jets \ggH SF \\
        \hline
                   \ggH &     780 $\pm$ 31 (891) &     397 $\pm$ 18 (422) &       86 $\pm$ 7 (89) \\
                    VBF &          5 $\pm$ 1 (7) &        29 $\pm$ 4 (42) &       10 $\pm$ 1 (13) \\
                    \WH &        24 $\pm$ 3 (11) &        34 $\pm$ 4 (16) &        12 $\pm$ 1 (6) \\
                    \ZH &         14 $\pm$ 1 (7) &         16 $\pm$ 2 (8) &         7 $\pm$ 1 (3) \\
                   \ttH &                    \NA &                    \NA &         1 $\pm$ 1 (1) \\
  \textit{Total signal} &     823 $\pm$ 31 (915) &     476 $\pm$ 18 (489) &     114 $\pm$ 7 (112) \\ [\cmsTabSkip]

                     \WW &  7034 $\pm$ 184 (6464) &  2711 $\pm$ 128 (3064) &    276 $\pm$ 61 (480) \\
              Top quark &   1345 $\pm$ 42 (1294) &   3711 $\pm$ 75 (3524) &  1879 $\pm$ 51 (1758) \\
              Nonprompt &     641 $\pm$ 88 (701) &     366 $\pm$ 54 (412) &    103 $\pm$ 18 (119) \\
                     DY &  3149 $\pm$ 271 (2706) &  4098 $\pm$ 197 (3284) &   1403 $\pm$ 83 (829) \\
         $\PV\PZ$/$\PV\PGg^*$ &     327 $\pm$ 13 (371) &     270 $\pm$ 10 (301) &       63 $\pm$ 4 (70) \\
              $\PV\PGg$ &      138 $\pm$ 6 (145) &     193 $\pm$ 15 (201) &       48 $\pm$ 5 (47) \\
               Triboson &          4 $\pm$ 1 (5) &        10 $\pm$ 1 (11) &         6 $\pm$ 1 (6) \\
\textit{Total background} &  12639 $\pm$ 342 (11684) &  11359 $\pm$ 253 (10797) &  3777 $\pm$ 117 (3309) \\ [\cmsTabSkip]

\textit{Total prediction} &  13462 $\pm$ 343 (12599) &  11835 $\pm$ 254 (11286) &  3891 $\pm$ 117 (3421) \\
          \textit{Data} &  13507                   &                    11976 &                   3950 \\
        \hline
    \end{tabular}
    \label{tab:yields_2}
\end{table*}

\begin{table*}
    \centering
    \topcaption{Number of events by process in the VBF and {\VH}2j categories after the fit to the data, scaling the \ggH, VBF, \WH, and \ZH production modes separately. The \ttH contribution is fixed to its SM expectation. Numbers in parenthesis indicate expected yields.}
    \begin{tabular}{lllll}
        \hline
                Process &                VBF DF &              VBF SF &             {\VH}2j DF &             {\VH}2j SF \\
        \hline
                   \ggH &     114 $\pm$ 8 (115) &     21 $\pm$ 2 (21) &     36 $\pm$ 3 (39) &     27 $\pm$ 2 (29) \\
                    VBF &      62 $\pm$ 11 (91) &     39 $\pm$ 5 (57) &       2 $\pm$ 1 (3) &       2 $\pm$ 1 (2) \\
                    \WH &        14 $\pm$ 1 (7) &       1 $\pm$ 1 (1) &     26 $\pm$ 4 (13) &      16 $\pm$ 2 (8) \\
                    \ZH &         5 $\pm$ 1 (2) &       1 $\pm$ 1 (0) &      13 $\pm$ 2 (7) &       8 $\pm$ 1 (4) \\
                   \ttH &                   \NA &                 \NA &                 \NA &                 \NA \\
  \textit{Total signal} &    195 $\pm$ 14 (215) &     62 $\pm$ 6 (79) &     77 $\pm$ 5 (62) &     53 $\pm$ 3 (43) \\ [\cmsTabSkip]

                    \WW &  1319 $\pm$ 57 (1368) &  109 $\pm$ 17 (102) &   98 $\pm$ 44 (205) &   56 $\pm$ 22 (134) \\
              Top quark &  2875 $\pm$ 65 (3148) &   267 $\pm$ 8 (249) &  743 $\pm$ 32 (730) &  539 $\pm$ 16 (514) \\
              Nonprompt &    404 $\pm$ 36 (399) &     28 $\pm$ 4 (32) &   81 $\pm$ 13 (113) &    62 $\pm$ 10 (72) \\
                     DY &     249 $\pm$ 4 (241) &  402 $\pm$ 27 (465) &     77 $\pm$ 3 (77) &  555 $\pm$ 48 (479) \\
   $\PV\PZ$/$\PV\PGg^*$ &     184 $\pm$ 9 (221) &     11 $\pm$ 1 (12) &     49 $\pm$ 3 (55) &     23 $\pm$ 2 (27) \\
              $\PV\PGg$ &     110 $\pm$ 4 (117) &     10 $\pm$ 1 (10) &     26 $\pm$ 3 (25) &     16 $\pm$ 5 (17) \\
               Triboson &       11 $\pm$ 1 (11) &       1 $\pm$ 1 (1) &       6 $\pm$ 1 (7) &       4 $\pm$ 1 (3) \\
\textit{Total background} & 5154 $\pm$ 94 (5505) &  827 $\pm$ 33 (871) & 1080 $\pm$ 56 (1212) &  1255 $\pm$ 56 (1245) \\ [\cmsTabSkip]

\textit{Total prediction} & 5349 $\pm$ 95 (5720) &  889 $\pm$ 34 (950) & 1157 $\pm$ 56 (1274) &  1308 $\pm$ 56 (1288) \\
          \textit{Data} &                 5254 &                 862 &                 1164 &                  1318 \\
        \hline
    \end{tabular}
    \label{tab:yields_3}
\end{table*}

\begin{table*}
    \centering
    \topcaption{Number of events by process in the \WH{}SS, \WH{}3\Pell, \ZH{}3\Pell, and \ZH{}4\Pell categories after the fit to the data, scaling the \ggH, VBF, \WH, and \ZH production modes separately. The \ttH contribution is fixed to its SM expectation. Numbers in parenthesis indicate expected yields.}
    \begin{tabular}{lllll}
        \hline
                Process &               \WH{}SS &        \WH{}3\Pell &           \ZH{}3\Pell &       \ZH{}4\Pell \\
        \hline
                   \ggH &         1 $\pm$ 1 (1) &                \NA &                   \NA &               \NA \\
                    VBF &                   \NA &                \NA &                   \NA &               \NA \\
                    \WH &     148 $\pm$ 12 (69) &    44 $\pm$ 5 (20) &         2 $\pm$ 1 (1) &               \NA \\
                    \ZH &       10 $\pm$ 11 (5) &      3 $\pm$ 1 (2) &       74 $\pm$ 7 (36) &   19 $\pm$ 2 (10) \\
                   \ttH &         1 $\pm$ 1 (1) &                \NA &         1 $\pm$ 1 (1) &               \NA \\
  \textit{Total signal} &     159 $\pm$ 12 (76) &    48 $\pm$ 5 (22) &       76 $\pm$ 7 (38) &   19 $\pm$ 2 (10) \\ [\cmsTabSkip]

                    \WW &       40 $\pm$ 1 (39) &                \NA &                   \NA &               \NA \\
              Top quark &       62 $\pm$ 1 (62) &                \NA &                   \NA &               \NA \\
              Nonprompt &    596 $\pm$ 37 (805) &    55 $\pm$ 6 (85) &    166 $\pm$ 16 (215) &               \NA \\
                     DY &       28 $\pm$ 7 (35) &                \NA &       30 $\pm$ 1 (29) &     1 $\pm$ 1 (1) \\
   $\PV\PZ$/$\PV\PGg^*$ &  1309 $\pm$ 26 (1355) & 311 $\pm$ 10 (276) &  1905 $\pm$ 25 (1796) &   45 $\pm$ 1 (39) \\
              $\PV\PGg$ &    135 $\pm$ 11 (162) &    14 $\pm$ 3 (20) &       36 $\pm$ 6 (40) &               \NA \\
               Triboson &       41 $\pm$ 1 (41) &    15 $\pm$ 1 (15) &       30 $\pm$ 1 (30) &     3 $\pm$ 1 (3) \\
\textit{Total background} &  2211 $\pm$ 47 (2498) &  396 $\pm$ 12 (397) &  2167 $\pm$ 30 (2110) &  50 $\pm$ 1 (44) \\ [\cmsTabSkip]
\textit{Total prediction} &  2370 $\pm$ 49 (2574) &  444 $\pm$ 13 (419) &  2243 $\pm$ 31 (2148) &  69 $\pm$ 2 (54) \\
          \textit{Data} &                  2359 &                 423 &                  2315 &               69 \\
        \hline
    \end{tabular}
    \label{tab:yields_4}
\end{table*}

The \hww selection is subject to some degree of contamination from events in which the Higgs boson decays to a pair of \PGt leptons that themselves decay leptonically. These events are included in the signal definition, and their contribution ranges from below 1\% in the ggH and VBF categories up to $\approx$10\% in some of the \WH categories. As described in previous sections, CRs are used to fix the normalization of dominant backgrounds from data. This is achieved by scaling the corresponding background contributions jointly in the CR and SR. Given that the procedure effectively amounts to a measurement of the cross section of the background in question, the contributions from the 2017 and 2018 data sets are scaled together. The 2016 data set is kept separate in this regard because a different \PYTHIA tune was used.

For inclusive measurements, results are extracted in the form of signal strength modifiers $\mu$. These are defined as the product of the production cross section and the branching ratio to a W boson pair, normalized to the SM prediction ($\sigma\mathcal{B}/(\sigma\mathcal{B})_{\mathrm{SM}}$). Couplings of the Higgs boson to fermions and vector bosons are measured in the $\kappa$ framework~\cite{Heinemeyer:1559921}, while STXS results are provided as cross sections.

\subsection{Signal strength modifiers}
\label{subsec:signal_strenghts}

\begin{figure}
    \centering
    \includegraphics[width=0.49\textwidth]{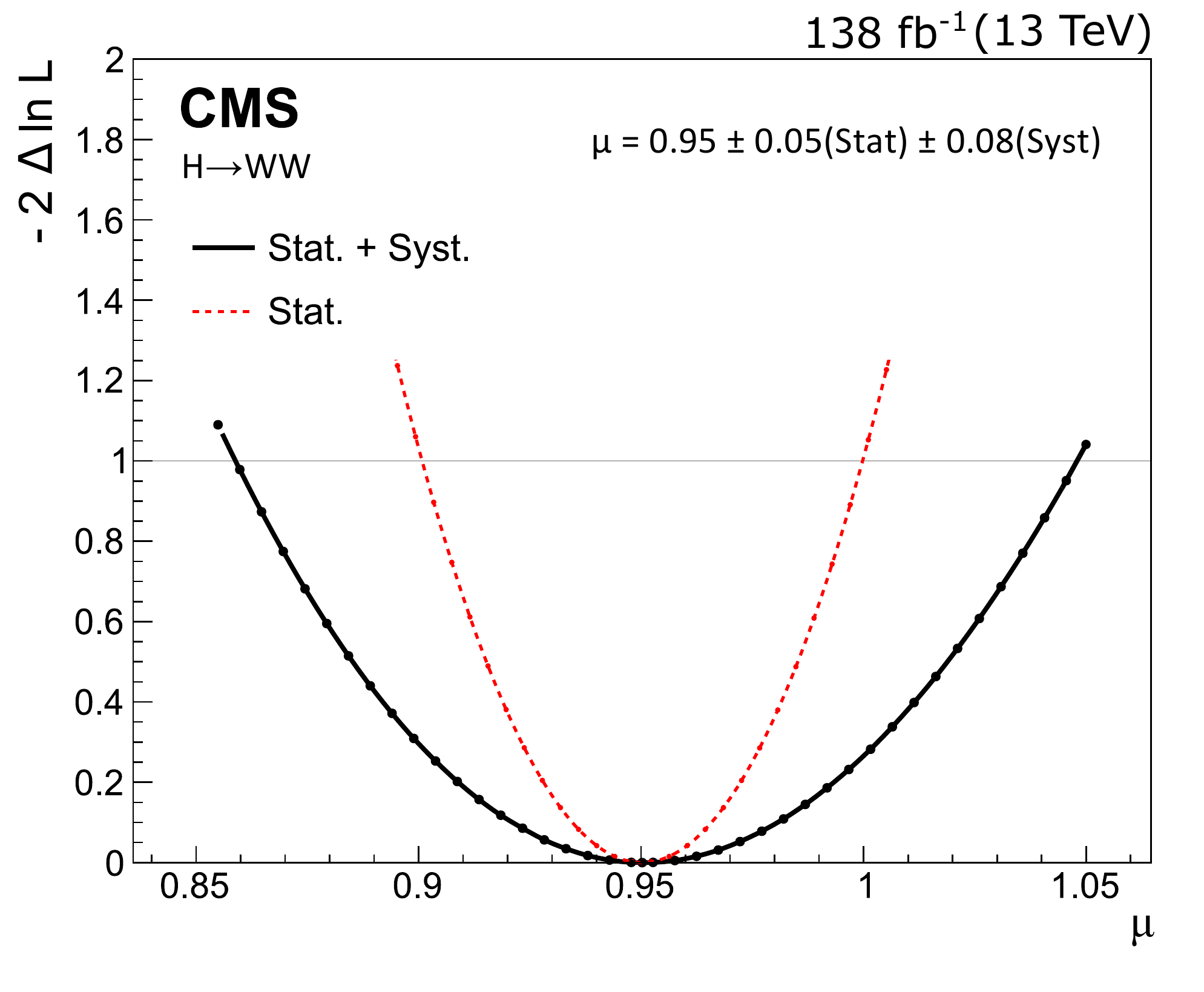}
    \caption{Observed profile-likelihood function for the global signal strength modifier $\mu$. The dashed curve corresponds to the profile-likelihood function obtained considering statistical uncertainties only.}
    \label{fig:LH_scan_mu}
\end{figure}

The global signal strength modifier is extracted by fitting the template to data leaving all contributions coming from the Higgs boson free to float, but keeping the relative importance of the different production modes fixed to the values predicted by the SM. As such, this measurement gives information on the compatibility of the SM with the LHC Run 2 data set. The observed signal strength modifier is:
\begin{linenomath}
\begin{equation}
    \mu = 0.95^{+0.10}_{-0.09} = 0.95\pm 0.05\stat\pm 0.08\syst,
\end{equation}
\end{linenomath}
where the uncertainty has been broken down into its statistical and systematic components. The purely statistical component is extracted by fixing all nuisance parameters in the likelihood function to their best fit values and extracting the corresponding profile. The systematic component is obtained by the difference in quadrature between the total uncertainty and the statistical one. The observed and expected profile likelihood functions, both with the full set of uncertainty sources as well as with statistical ones only, are shown in Fig.~\ref{fig:LH_scan_mu}.

\begin{figure}
    \centering
    \includegraphics[width=\cmsFigureWidth]{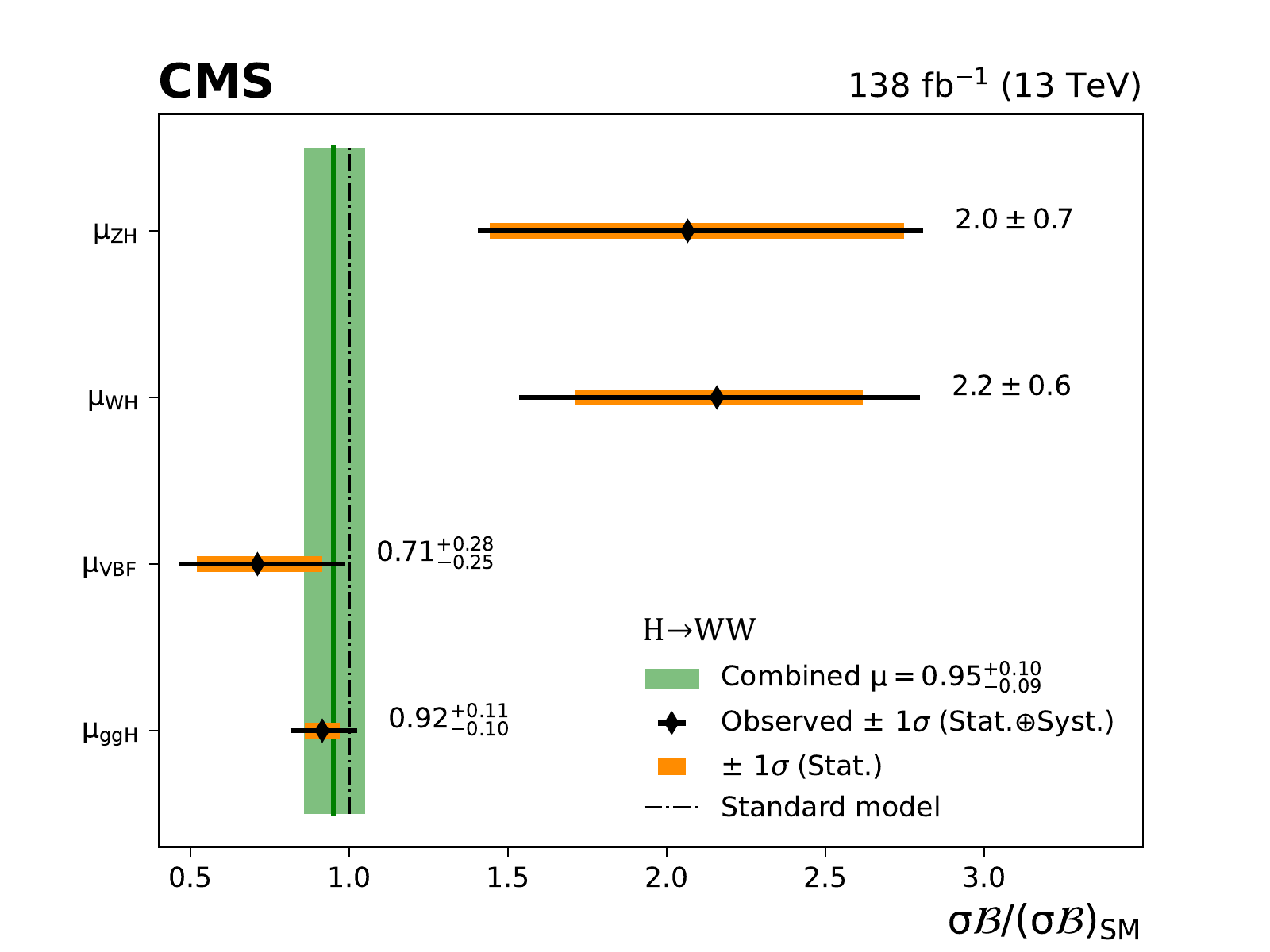}
    \caption{Observed signal strength modifiers for the main SM production modes.}
    \label{fig:prodModes_mus}
\end{figure}

\begin{figure}
    \centering
    \includegraphics[width=\cmsFigureWidth]{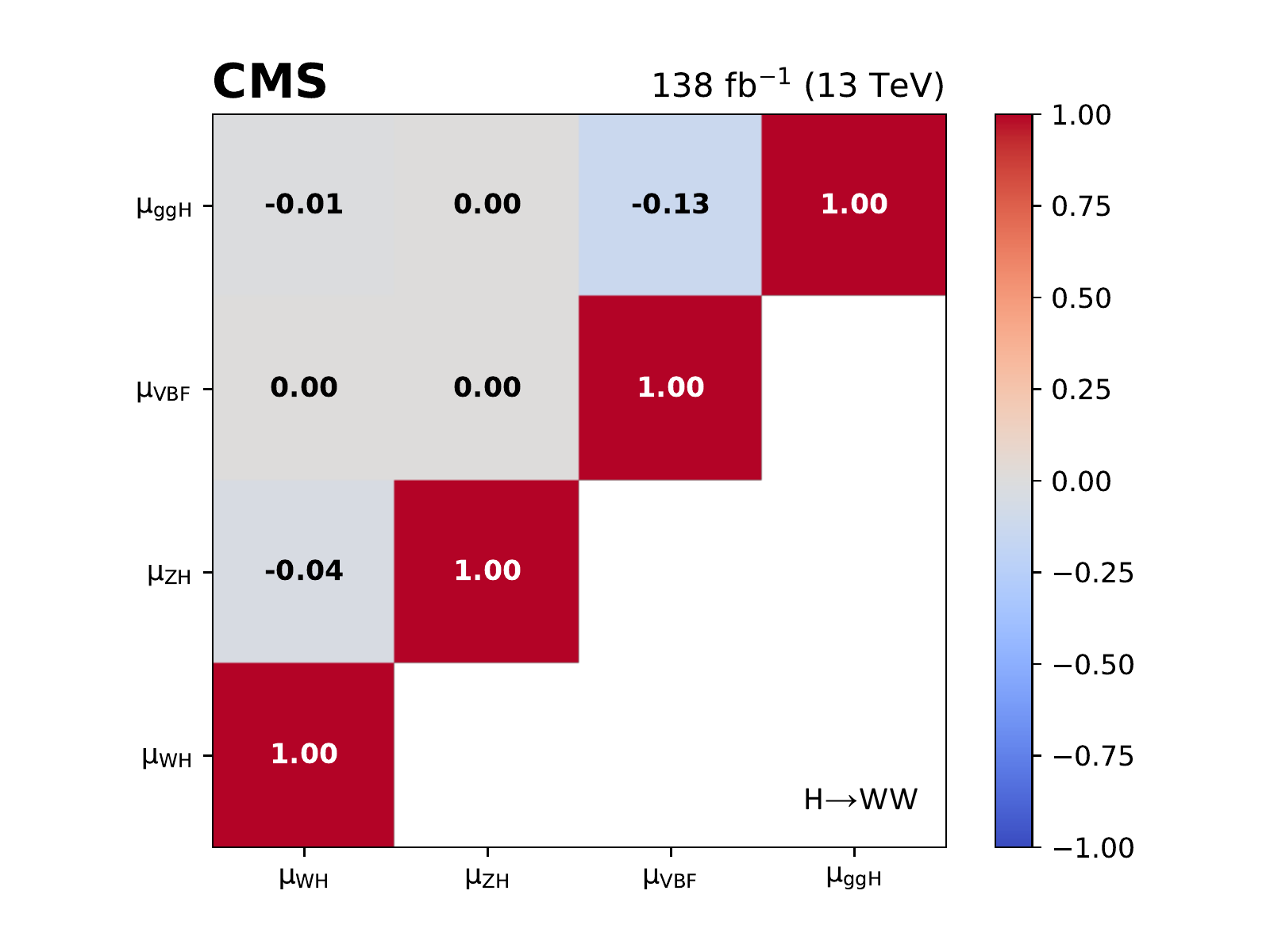}
    \caption{Correlation matrix between the signal strength modifiers of the main production modes of the Higgs boson.}
    \label{fig:corr_matrix_prodModes}
\end{figure}

Results are also extracted for individual production modes, by performing a 4-parameter fit in which contributions from the \ggH, VBF, \WH, and \ZH modes are left free to float independently. Contributions from the \ttH and \bbH production modes are fixed to their SM expected values within uncertainties, given that this analysis has little sensitivity to them. Results are summarized in Fig.~\ref{fig:prodModes_mus}, where the separate contributions of statistical and systematic sources of uncertainty are also shown. Results correspond to observed (expected) significances of 10.5 (11.8)$\sigma$, 3.15 (4.74)$\sigma$, 3.61 (1.82)$\sigma$, and 3.73 (2.19)$\sigma$ for the \ggH, VBF, \WH, and \ZH modes, respectively. The correlation matrix among the signal strengths is given in Fig.~\ref{fig:corr_matrix_prodModes}. The compatibility of the result with the SM is found to be 7\%.

\subsection{Higgs boson couplings}
\label{subsec:kappas}

\begin{figure}
    \centering
    \includegraphics[width=\cmsFigureWidth]{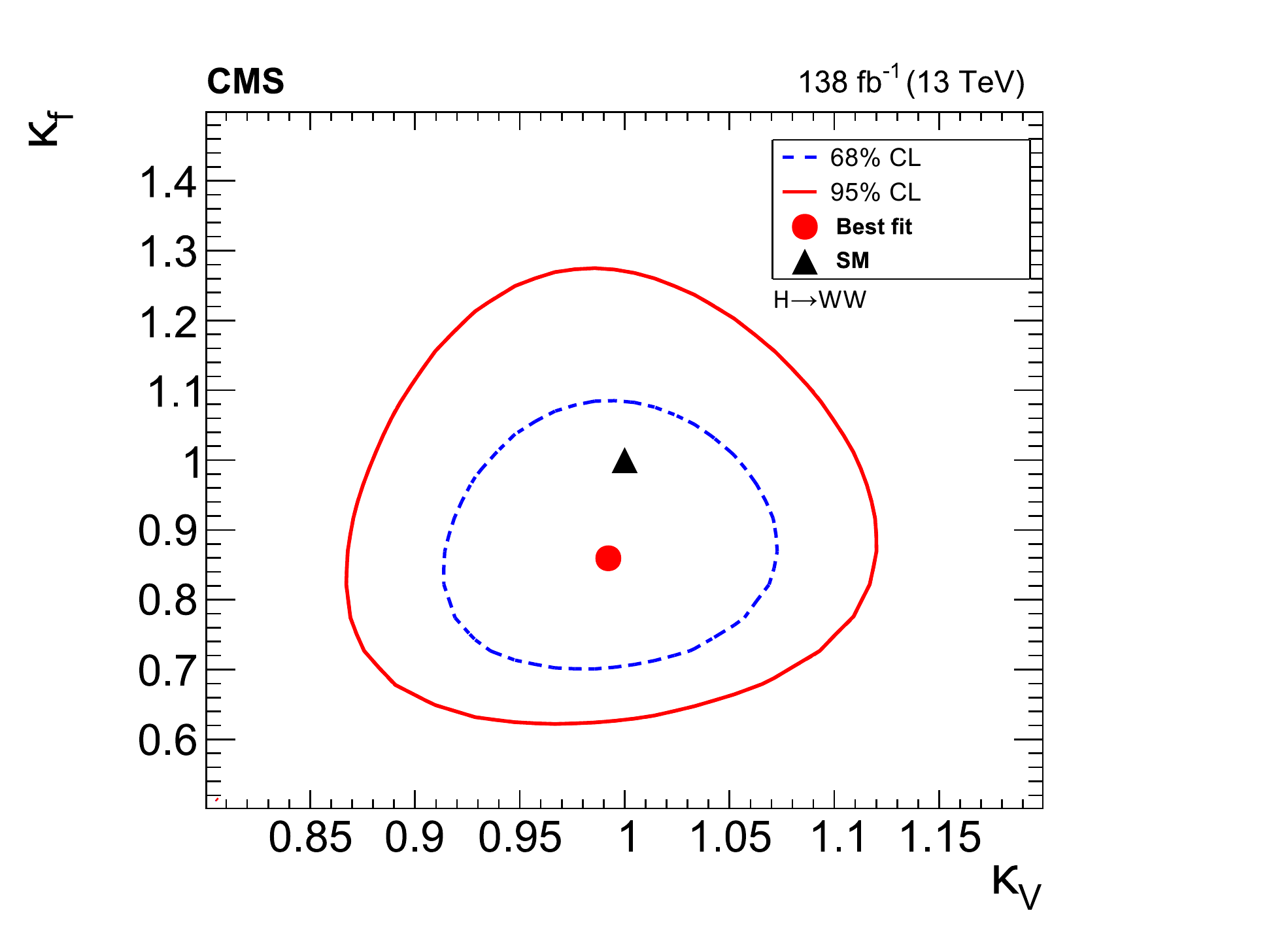}
    \caption{Two-dimensional likelihood profile as a function of the coupling modifiers $\kappa_{\PV}$ and $\kappa_\mathrm{f}$, using the $\kappa$-framework parametrization. The 95 and 68\% confidence level contours are shown as continuous and dashed lines, respectively.}
    \label{fig:kappas}
\end{figure}

Given its large branching fraction and relatively low background, the \hww channel is a good candidate to measure the couplings of the Higgs boson to fermions and vector bosons. This is performed in the so-called $\kappa$ framework. Two coupling modifiers $\kappa_{\PV}$ and $\kappa_\mathrm{f}$ are defined, for couplings to vector bosons and fermions respectively. These scale the signal yield of the \hww channel as follows:
\begin{linenomath}
\ifthenelse{\boolean{cms@external}}{\begin{multline*}
    \sigma\mathcal{B}(\PX_i\to\PH\to\WW) = \\
    \kappa_i^2\frac{\kappa_{\PV}^2}{\kappa_{\PH}^2}\sigma_{\mathrm{SM}}\mathcal{B}_{\mathrm{SM}}(\PX_i\to\PH\to\WW),
\end{multline*}}{\begin{equation}
    \sigma\mathcal{B}(\PX_i\to\PH\to\WW) = \kappa_i^2\frac{\kappa_{\PV}^2}{\kappa_{\PH}^2}\sigma_{SM}\mathcal{B}_{SM}(\PX_i\to\PH\to\WW),
\end{equation}}
\end{linenomath}
where $\kappa_{\PH} = \kappa_{\PH}(\kappa_{\PV}, \kappa_{\mathrm{f}})$ is the modifier to the total Higgs boson width, and $\PX_i$ are the different production modes. The corresponding coupling modifiers $\kappa_i$ equal $\kappa_{\mathrm{f}}$ for the \ggH, \ttH, and \bbH modes, and $\kappa_{\PV}$ for the VBF and \VH modes. Possible contributions to the total width of the Higgs boson coming from outside of the SM are neglected. The best fit values for the coupling modifiers are found to be $\kappa_{\PV} = 0.99\pm 0.05$ and $\kappa_\mathrm{f} = 0.86^{+0.14}_{-0.11}$, where the better sensitivity to $\kappa_{\PV}$ is due to the \hww decay vertex. The two-dimensional likelihood profile for the fit is shown in Fig.~\ref{fig:kappas}.

\subsection{STXS}
\label{subsec:STXS_results}

\begin{figure*}[hbt]
    \centering
    \includegraphics[width=\textwidth]{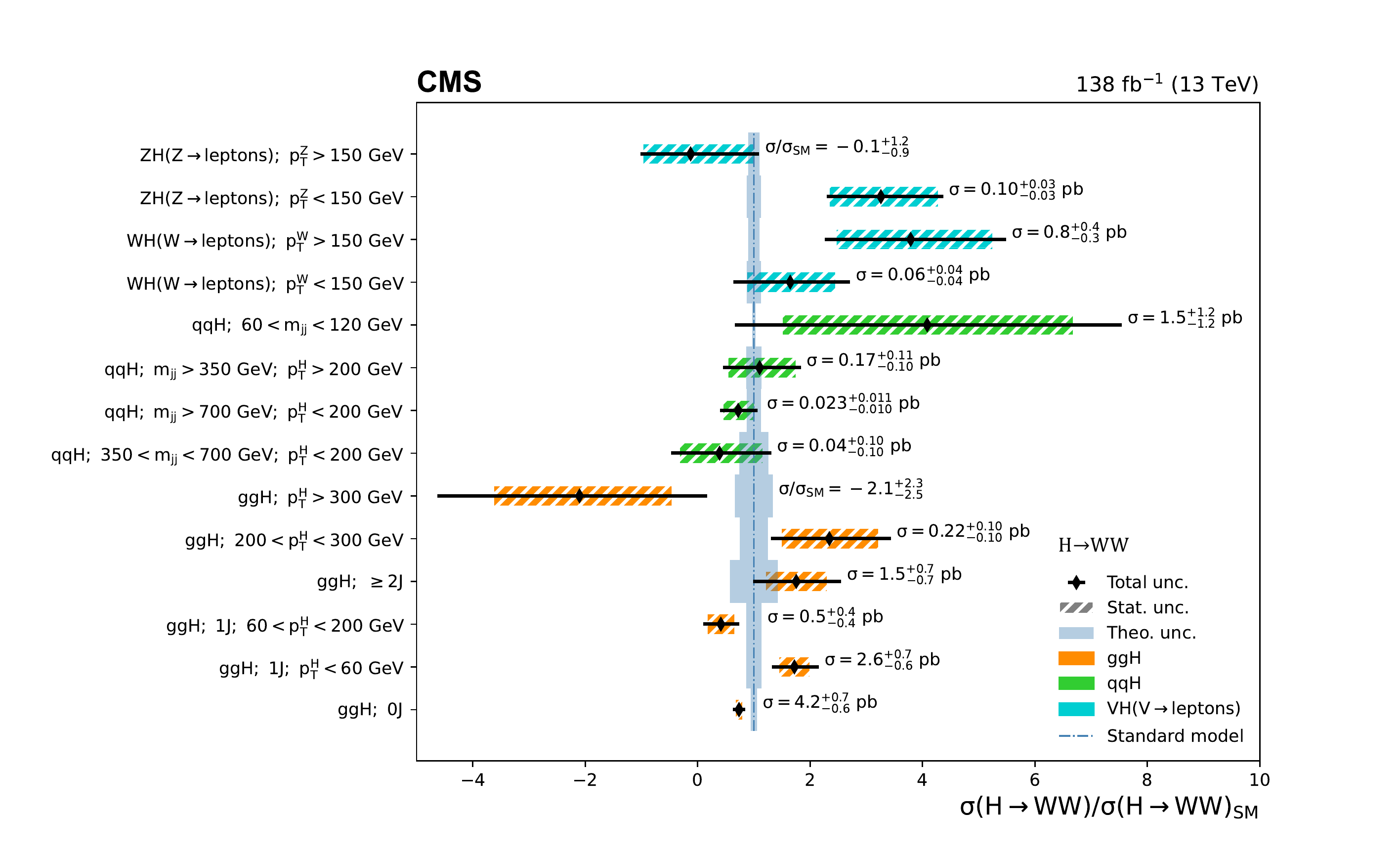}
    \caption{Observed cross sections of the \hww process in each STXS bin, normalized to the SM expectation.}
    \label{fig:STXS_sigmas}
\end{figure*}

\begin{figure*}[h]
    \centering
    \includegraphics[width=\textwidth]{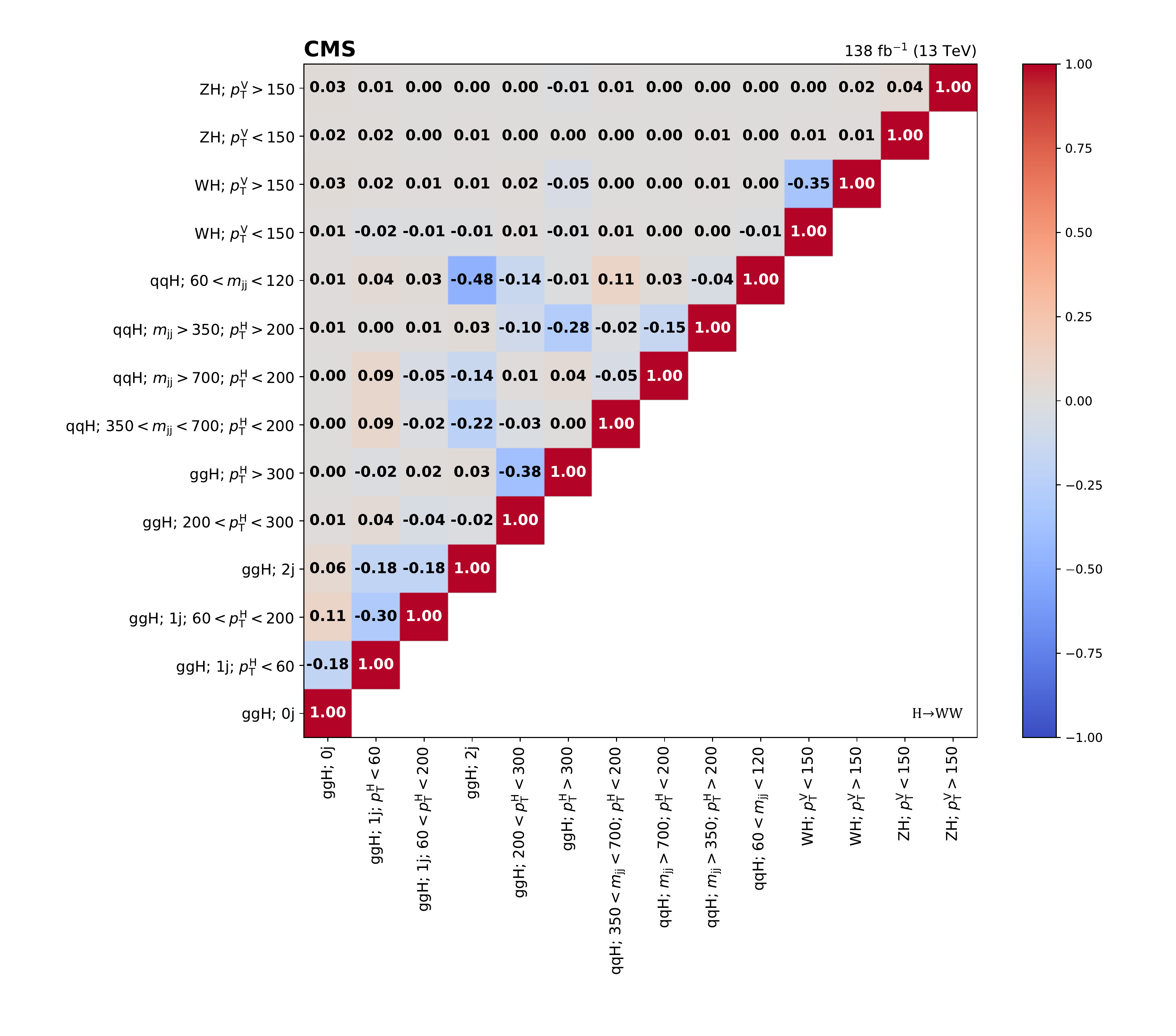}
    \caption{Correlation matrix between the measured STXS bins. All dimensional quantities in bin definitions are measured in \GeVns.}
    \label{fig:STXS_corr}
\end{figure*}

As explained in Section~\ref{sec:STXS}, the STXS measurement is carried out under the Stage 1.2 framework, although not all STXS bins are measured independently because of sensitivity limitations. Results are shown in Table~\ref{tab:STXS_mus} and in Fig.~\ref{fig:STXS_sigmas}, for the signal strength modifiers and cross sections. The uncertainties are reported separately for statistical (stat), theoretical (theo), and experimental (exp) systematic sources. The correlation matrix for the measured STXS bins is shown in Fig.~\ref{fig:STXS_corr}. Since final results are reported as cross sections, the effect of theoretical uncertainties in the normalization of signal templates is dropped, while uncertainties in the shape of the templates, such as STXS bin migration, are accounted for. In cases where cross sections are measured to be zero, an upper limit is reported instead of a symmetric confidence interval, so that all intervals reported correspond to a 68\% confidence level. The compatibility of the STXS fit with the SM is found to be 1\%.

\begin{table*}
    \centering
    \topcaption{Observed cross sections of the \hww process in each STXS bin. The uncertainties in the observed cross sections and their ratio to the SM expectation do not include the theoretical uncertainties on the latter. In cases where the ratio to the SM cross section is measured below zero, an upper limit at 68\% confidence level on the observed cross section is reported. All dimensional quantities in STXS bin definitions are measured in \GeVns.}
    \cmsTable{
    \begin{tabular}{llll}
        \hline
        STXS bin & $\sigma(\hww) /\sigma(\hww)_\mathrm{SM}$ & $\sigma(\hww)\ \mathrm{[pb]}$ & $\sigma(\hww)_{\mathrm{SM}}$ [pb] \\
        \hline
        \ZH($\PZ\to \text{leptons}$); $\pt^{\PV}>150$ & $-0.1^{+1.2}_{-0.9}\stat\pm 0.1\thy^{+0.4}_{-0.3}\expt$ & $<$0.03 & $0.139\pm 0.013$ \\
        \ZH($\PZ\to \text{leptons}$); $\pt^{\PV}<150$ & $3.3^{+1.0}_{-0.9}\stat\pm 0.1\thy^{+0.4}_{-0.3}\expt$ & $0.10\pm 0.03$ & $0.030\pm 0.004$ \\
        \WH($\PW\to \text{leptons}$); $\pt^{\PV}>150$ & $3.8^{+1.5}_{-1.3}\stat\pm 0.1\thy^{+0.8}_{-0.7}\expt$ & $0.8^{+0.4}_{-0.3}$ & $0.22\pm 0.02$ \\
        \WH($\PW\to \text{leptons}$); $\pt^{\PV}<150$ & $1.6\pm 0.8\stat\pm 0.1\thy^{+0.7}_{-0.6}\expt$ & $0.06\pm 0.04$ & $0.035\pm 0.005$ \\ [\cmsTabSkip]

        qqH; $60<\mjj<120$ & $4.1\pm 2.6\stat^{+0.7}_{-0.6}\thy\pm 2.2\expt$ & $1.5\pm 1.2$ & $0.36\pm 0.01$ \\
        qqH; $\ptH>200$ & $1.1^{+0.7}_{-0.6}\stat\pm 0.1\thy\pm 0.3\expt$ & $0.17^{+0.11}_{-0.10}$ & $0.15\pm 0.02$ \\
        qqH; $\ptH<200$; $\mjj>700$ & $0.7\pm 0.3\stat\pm 0.1\thy\pm 0.2\expt$ & $0.023^{+0.011}_{-0.010}$ & $0.032\pm 0.004$ \\
        qqH; $\ptH<200$; $350<\mjj<700$ & $0.4^{+0.8}_{-0.7}\stat\pm 0.2\thy\pm 0.5\expt$ & $0.04\pm 0.10$ & $0.11\pm 0.03$ \\ [\cmsTabSkip]

        \ggH; $\ptH>300$ & $-2.1^{+1.7}_{-1.5}\stat^{+0.2}_{-0.3}\thy^{+1.6}_{-2.0}\expt$ & $<$0.04 & $0.028\pm 0.009$ \\
        \ggH; $200<\ptH<300$ & $2.3\pm 0.9\stat\pm 0.1\thy\pm 0.6\expt$ & $0.22\pm 0.10$ & $0.09\pm 0.02$ \\
        \ggH; $\geq$2j & $1.8\pm 0.6\stat\pm 0.4\thy\pm 0.4\expt$ & $1.5 \pm 0.7$ & $0.9\pm 0.4$ \\
        \ggH; 1j; $\ptH>60$ & $0.41\pm 0.25\stat^{+0.10}_{-0.06}\thy\pm 0.17\expt$ & $0.5\pm 0.4$ & $1.15\pm 0.16$ \\
        \ggH; 1j; $\ptH<60$ & $1.7\pm 0.3\stat\pm 0.2\thy\pm 0.2\expt$ & $2.6^{+0.7}_{-0.6}$ & $1.5\pm 0.2$ \\
        \ggH; 0j & $0.74\pm 0.07\stat\pm 0.04\thy^{+0.08}_{-0.07}\expt$ & $4.2^{+0.7}_{-0.6}$ & $5.8\pm 0.3$ \\
        \hline
    \end{tabular}
    }
    \label{tab:STXS_mus}
\end{table*}

\ifthenelse{\boolean{cms@external}}{\clearpage}{}
\section{Summary}
\label{sec:conclusions}

A measurement of production cross sections for the Higgs boson has been performed targeting the gluon fusion, vector boson fusion, and \PZ or \PW associated production processes in the \hww decay channel. Results are presented as signal strength modifiers, coupling modifiers, and differential cross sections in the simplified template cross section Stage 1.2 framework. The measurement has been performed on data from proton-proton collisions recorded by the CMS detector at a center-of-mass energy of 13\TeV in 2016--2018, corresponding to an integrated luminosity of 138\fbinv. Specific event selections targeting different final states have been employed, and results have been extracted via a simultaneous maximum likelihood fit to all analysis categories. The overall signal strength for production of a Higgs boson is found to be $\mu = 0.95^{+0.10}_{-0.09}$. All results are in good agreement with the standard model expectation.

\begin{acknowledgments}
We congratulate our colleagues in the CERN accelerator departments for the excellent performance of the LHC and thank the technical and administrative staffs at CERN and at other CMS institutes for their contributions to the success of the CMS effort. In addition, we gratefully acknowledge the computing centers and personnel of the Worldwide LHC Computing Grid and other centers for delivering so effectively the computing infrastructure essential to our analyses. Finally, we acknowledge the enduring support for the construction and operation of the LHC, the CMS detector, and the supporting computing infrastructure provided by the following funding agencies: BMBWF and FWF (Austria); FNRS and FWO (Belgium); CNPq, CAPES, FAPERJ, FAPERGS, and FAPESP (Brazil); MES and BNSF (Bulgaria); CERN; CAS, MoST, and NSFC (China); MINCIENCIAS (Colombia); MSES and CSF (Croatia); RIF (Cyprus); SENESCYT (Ecuador); MoER, ERC PUT and ERDF (Estonia); Academy of Finland, MEC, and HIP (Finland); CEA and CNRS/IN2P3 (France); BMBF, DFG, and HGF (Germany); GSRI (Greece); NKFIH (Hungary); DAE and DST (India); IPM (Iran); SFI (Ireland); INFN (Italy); MSIP and NRF (Republic of Korea); MES (Latvia); LAS (Lithuania); MOE and UM (Malaysia); BUAP, CINVESTAV, CONACYT, LNS, SEP, and UASLP-FAI (Mexico); MOS (Montenegro); MBIE (New Zealand); PAEC (Pakistan); MES and NSC (Poland); FCT (Portugal); MESTD (Serbia); MCIN/AEI and PCTI (Spain); MOSTR (Sri Lanka); Swiss Funding Agencies (Switzerland); MST (Taipei); MHESI and NSTDA (Thailand); TUBITAK and TENMAK (Turkey); NASU (Ukraine); STFC (United Kingdom); DOE and NSF (USA).
         
\hyphenation{Rachada-pisek} Individuals have received support from the Marie-Curie program and the European Research Council and Horizon 2020 Grant, contract Nos.\ 675440, 724704, 752730, 758316, 765710, 824093, 884104, and COST Action CA16108 (European Union); the Leventis Foundation; the Alfred P.\ Sloan Foundation; the Alexander von Humboldt Foundation; the Belgian Federal Science Policy Office; the Fonds pour la Formation \`a la Recherche dans l'Industrie et dans l'Agriculture (FRIA-Belgium); the Agentschap voor Innovatie door Wetenschap en Technologie (IWT-Belgium); the F.R.S.-FNRS and FWO (Belgium) under the ``Excellence of Science -- EOS" -- be.h project n.\ 30820817; the Beijing Municipal Science \& Technology Commission, No. Z191100007219010; the Ministry of Education, Youth and Sports (MEYS) of the Czech Republic; the Hellenic Foundation for Research and Innovation (HFRI), Project Number 2288 (Greece); the Deutsche Forschungsgemeinschaft (DFG), under Germany's Excellence Strategy -- EXC 2121 ``Quantum Universe" -- 390833306, and under project number 400140256 - GRK2497; the Hungarian Academy of Sciences, the New National Excellence Program - \'UNKP, the NKFIH research grants K 124845, K 124850, K 128713, K 128786, K 129058, K 131991, K 133046, K 138136, K 143460, K 143477, 2020-2.2.1-ED-2021-00181, and TKP2021-NKTA-64 (Hungary); the Council of Science and Industrial Research, India; the Latvian Council of Science; the Ministry of Education and Science, project no. 2022/WK/14, and the National Science Center, contracts Opus 2021/41/B/ST2/01369 and 2021/43/B/ST2/01552 (Poland); the Funda\c{c}\~ao para a Ci\^encia e a Tecnologia, grant CEECIND/01334/2018 (Portugal); the National Priorities Research Program by Qatar National Research Fund; MCIN/AEI/10.13039/501100011033, ERDF ``a way of making Europe", and the Programa Estatal de Fomento de la Investigaci{\'o}n Cient{\'i}fica y T{\'e}cnica de Excelencia Mar\'{\i}a de Maeztu, grant MDM-2017-0765 and Programa Severo Ochoa del Principado de Asturias (Spain); the Chulalongkorn Academic into Its 2nd Century Project Advancement Project, and the National Science, Research and Innovation Fund via the Program Management Unit for Human Resources \& Institutional Development, Research and Innovation, grant B05F650021 (Thailand); the Kavli Foundation; the Nvidia Corporation; the SuperMicro Corporation; the Welch Foundation, contract C-1845; and the Weston Havens Foundation (USA).  
\end{acknowledgments}

\bibliography{auto_generated}
\cleardoublepage \appendix\section{The CMS Collaboration \label{app:collab}}\begin{sloppypar}\hyphenpenalty=5000\widowpenalty=500\clubpenalty=5000
\cmsinstitute{Yerevan Physics Institute, Yerevan, Armenia}
{\tolerance=6000
A.~Tumasyan\cmsAuthorMark{1}\cmsorcid{0009-0000-0684-6742}
\par}
\cmsinstitute{Institut f\"{u}r Hochenergiephysik, Vienna, Austria}
{\tolerance=6000
W.~Adam\cmsorcid{0000-0001-9099-4341}, J.W.~Andrejkovic, T.~Bergauer\cmsorcid{0000-0002-5786-0293}, S.~Chatterjee\cmsorcid{0000-0003-2660-0349}, K.~Damanakis\cmsorcid{0000-0001-5389-2872}, M.~Dragicevic\cmsorcid{0000-0003-1967-6783}, A.~Escalante~Del~Valle\cmsorcid{0000-0002-9702-6359}, P.S.~Hussain\cmsorcid{0000-0002-4825-5278}, M.~Jeitler\cmsAuthorMark{2}\cmsorcid{0000-0002-5141-9560}, N.~Krammer\cmsorcid{0000-0002-0548-0985}, L.~Lechner\cmsorcid{0000-0002-3065-1141}, D.~Liko\cmsorcid{0000-0002-3380-473X}, I.~Mikulec\cmsorcid{0000-0003-0385-2746}, P.~Paulitsch, F.M.~Pitters, J.~Schieck\cmsAuthorMark{2}\cmsorcid{0000-0002-1058-8093}, R.~Sch\"{o}fbeck\cmsorcid{0000-0002-2332-8784}, D.~Schwarz\cmsorcid{0000-0002-3821-7331}, S.~Templ\cmsorcid{0000-0003-3137-5692}, W.~Waltenberger\cmsorcid{0000-0002-6215-7228}, C.-E.~Wulz\cmsAuthorMark{2}\cmsorcid{0000-0001-9226-5812}
% ...
\par}
\vskip\cmsinstskip
\dag:~Deceased\\
$^{1}$Also at Yerevan State University, Yerevan, Armenia\\
$^{2}$Also at TU Wien, Vienna, Austria\\
% ...
\end{sloppypar}
\end{document}